\documentclass[twocolumn,pra,floatfix,amsmath,aps,amssymb,longbibliography,nofootinbib]{revtex4-2}
\usepackage{graphicx}
\usepackage{xcolor}
\usepackage{hyperref}
\usepackage{txfonts}
\usepackage{amsmath,amsfonts,amssymb,verbatim,float,mathtools}
\usepackage{physics}
\usepackage{soul}

\hypersetup{
    unicode=true, 
    plainpages=false,
    colorlinks=true,
    linkcolor=blue,
    citecolor=blue,
    filecolor=blue,
    urlcolor=blue
}
\usepackage{bbold}
\usepackage{amsmath,amsfonts,bm}
\usepackage{amssymb}
\usepackage{braket}

\usepackage{xifthen}

\newcommand{\inline}[1]{${#1}$}

\newcommand{\boldvec}[1]{\ensuremath{\boldsymbol{\mathbf{#1}}}}

\newcommand{\boldgreek}[1]{\ensuremath{{ \boldsymbol{\bm{#1}}}}}

\newcommand{\pol}{\hat{\boldvec e}}
\newcommand{\rv}{{\bf r}}
\newcommand{\Ev}{{\boldvec E}}

\newcommand{\Pv}{{\bf P}}
\newcommand{\Ec}{{\cal E}}

\newcommand{\Dc}{{\cal D}}

\newcommand{\kv}{{\bf k}}
\newcommand{\eo}{\epsilon_0}
\newcommand{\beq}{\begin{equation}}
\newcommand{\eeq}{\end{equation}}
\newcommand{\bea}{\begin{eqnarray}}
\newcommand{\eea}{\end{eqnarray}}
\newcommand{\BEQAL}{\begin{align}}
\newcommand{\EEQAL}{\end{align}}

\newcommand{\<}{\langle}
\renewcommand{\>}{\rangle}

\newcommand{\Pc}{{\cal P}}

\newcommand{\commentout}[1]{{}}

\newcommand{\br}{{\bf r}}

\newcommand{\cbE}{\boldsymbol{\mathbf{\cal E}}}

\newcommand{\spvec}[1]{\ensuremath{\boldvec{#1}}}

\newcommand{\G}{{\sf G}}

\newcommand{\stochx}{\ensuremath{r}}
\newcommand{\stochxv}{\spvec{\stochx}}

\newcommand{\psihat}[1]{\hat{\psi}_{#1}}
\newcommand{\psihatdag}[1]{\hat{\psi}_{#1}^{\dag}}

\newcommand{\eqcaption}[3][]{%
    \ifthenelse{\equal{#1}{true}}{%
        \ifthenelse{\isempty{#3}}{%
           \textcolor{red}{[}{#2}\textcolor{red}{]}
        }{%
            #2~[Eq.~\eqref{#3}]%
        }%
    }{%
        \textcolor{yellow}{[}{#2}\textcolor{yellow}{]}
    }%
}

\begin{document}

\title{Propagation of light in cold emitter ensembles with quantum position correlations due to static long-range dipolar interactions}
\author{G. J. Bean, N.\ D.\ Drummond, and J.\ Ruostekoski}
\affiliation{Department of Physics, Lancaster University, Lancaster LA1 4YB, United Kingdom}
\begin{abstract}
We analyze the scattering of light from dipolar emitters whose disordered positions exhibit correlations induced by static, long-range dipole-dipole interactions. The quantum-mechanical position correlations are calculated for zero temperature bosonic atoms or molecules using variational and diffusion quantum Monte Carlo methods. For stationary atoms in dense ensembles in the limit of low light intensity, the simulations yield solutions for the optical responses to all orders of position correlation functions that involve electronic ground and excited states.  We calculate how coherent and incoherent scattering, collective linewidths, line shifts, and eigenmodes, and disorder-induced excitation localization are influenced by the static interactions and the density. We find that dominantly repulsive static interactions in strongly confined oblate and prolate traps introduce short-range ordering among the dipoles which curtails large fluctuations in the light-mediated resonant dipole-dipole interactions. This typically results in an increase in coherent reflection and optical depth, accompanied by reduced incoherent scattering. The presence of static dipolar interactions permits the highly selective excitation of subradiant eigenmodes in dense clouds. This effect becomes even more pronounced in a prolate trap, where the resonances narrow below the natural linewidth. When the static dipolar interactions affect the optical transition frequencies, the ensemble exhibits inhomogeneous broadening due to the nonuniformly experienced static dipolar interactions that suppress cooperative effects, but we argue that, e.g., for Dy atoms such inhomogeneous broadening is negligible.
\end{abstract}
\date{\today} 
\maketitle

\section{Introduction}

Ultracold matter exhibiting static long-range dipole-dipole (DD) interactions~\cite{Baranov12} has in recent years witnessed significant breakthroughs in the preparation of atomic gases with large magnetic dipole moments, such as Cr~\cite{Griesmaier05}, Er~\cite{Aikawa12}, and Dy~\cite{Lu11}, as well as heteronuclear polar molecules~\cite{DeMarco19,Valtolina20,Duda23}. Long-range dipolar interactions can also be engineered using highly excited long-lived Rydberg states~\cite{Saffman10} and polar molecules can be used to influence atomic transitions~\cite{Guttridge23}. The DD interactions can give rise to exciting effects, including the formation of self-bound droplets and supersolids~\cite{Schmitt16,Baillie16,Bottcher19,Tanzi21,Norcia21,Bottcher2021,Schmidt22,Bland22}, the emergence of rotonic excitations~\cite{Santos03,Astrakharchik07,Chomaz18} and crystallization~\cite{Pupillo10,Kadau16,Baillie18}. 

Disordered media of resonant light emitters provide rich mesoscopic systems~\cite{Ishimaru1978,Lagendijk,vanRossum} where light can mediate strong interactions. Cold and dense atomic ensembles in the limit of low light intensity (LLI) represent systems of dipolar emitters where light can induce strong position-dependent correlations, even though each individual atom's response to a coherent light field behaves like a linear classical oscillator~\cite{Morice1995a,Ruostekoski1997a,Javanainen2014a}. The light-mediated resonant DD interactions are highly sensitive to atomic positions, leading to radiative excitations in each atom that are correlated to the positions of every other atom in the sample. This granularity of atoms, which determines their optical response, cannot be captured by standard electrodynamics of continuous media, where light-induced interactions between atoms are treated in an averaged sense~\cite{Javanainen2014a,JavanainenMFT,Javanainen17,Andreoli21}. In the limit where the atoms are considered stationary during imaging, the collective optical response of $N$ atoms is dependent upon the position correlation functions $\rho_j(\br_1,\ldots,\br_j)$, for $j=1,\ldots,N$, that are determined before the light interacts with the sample~\cite{Ruostekoski1997a}. When atomic positions are independent and uncorrelated, all these initial correlations factorize $\rho_j(\br_1,\ldots,\br_j)=\rho_1(\br_1)\ldots\rho_1(\br_j)$, and solving for the optical response involves calculating the radiative excitations of the atoms that can, nevertheless, still be correlated in a disordered system~\cite{Javanainen1999a,Lee16}. However, highly non-trivial position correlation functions between electronic ground-state atoms $\rho_j(\br_1,\ldots,\br_j)$ can exist due to quantum statistics and atomic interactions. Aside from the fundamental interest in cooperative responses in such systems, the scattered light then also directly conveys information about these correlations, serving as a diagnostic method.

Here we solve the collective optical responses of stationary dipolar emitters experiencing static long-range DD interparticle interactions and discuss experimental consequences of the findings. The emitters can represent atoms or molecules, provided that radiative coupling between molecular vibrational states can be neglected~\cite{Humphries23}. The positions of dipoles, determined in the absence of light at zero temperature, are sampled using quantum Monte Carlo methods~\cite{Ceperley_1980,Needs_2020} which approximately generate the position correlation functions $\rho_j(\br_1,\ldots,\br_j)$, encompassing all orders $j=1,\ldots,N$. We consider strongly confined traps that take the shape of oblate (pancake-shaped) and prolate (cigar-shaped) geometries, where the static DD interaction is dominantly repulsive. We focus on the interaction regimes that are sufficiently weak, such that the density distributions are not crystallized. The configurations of dipoles where the positions do not fluctuate independently dramatically influence the optical response in situations where the system is homogeneously broadened, as previously also demonstrated in simulations of light propagation using the Metropolis algorithm in the presence of non-trivial correlations $\rho_j(\br_1,\ldots,\br_j)$ for quantum degenerate ideal fermionic atoms in a one-dimensional optical waveguide~\cite{Ruostekoski_waveguide}. The effect of static DD interactions on light scattering has recently been studied in Ref.~\cite{Bassler23b} without including position correlations $\rho_j(\br_1,\ldots,\br_j)$ of the atoms in the analysis.

The strength of recurrent scattering in cooperative optical responses depends on the interparticle separation in units of the resonance wavenumber of light. Increased density leads to more pronounced deviations from the single-atom Lorentzian lineshape. To characterize the impact of the static DD interactions, we systematically vary the strength of these interactions while approximately maintaining a constant peak density.
Our observations reveal that, in both prolate and oblate traps, repulsive static interactions lead to short-range ordering among the dipoles, which, in turn, curtails fluctuations in the light-induced resonant DD interactions, particularly as these interactions become very large at short interparticle separations. This phenomenon is identified in increased coherent reflection and optical depth that are accompanied by reduced incoherent scattering. 
In an oblate trap, coherent transmission and reflection resonances narrow at low density but broaden at high density. In a prolate trap, the scattered light resonance can narrow even below the natural linewidth.
The collective resonance shifts are substantially smaller than those predicted by the collective Lamb shift in continuous medium electrodynamics, underscoring the violation of the standard optics in the system~\cite{Javanainen2014a,JavanainenMFT,Jenkins_thermshift,Dalibard_slab}.
Intriguingly, especially in a prolate trap, the presence of the static DD interactions enables better targeted excitation of subradiant eigenmodes at high densities, although the linewidth of the eigenmodes can exhibit significant variation  between different realizations due to the position fluctuations in the ensemble.
Additionally, we find that the main difference between the two-level and the isotropic $J=0\rightarrow J'=1$ transitions is the emergence of resonances where the orientations of the dipoles in an oblate trap, in the latter case, are parallel to the normal of the trap plane. 

In disordered dense samples, the excitations can become very localized leading to high concentrations of energy. We analyze the excitations peaks and their widths in individual realizations and find that the static interactions enhance the localization peak strengths, providing control and manipulation of optical fields on a subwavelength scale.
We also study the effect of the static DD interactions on optical transition frequencies. When these interactions start substantially influencing the resonances, atoms become inhomogeneously broadened due to the effect of the atoms experiencing the static DD interactions nonuniformly. This broadens the resonances and reduces cooperativity. However, the simulations indicate that such effects for Dy atoms are negligible.

The article is organized as follows: Section~\ref{sec:theory} provides the theoretical background for describing static interactions and light-matter coupling, while Sec.~\ref{sec:stoch} gives an overview of stochastic electrodynamics and the quantum Monte Carlo methods. The optical responses in an oblate trap, when the strength of the static interactions is varied for a constant peak density, are in Sec.~\ref{sec:dipolaroblate}, the differences between two-level and isotropic transitions in Sec.~\ref{sec:iso}, inhomogeneous broadening in Sec.~\ref{sec:Zeeman}, and the results in a prolate trap in Sec.~\ref{sec:dipolarprolate}. The localization of excitations is presented in Sec.~\ref{sec:spots} and the optical responses for variable densities in Sec.~\ref{sec:density} before concluding remarks in Sec.~\ref{sec:conc}.

\section{Theoretical background}
\label{sec:theory}

\subsection{Static long-range dipole-dipole interactions}
\label{sec:atoms}

\begin{figure}
    \centering
    \includegraphics[width=0.99\columnwidth]{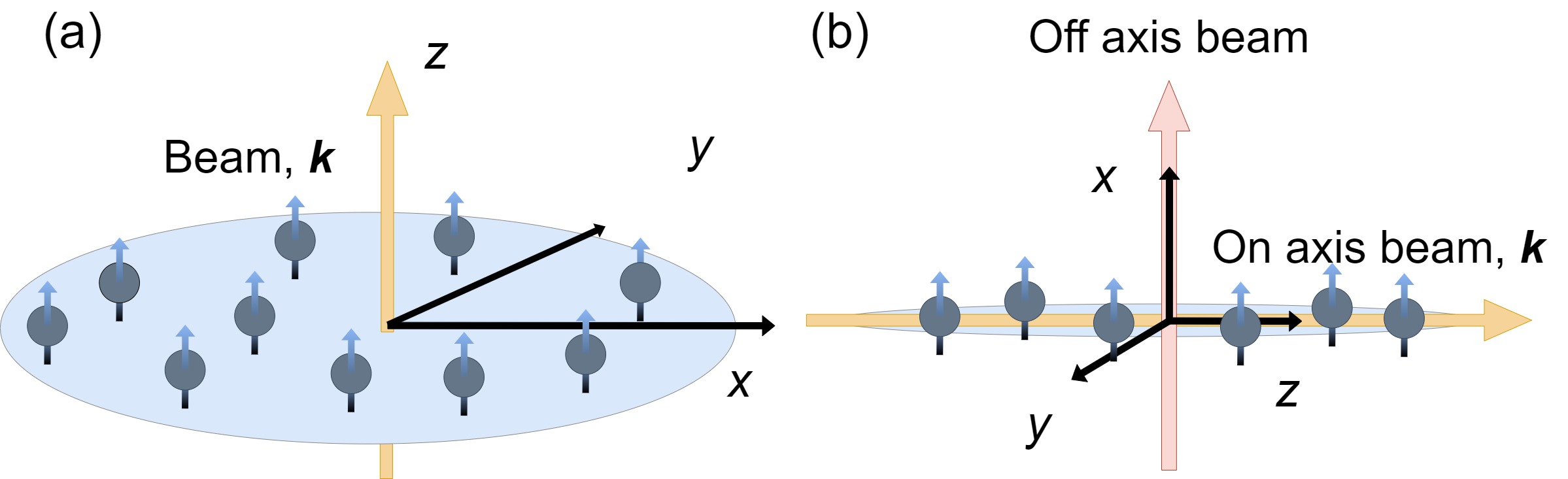}
    \caption{Schematic illustration of the trapping geometries illuminated by coherent light. (a) An oblate (pancake-shaped) trap in the $xy$ plane ($\ell_z\ll \ell_x=\ell_y$), with the static atomic dipoles aligned perpendicular to the plane along the light propagation direction. An elongated prolate (cigar-shaped) trap ($\ell_z\gg \ell_x=\ell_y$) with the static dipoles aligned perpendicular to the long axis.
    The incident light in (b) propagates along either the long or short axis of the trap.
        }
    \label{fig:trap}
\end{figure}

We assume particles, henceforth referred to as atoms, with the mass $M$ that are harmonically trapped and experience long-range DD interactions
\begin{equation}
    \hat{H} = \sum_{j=1}^N \left[-\frac{\hbar^2 \boldvec{\nabla}_j^2}{2M}  + V_{\rm{trap}}(\boldvec{r}_j)\right] + \sum_{j=1}^N \sum_{l> j}^N [V_{\rm{dd}}(\boldvec{r}_{lj}) +V_{\rm{sr}}(\boldvec{r}_{lj})],
    \label{eq: Hamiltonian}
    \end{equation}   
where the trapping potential with the frequency $\omega_j$,
\begin{align}
    V_{\rm{trap}}(\boldvec{r}_j) = \frac{1}{2}M (\omega_x^2 x_j^2 + \omega_y^2 y_j^2+ \omega_z^2 z_j^2 ),
    \label{eq; Trap Potnetial}
\end{align}
is associated with the characteristic length scale $ \ell_j = {[\hbar /( M\omega_j )]}^{1/2} $. 
We consider the atoms confined in an oblate or prolate trap (Fig.~\ref{fig:trap}) with all the static (e.g.\ magnetic) dipoles $\boldgreek{ \mu }_j $ oriented in the same direction.
The static DD interaction potential $V_{\rm{dd}}$ between the atoms, which is independent of the light-induced radiative optical DD coupling between the atoms, is then given by
\begin{equation}
V_{\rm{dd}}(\boldvec{r}_{\ell j})=-\frac{C\mu^2}{4 \pi |\boldvec{r}_{\ell j}|^3} [3 (\hat{\boldgreek{\mu}}\cdot \hat{\boldvec{r}}_{\ell j})^2 - 1] - \frac{2C\mu^2}{3} \delta(\boldvec{r}_{\ell j}),
\end{equation} 
where $\boldvec{r}_{\ell j}= \boldvec{r}_j - \boldvec{r}_{\ell}$ is the vector joining the atoms $j$ and $\ell$ and $\hat{{\bf r}}_{\ell j}={\bf r}_{\ell j}/|{\bf r}_{\ell j}|$. For magnetic dipoles, we have $C=\mu _0$.
The DD interaction can be characterized by the interaction length 
\begin{equation}
    R_{\rm{dip}} = \frac{M C \mu^2 }{4 \pi \, \hbar^2}.
    \label{eq:Rdip}
\end{equation}
The DD interaction diverges at small interatomic separations and, following Ref.~\cite{Bottcher19}, we introduce an additional repulsive short-range interaction potential, similar to the Lennard-Jones potential,
\begin{equation}
V_{\rm{sr}}(\boldvec{r}_{\ell j}) = -\frac{c_6}{|\boldvec{r}_{\ell j}|^6} + \frac{c_{12}}{|\boldvec{r}_{\ell j}|^{12}} .
\end{equation}
We maintain in the simulations fixed ratios $c_6= 0.0271 R_{\rm{dip}}^6 E_{\rm{dip}}$ and $c_{12}= 4.47\times 10^{-4} R_{\rm{dip}}^{12} E_{\rm{dip}}$ [$E_{\rm dip} = \hbar^2/(2M R_{\rm dip}^2)$] where $V_{\rm{dd}}$ is dominantly repulsive.

\subsection{Collective atom-light interactions}
\label{sec:collective}
We solve the optical response of the atoms in the limit of LLI where the individual atoms respond to light linearly and behave as a set of coupled, driven harmonic oscillators.  In this section, we first consider the commonly employed basic formalism~\cite{Ruostekoski23} for the coupled dynamics between the atoms and light for a fixed set of atomic positions $\{\textbf{r}_1,\textbf{r}_2,\ldots, \textbf{r}_N\}$. When the positions fluctuate, these equations are then utilized in stochastic electrodynamics simulations for a disordered system in Sec.~\ref{sec:sed}. 

The atoms are illuminated by a near monochromatic Gaussian incident light field with the positive frequency component of the amplitude $\boldvec{\Ec}(\rv)$, the wavevector $\kv$, and frequency $\omega=kc=2\pi c/\lambda$. The incident field drives a two-level or a $\ket{J=0, m_J=0}\rightarrow\ket{J' = 1, m_{J'}=\mu}$ atomic transition and is detuned
$\Delta_\mu^{(j)} = \omega - \omega_\mu^{(j)}$ from the resonance $\omega_\mu^{(j)}$ of the level $\mu$ in the atom $j$ that may vary between the atoms due to the interactions.
The system dynamics is described by the positive frequency components of the atomic polarization amplitudes $\boldvec{\Pc}$ of the dipole matrix elements
$
  \boldvec{d}_j  =  \Dc \boldvec{\Pc}^{(j)} 
$.
All observables are expressed in terms of slowly varying field amplitudes and atomic variables by factoring out the rapidly oscillating dominant frequency of the driving light
$     \boldvec{\Ec} e^{-i\omega t} \rightarrow \boldvec{\Ec} $, $      \boldvec{\Pc}^{(j)}  e^{-i\omega t} \rightarrow  \boldvec{\Pc}^{(j)} $~\cite{Ruostekoski1997a}.

The positive frequency component of the scattered field $\boldvec{E}_{s}^{(\ell)}$ from the atom $\ell$ at $\rv_\ell$  is given by
\begin{equation}
    \eo   \boldvec{E}_{s} ^{(\ell)} (\boldvec{r}) =  {\sf G}(\boldvec{r} - \boldvec{r}_\ell) \boldvec{d}_\ell,
\label{eq: Scattered Field kernel}
\end{equation}
where ${\sf G}(\rv)$ is the monochromatic dipole radiation kernel \cite{Jackson} describing the radiated field at point $\boldvec{r}$ due to an oscillating dipole at the origin with the dipole ${\boldvec d}$
\begin{align}\label{Gdef}
\mathsf{G}(\mathbf{r})\mathbf{d}&=-\frac{\mathbf{d}\delta(\mathbf{r})}{3}+\frac{k^3}{4\pi}\Bigg\{\left(\hat{\mathbf{r}}\times\mathbf{d}\right)\times\hat{\mathbf{r}}\frac{e^{ikr}}{kr}\nonumber\\
&\phantom{==}-\left[3\hat{\mathbf{r}}\left(\hat{\mathbf{r}}\cdot\mathbf{d}\right)-\mathbf{d}\right]\left[\frac{i}{(kr)^2}-\frac{1}{(kr)^3}\right]e^{ikr}\Bigg\},
\end{align}
where $r=|\rv|$ and $\boldvec{\hat{r}}=\rv/r$.

The positive frequency component of the total external field driving the atom $j$ is given by the incident field plus the scattered fields from all the other atoms
\begin{equation}
 \Ev_{\rm ext}(\stochxv_j) = \boldvec{\Ec}(\stochxv_j)+\sum_{\ell\neq j}
 \Ev^{(\ell)}_s(\stochxv_j).  \label{eq: External Field}
\end{equation}
Thus, the dipole amplitude of each atom depends on the fields from all other atoms in the system, 
leading to a set of coupled equations for $N$ atoms with a configuration of positions $\{\textbf{r}_1,\textbf{r}_2,\ldots, \textbf{r}_N\}$.
In the LLI limit of the coherently driven system, these equations determine the polarization of each atom 
and the population of the excited state can be neglected \cite{Javanainen1999a}. The dynamics then corresponds to a damped classical linear oscillator amplitude driven by ${\boldvec{E}}_{\rm{ext}}$
\begin{align}
  \frac{d {\Pc}_\mu^{(j)}}{ dt}  &= (i {\Delta}_\mu^{(j)}-\gamma) {\Pc}_\mu^{(j)}
  + i\frac{\Dc}{\hbar }\,\pol_\mu^\ast\cdot {\boldvec{E}}_{\rm{ext}}(\stochxv_j)\nonumber\\
  & = (i\Delta_\mu^{(j)}-\gamma) {\Pc}_\mu^{(j)} + {i\frac{\Dc}{\hbar } \pol_\mu^\ast\cdot {\boldvec \Ec}(\stochxv_j)}  +i\xi \sum_{\nu,\ell\neq j}  {\sf{G}}_{\mu\nu}^{(j\ell)} {\Pc}_\nu^{(\ell)}\, ,
  \label{eq:poldynamics}
\end{align}
where $ {\sf{G}}_{\mu\nu}^{(j\ell)} = \pol_\mu^\ast \cdot   {\sf{G}}(\rv_j-\rv_\ell) \pol_\nu$ and  the single atom linewidth $\gamma$
\begin{equation}
    \gamma = \frac{\Dc^2 k^3}{6 \pi \hbar \epsilon_0}, \quad \xi  = \frac{6 \pi \gamma}{k^3} = \frac{\Dc^2}{\hbar\epsilon_0}.
    \label{eq: Gamma}
\end{equation}
We solve Eq.~\eqref{eq:poldynamics} in the steady state, assuming that the time scale $1/\gamma$ is short compared with the time scale of any center-of-mass motion of the atoms.
The dynamics of the atomic polarizations can be compactly represented in the matrix form~\cite{Lee16}
\begin{equation}
             \dot{\boldvec{b}} =  i (\mathcal{H}+\delta\mathcal{H})\boldvec{b} + \boldvec{f}, 
             \label{eq: Matrix Equation}
\end{equation}
where 
$b_{3j -1 + \nu} = \Pc_\nu^{(j)}$, 
$
     f_{3j -1 + \nu} = i {\Dc} \pol_\nu^\ast \cdot  {\boldvec \Ec}({{\stochxv}}_j)/\hbar
$.
The off-diagonal elements of the matrix $\mathcal{H}$ describe the light-mediated interactions between the atoms and are given by
\begin{align}
\mathcal{H} _{3j -1 + \mu,3\ell -1 + \nu}= \xi  {\sf{G}}_{\mu\nu}^{(j\ell)}, \quad j\neq\ell .
    \end{align}
The diagonal elements are $i\gamma$, while the diagonal matrix $\delta\mathcal{H}$ contains the detunings $\Delta^{(j)}_\mu$.
Finally, the total field can be calculated from the sum of the incident light field and scattered field from all the atoms
\begin{equation}
  \Ev(\boldvec{r}) =\boldvec{\Ec}(\boldvec{r})+   \Ev_s(\boldvec{r})=  \boldvec{\Ec}(\boldvec{r})+\sum_{\ell}
  \Ev^{(\ell)}_s(\boldvec{r}).
  \label{eq: Total Field}
\end{equation}

We describe the response in terms of the collective excitation eigenmodes $\boldvec{v}_j$ which correspond to the eigenvectors of ${\cal H}$~\cite{Rusek96,JenkinsLongPRB,Jenkins_long16}, with eigenvalues $\delta_j+i\nu_j$. These represent
the collective line shift $\delta_j=\omega_0-\omega_j$ from the single atom resonance $\omega_0$ (here assumed degenerate) and  the collective linewidth $\nu_j$. 
When $\nu_j<\gamma$, excitations are called subradiant, whereas $\nu_j>\gamma$, they are super-radiant~\cite{Dicke54}. 
We determine the occupation of the eigenmode ${\bf b}$ from
\cite{Facchinetti16,Ruostekoski23}
\begin{align}
    L_j = \frac{| \boldvec{v}_j^T \boldvec{b}|^2}{\sum_\ell | \boldvec{v}_\ell^T \boldvec{b}|^2}.
    \label{eq: occupation}   
\end{align}

\section{Stochastic simulations}
\label{sec:stoch}

\subsection{Stochastic electrodynamics}
\label{sec:sed}

We have discussed how the coupled dynamics between atoms at a fixed set of positions $\{\textbf{r}_1,\textbf{r}_2,\ldots, \textbf{r}_N\}$ and coherent light in the limit of LLI can be solved using Eqs.~\eqref{eq:poldynamics}, \eqref{eq: Total Field}, and~\eqref{eq: Scattered Field kernel}. When atomic positions fluctuate, the system becomes a disordered medium where light can establish correlations among the atoms in cold and dense ensembles that significantly modify the optical response, even within the classical regime~\cite{Javanainen2014a}. Here we briefly review how position fluctuations are conveniently described by employing second-quantized atomic field operators for the ground and excited states $\psihat{g,e}(\mathbf{r})$~\cite{Ruostekoski1997a}, where the label $e$ implicitly incorporates the Zeeman levels of the $J=0\rightarrow J'=1$ transition. The positive frequency component of the atomic polarization density operator then takes the form
$\hat{\mathbf{P}}(\mathbf{r})=  \mathop{\psihatdag{g}(\mathbf{r})}\textbf{d}_{ge}\mathop{\psihat{e}(\mathbf{r})}$, with $\textbf{d}_{ge}$ representing the dipole matrix element between the electronic ground and excited levels. We adopt the convention that the repeated level index $e$ is summed over. The expectation value of $\hat{\mathbf{P}}(\mathbf{r})$ with respect to the fixed atomic positions $\{\textbf{r}_1,\textbf{r}_2,\ldots, \textbf{r}_N\}$ is related to $\boldvec{\Pc}^{(j)}$ in Eq.~\eqref{eq:poldynamics} via~\cite{Lee16}
\beq
\label{eq:pabExp} 
\langle \hat{\mathbf{P}}(\mathbf{r}) \rangle_{\{\mathbf{r}_1,\dots,\mathbf{r}_N\}}
 = 
\sum_{j=1}^N {\cal D} \boldvec{\Pc}^{(j)} \mathop{\delta(\mathbf{r}-\mathbf{r}_j)} .
\eeq
Solving for the light field from Eqs.~\eqref{eq: Total Field} and~\eqref{eq: Scattered Field kernel} in the general case of fluctuating positions requires an integral of the radiation kernel and the expectation value $\<\hat{\textbf{P}}(\mathbf{r})\>$. Multiply scattered light establishes correlations between the atoms, leading to a hierarchy of equations of motion for the correlation functions of atomic density and polarization~\cite{Morice1995a,Ruostekoski1997a}. Specifically, in the limit of LLI, we introduce normally ordered correlation functions 
\bea
\rho_1(\br_1) &=& \< {\hat\psi^\dagger_g(\br_1)\hat\psi_g(\br_1)} \>,\nonumber\\
\rho_2(\br_1,\br_2) &=& \<{\hat\psi_g^\dagger(\br_1)\hat\psi_g^\dagger(\br_2)\hat\psi_g(\br_2)\hat\psi_g(\br_1)}\> ,\nonumber\\
&&\ldots \\
\Pv_2(\br_1;\br_2) &=&\langle{\hat\psi_g^\dagger(\br_1)\hat\Pv(\br_2)\hat\psi_g(\br_1)}\rangle,\nonumber\\
\Pv_3(\br_1,\br_2;\br_3) &=&\langle{\hat\psi_g^\dagger(\br_1)\hat\psi_g^\dagger(\br_2)\hat\Pv(\br_3)\hat\psi_g(\br_2)\hat\psi_g(\br_1)}\rangle,\nonumber\\
&&\ldots 
\eea
The LLI is indicated by including at most one $\Pv$ factor in each expression and by the fact that the ground-state correlations $\rho_k$ are not perturbed by light. Then $\Pv_k$  ($k=1,\ldots,N$) satisfy the dynamics for degenerate, uniform transition frequencies ($\Delta^{(j)}_\mu=\Delta$)~\cite{Ruostekoski1997a}
\begin{align}
&\dot{\bf P}_p(\br_1,\ldots,\br_{p-1};\br_p) =\nonumber\\
&(i\Delta-\gamma){\bf P}_p(\br_1,\ldots,\br_{p-1};\br_p)
+i\xi{\cbE}(\br_p)\rho_p(\br_1,\ldots,\br_p )\nonumber\\
&+i\xi\sum_{q\ne p}\G(\br_p-\br_q){\bf P}_p(\br_1,\ldots,\br_{q-1},\br_{q+1},\dots,\br_p;\br_q)\nonumber\\
&+i\xi\int d^3r_{p+1}\G(\br_p-\br_{p+1}){\bf P}_{p+1}(\br_1,\ldots,\br_p;\br_{p+1}).
\label{EXACT}
\end{align}
The ground-state correlations $\rho_p$ exist before the light enters the sample. ${\bf P}_p(\br_1,\ldots,\br_{p-1};\br_p) $ represents the correlation function for ground state atoms at $\br_1,\ldots,\br_{p-1}$, given that there is polarization at $\br_p$. The strong coupling emerges from the third term on the right-hand side, representing repeated exchanges of a photon between the same atoms, giving rise to recurrent scattering processes.

The challenge of solving Eq.~\eqref{EXACT} arises from the complex hierarchy of $N$ equations for $N$ atoms. Standard optics of continuous media is recovered by factorizing all correlations~\cite{Ruostekoski1997a,Javanainen2014a,JavanainenMFT,Javanainen17}, i.e., disregarding any light-established correlations between the atoms  $\Pv_k(\br_1,\ldots,\br_{k-1};\br_k)= \rho^{k-1}\Pv(\br_k)$ and considering an initial classical uncorrelated distribution of atoms  $\rho_n=\rho_1^n$. At very low densities, $\rho/k^3\ll 1$, the hierarchy can be truncated, e.g., by including only correlations between pairs of atoms, resulting in a closed equation for $\Pv_2$. In this case, the optical response also depends on the ground-state pair correlation function $\rho_2(\br_1,\br_2)$. The behavior of a resonance linewidth is strongly geometry-dependent but can be integrated for semi-infinite quantum degenerate ensembles, already exhibiting through $\rho_2(\br_1,\br_2)$ linewidth broadening for bosonic atoms~\cite{Morice1995a}, narrowing for fermionic atoms~\cite{Ruostekoski1999a}, and the effects of quasiparticle pairing~\cite{bcs1,BerhaneKennedybcs}.

An alternative approach to solving the LLI response of Eq.~\eqref{EXACT} is through a stochastic Langevin-type method. This method considers the dynamics for excitations for a given set of fixed atomic positions $\{\textbf{r}_1,\textbf{r}_2,\ldots, \textbf{r}_N\}$ while treating the atomic positions as stochastic variables. In each stochastic realization, the positions of the atoms are sampled to match the proper correlation functions $\rho_p(\br_1,\ldots,\br_p)$, for $p=1,\ldots,N$ in the absence of light. Subsequently, the excitations and the scattered light for the specific set of atomic positions $\{\textbf{r}_1,\textbf{r}_2,\ldots, \textbf{r}_N\}$ are solved using Eqs.~\eqref{eq:poldynamics} and~\eqref{eq: Total Field}. The expectation values are obtained by
ensemble-averaging over many realizations. This stochastic classical-electrodynamics approach of coherent scattering can be formally shown to converge to an exact solution for stationary atoms at arbitrary densities by reproducing the correct hierarchy of correlation functions for both 1D scalar theory~\cite{Javanainen1999a} and 3D vector-electrodynamics with a single electronic ground level~\cite{Lee16}. 
The probability distribution required to sample the normally-ordered ground-state atom correlation functions $\rho_p(\br_1,\ldots,\br_p)$ is determined by  the many-body wave function $P(\rv_1,\rv_2, \ldots, \rv_N) = |\Psi(\rv_1,\rv_2, \ldots, \rv_N) |^2$~\cite{Javanainen1999a}. 

For uncorrelated atoms, where $\rho_n=\rho_1^n$, each atom can be sampled independently. Such scenarios include an ideal Bose-Einstein condensate, a Mott-insulator ground state in an optical lattice, and classical atoms.
In the case of atoms obeying Fermi-Dirac statistics in a 1D waveguide, the optical response has been solved by stochastically sampling the atomic positions~\cite{Ruostekoski_waveguide}. In Ref.~\cite{Ruostekoski_waveguide}, the atomic positions  are defined by the Slater determinant of the many-body wave function and sampled using the Metropolis algorithm. Additionally, to account for the influence of a hard-core radius of the classical atom distribution, simulations have been conducted in light scattering by excluding a hard-sphere volume around the already existing atoms in the sampling~\cite{Wang20}.

Here we calculate the optical response of atoms that exhibit quantum-mechanical position correlations due to static long-range DD interactions that are present before the light enters the sample. We consider an atomic gas in a strongly confined oblate or prolate trap at zero temperature where the DD interactions are dominantly repulsive. These repulsive static DD interactions inhibit small interatomic separations, therefore suppressing light-mediated resonant DD interactions through increased interatomic spacing. While this effect draws some parallels to Fermi-Dirac statistics~\cite{Ruostekoski1999a,Ruostekoski_waveguide}, the DD repulsion between the atoms quickly becomes more substantial and is more long-range, which obscures quantum-statistical characteristics of the atoms in the optical response. Although the simulations are performed with bosonic atoms, any signatures of Bose-Einstein statistics in the scattering are rapidly washed out by the DD interactions. The ensemble can represent a strongly correlated quantum many-body system with long-range static DD interactions where the position of every atom affects the positions of every other atom. 
The correlated atomic positions in the calculations are therefore sampled using quantum Monte Carlo methods, as described in the following section.

\subsection{Quantum Monte Carlo sampling}
\label{sec:qmc}

Solving the stochastic electrodynamics simulations of the optical response, described in Sec.~\ref{sec:sed}, requires sampling  the atomic positions in a dipolar gas that are
distributed according to the square modulus of the ground-state wave function at zero temperature. We have used the variational and diffusion quantum Monte
Carlo (VMC and DMC) methods to find approximate ground-state wave functions and stochastic realizations of atomic positions 
to solve for the radiative excitations in Eq.~\eqref{eq:poldynamics} and scattered light in Eqs.~\eqref{eq: Total Field} and~\eqref{eq: Scattered Field kernel} for each given set of fixed positions. 

In the VMC method, quantum mechanical expectation values are evaluated
using a trial many-body wave function that is an explicit function of
interparticle distances.  The Metropolis algorithm is used to sample position realizations from
the square modulus of the wave function for Eq.~\eqref{eq:poldynamics}, estimators of the energy
and other expectation values.  Free parameters in the trial wave function
are obtained by minimizing the energy expectation value~\cite{Umrigar_2007}.

In the DMC method, drift, diffusion and branching/dying processes
governed by the many-body Schr\"{o}dinger equation in imaginary time
are simulated in order to project out the ground-state component of a
trial wave function~\cite{Ceperley_1980}.  
The product of the trial wave function and the solution of the imaginary-time Schr\"odinger equation is represented as the ensemble average of a discrete population of walkers (weighted delta functions), and the Green's function for the imaginary-time Schr\"odinger equation is treated as a transition-probability density for walkers over discrete time steps.
After equilibration, the walkers are distributed as the product of the trial wave function and its ground-state components, and walkers provide atomic configurations for Eq.~\eqref{eq:poldynamics} and ensemble-averaging expectation values.
For a bosonic gas the
ground-state wave function is nodeless and hence there are in
principle no uncontrolled approximations in the DMC method.  In
practice, the trial wave function must have a sufficiently large
overlap with the ground-state wave function that the algorithm can be
equilibrated on a tractable timescale.  

The DMC algorithm generates atomic configurations distributed as the 
product of the trial wave function and its ground-state component. The error in the distribution of atomic 
configurations is therefore first order in the error in the trial wave 
function. Our trial wave functions were of form
\begin{equation} \Psi({\bf r}_1,\ldots,{\bf
  r}_N)=e^{J({\bf r}_1,\ldots,{\bf r}_N)} \prod_{i=1}^N \phi_i({\bf
    r}_i), 
\end{equation} 
where the orbitals $\phi({\bf r}_i)$ were
Gaussian functions that could be (initially) centered in the middle of the
trap obtained by a brute-force
minimization of the potential energy.  The position and the width of
each Gaussian orbital in each Cartesian direction were treated as
optimizable parameters. The Jastrow exponent $J$ contained polynomial
two- and three-body terms~\cite{Drummond_2004}.  In addition, we used
two-body Jastrow terms designed to impose physically appropriate
behavior on the wave function at short range.  If the interaction
between atoms were of the isotropic form $d^2/r^3$, where $d$ is a
constant, then the Jastrow exponent would have to contain a pairwise
term $u_d(r)=-\sqrt{8d^2\mu/r}$ to ensure that the local-energy
contribution from a coalescing pair of atoms,
$E_{L2}=-\nabla^2\Psi/(2\mu \Psi)+d^2/r^3$, diverges more slowly
than $1/r^3$ at short range, where $\mu$ is the reduced mass of a pair
of atoms.  This negative, divergent two-body Jastrow term $u_d(r)$ has
the effect of making the wave function go rapidly to zero at
coalescence points, and we have continued to use this term even in the
presence of anisotropic dipolar interactions.  Similarly, two-body
Jastrow terms going as $-1/r^5$ were used to impose the exact
short-range behavior on the wave function in the presence of a
repulsive $r^{-12}$ interaction in our calculations using
Lennard-Jones potentials.
All our VMC and DMC calculations were performed using the
\textsc{casino} software~\cite{Needs_2020}.

\subsection{Scattered light}
\label{sec:light}

Since we consider here the LLI limit where each atom responds to light classically, the scattered light from the atomic ensemble is directly evaluated from the stochastic simulation data by ensemble-averaging over many realizations~\cite{Bettles20}.
We then have for the transmission $T$ and reflection $R$ 
\begin{align}
    T &=  \frac{ \int_A  \braket{{(    \boldvec{\Ec}^\ast + \Ev^\ast_s  )}\cdot (  \boldvec{\Ec} + \Ev_{s}  )} d \Omega} { \int_A  \braket{{\boldvec{\Ec}}^\ast  \cdot {\boldvec{\Ec}} }  d \Omega },
    \label{eq: Transmission}\\
R & =  \frac{ \int_{A'}  \braket{{\Ev^\ast_{s}  \cdot \Ev_{s}  }} d \Omega} { \int_{A'} \braket{{\boldvec{\Ec}}^\ast  \cdot {\boldvec{\Ec}} }  d \Omega },
\label{eq: Reflection}
\end{align}
where $\Ev$ and $\Ev^\ast$ denote the positive and negative frequency components, respectively, and the intensity $ I = 2 c  \epsilon_0 \braket{{\Ev^\ast  \cdot \Ev   }} $.
The lens regions centered on the optical axis in the forward and backward directions are $A$ and $A'$, respectively.

The scattered field consists of a mean field $\braket{\Ev_{s}}$ and fluctuations $ \delta \Ev_{s}=\Ev_{s}-\braket{\Ev_{s}}$~\cite{meystre1998}. 
The coherent scattering originates from the mean field part, while the incoherent contribution is due to the position fluctuations of the atoms.
Then
\begin{equation}
 \braket{\Ev^\ast  \cdot \Ev  } = {\boldvec{\Ec}}^\ast \cdot  {\boldvec{\Ec}} +  {\boldvec{\Ec}}^\ast \cdot \braket{\Ev_{s}} + \braket{\Ev^\ast_{s}} \cdot  {\boldvec{\Ec}} + \braket{\Ev^\ast_{s}  \cdot \Ev_{s}  } ,
 \label{eq:Total Intensity Terms}
\end{equation}
where the last term
\begin{equation}
\braket{\Ev^\ast_{s}  \cdot \Ev_{s}  } =  \braket{\Ev^\ast_{s}} \cdot \braket{\Ev_{s}} + \braket{\delta \Ev^\ast_{s} \cdot \delta \Ev_{s}} 
\label{eq:Es expectation}
\end{equation}
includes both the coherent (the first term) and incoherent (the second term) contributions. The coherent scattering is mostly directed in a narrow cone along the incident field direction, while the incoherently scattered light is less directed, and we vary the numerical aperture (NA) of the lenses in these two cases.
The coherent transmission is frequently expressed as the optical depth $   {\rm{OD_{coh}}}=-\ln{T_{\rm{coh}}}$.

\section{Optical response}
\label{sec:optical}

\begin{figure}
    \centering
\includegraphics[width=0.49\columnwidth]{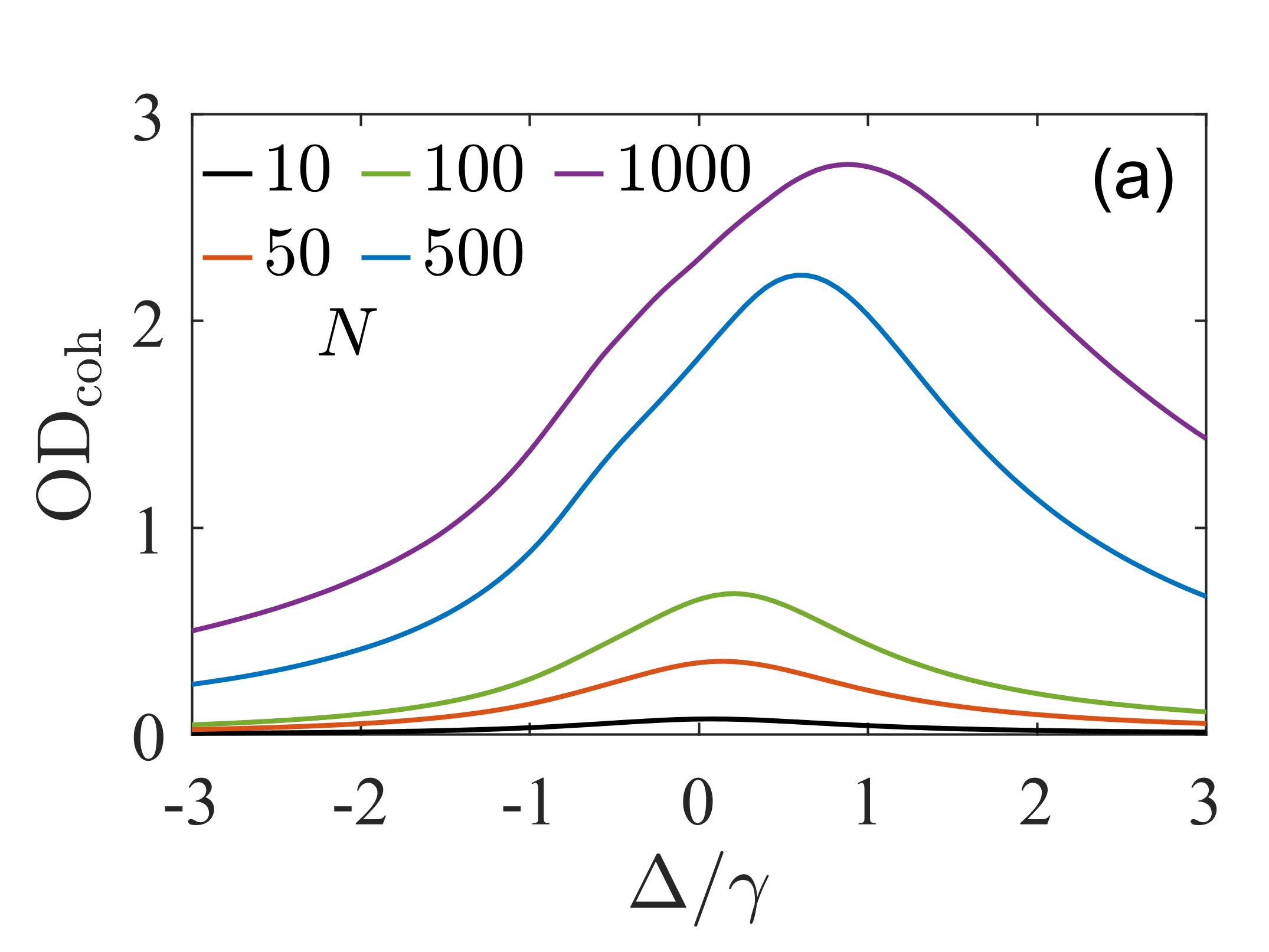}
\includegraphics[width=0.49\columnwidth]{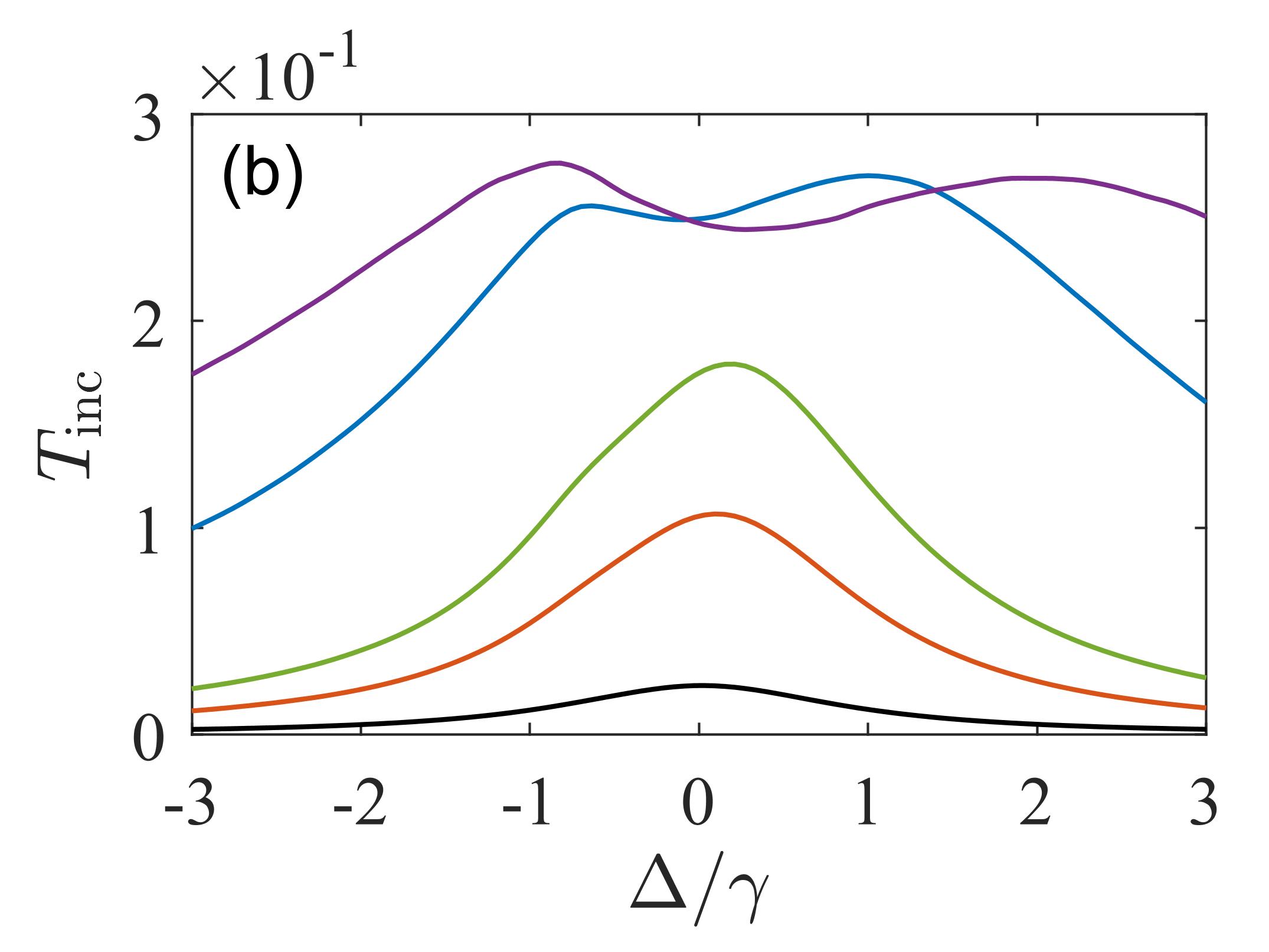}
\includegraphics[width=0.49\columnwidth]{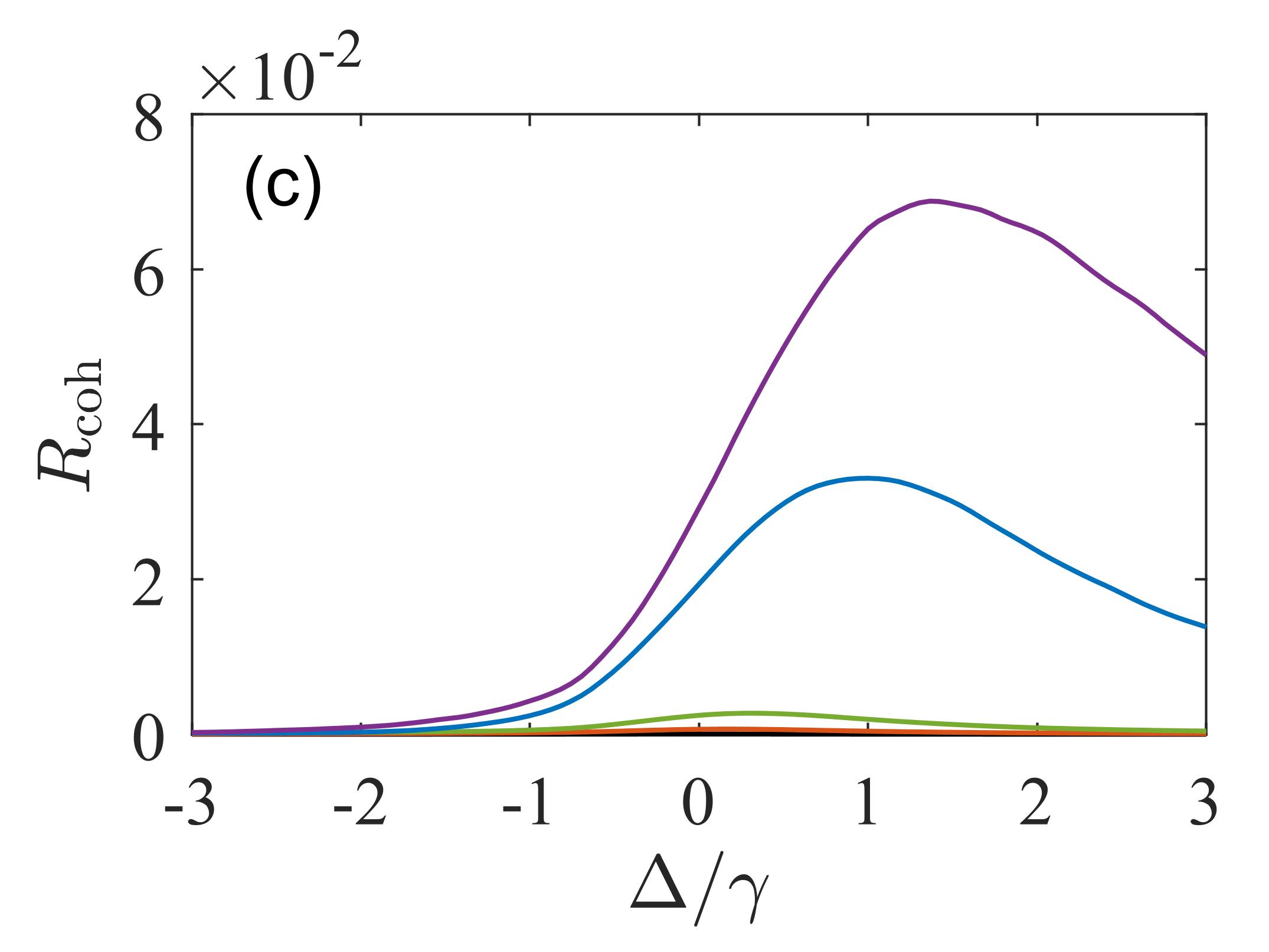}
\includegraphics[width=0.49\columnwidth]{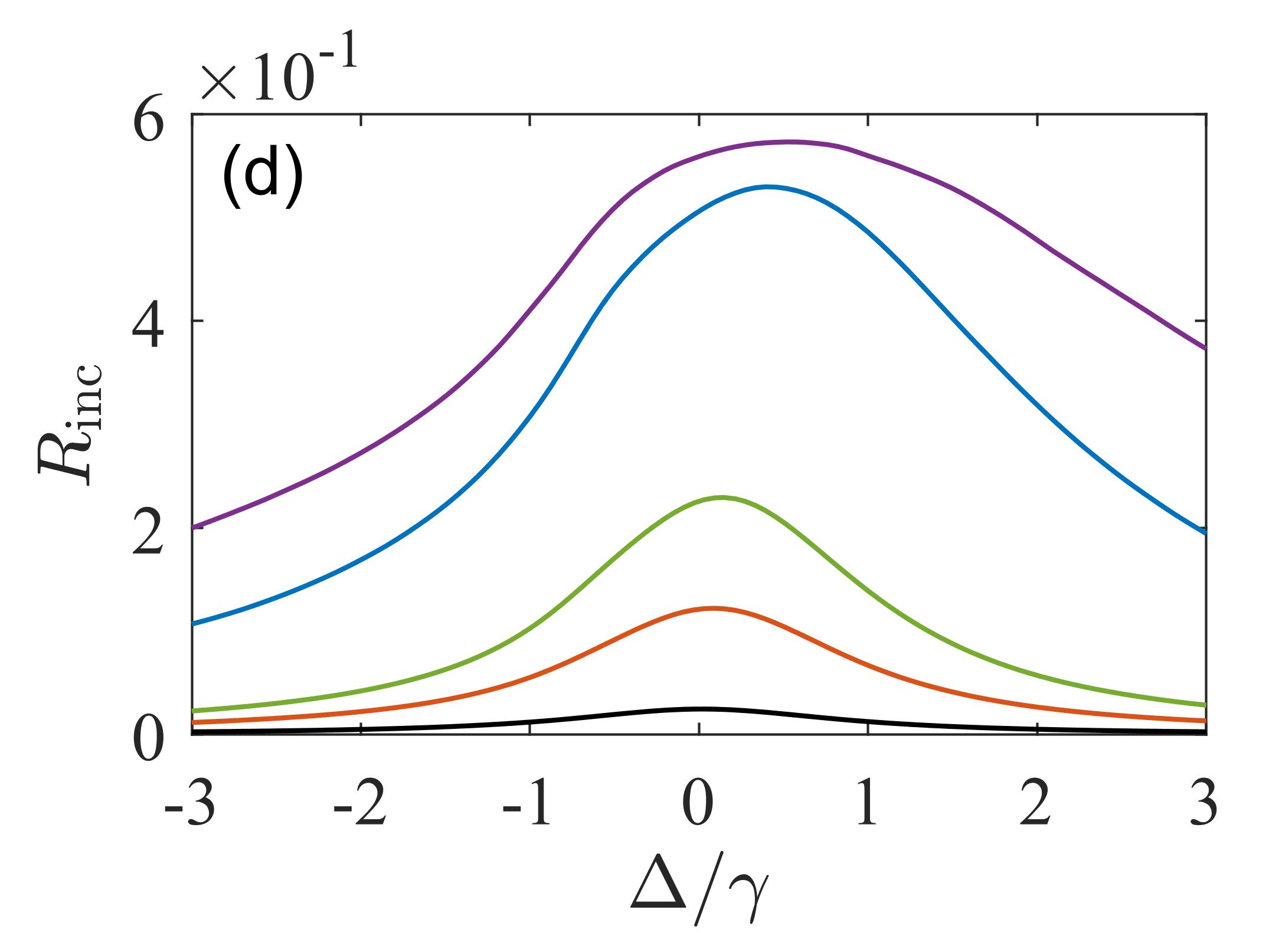}
\includegraphics[width=0.49\columnwidth]{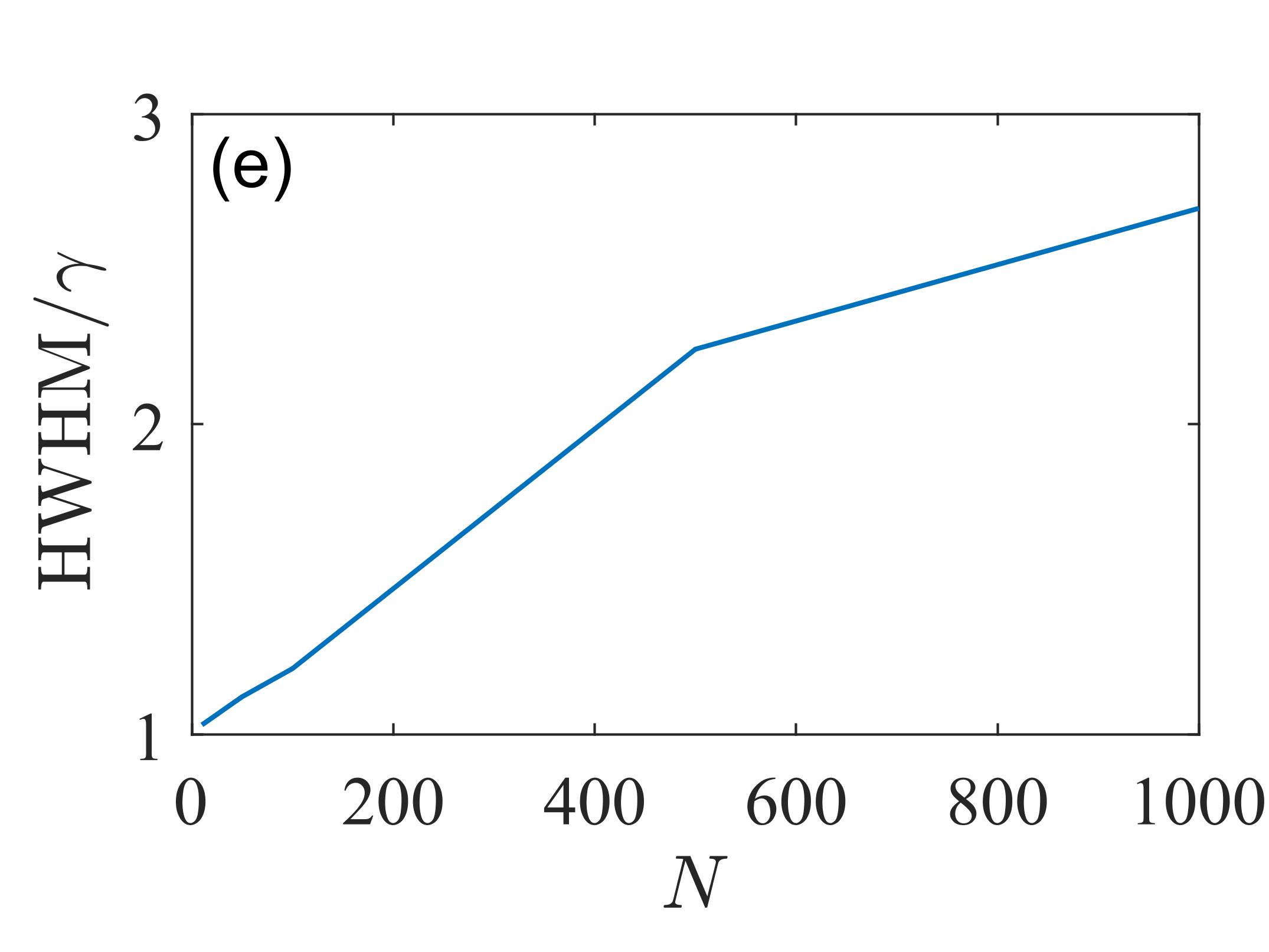}
\includegraphics[width=0.49\columnwidth]{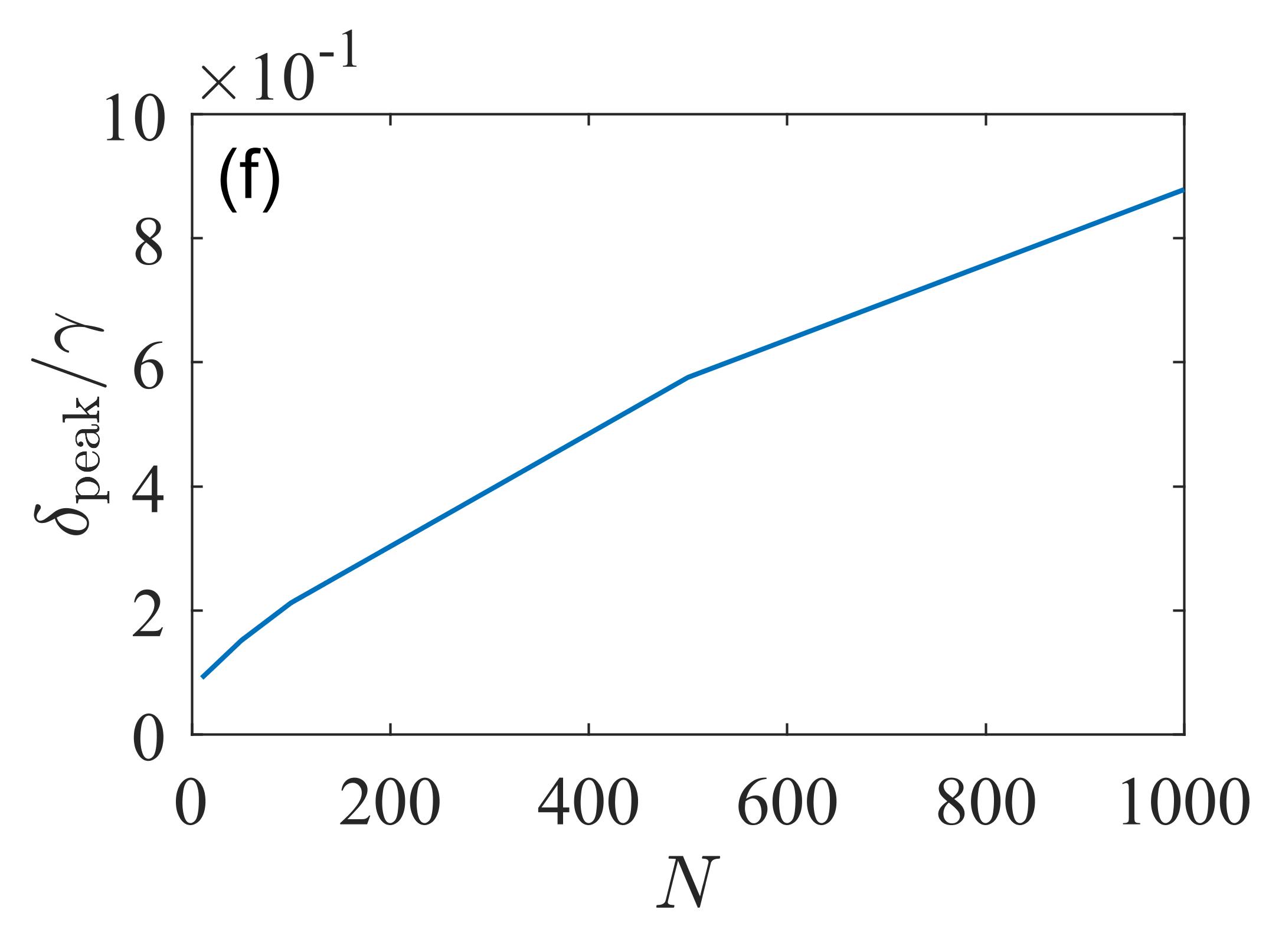}
    \vspace{-0.2cm}
    \caption{Transmission and reflection of $x$-polarized Gaussian beam from 200 atoms in an oblate trap \inline{\ell_x / \ell_z = 18}, with \inline{\ell_z=1/k}, as a function of the detuning for different atom numbers.  The peak densities for different $N$ are \inline{\bar{\rho}_{\rm{2D}}/k^2\simeq0.01,0.05,0.1,0.5,1}  (a) Coherent optical depth \inline{\rm{OD}_{\rm{coh}}},  (b) incoherent transmission \inline{T_{\rm{inc}}},  (c) coherent reflection \inline{R_{\rm{coh}}}, (d) incoherent reflection \inline{R_{\rm{inc}}}[Eqs.~\eqref{eq: Transmission}-\eqref{eq:Es expectation}].  Coherent transmission resonance (e)  HWHM and (f) shift. The lens NA 0.2 and 0.8 for coherent and incoherent light, respectively.}
    \label{fig: Non-int}
\end{figure}
We first consider the optical response of weakly excited, independently sampled trapped atoms. This corresponds to a cold disordered atomic ensemble where the atomic positions are not correlated, but light can establish correlations between the excitations of different atoms at small interatomic separations due to random position fluctuations. 
Variations of related scenarios have been investigated theoretically and several experiments have reached at least close to the regime where a random ensemble of atoms could show such correlations~\cite{Bender2010,Chomaz12,Bienaime2013,Balik2013,Javanainen2014a,Keaveney2012,Sokolov2011,SVI10,Pellegrino2014a,Havey_jmo14,JavanainenMFT,Skipetrov14,Sutherland16forward,Ye2016,antoine_polaritonic,Jenkins_thermshift,Jenkins_long16,ZHU16,Jennewein_trans,Javanainen17,Guerin_subr16,Kwong19,Binninger19,Robicheaux20,Dalibard_slab,Saint-Jalm2018,guerin16_review,Andreoli21,ma2022}.
The optical response  in an oblate trap is illustrated in Fig.~\ref{fig: Non-int},
while the atom number is varied for the $J=0\rightarrow J'=1$ transition. Even though the atoms are noninteracting before the light enters the sample, at increasing densities due to the light-mediated interactions the resonances are broadened and the lineshapes increasingly deviate from the independent-atom Lorentzian profile, exhibiting multiple peak resonances and asymmetric profiles.  Coherent transmission and reflection also show a pronounced density-dependent resonance shift. The dependence of the correlated response on the peak 2D density per wavenumber squared  \inline{\bar{\rho}_{\rm{2D}}/k^2} is clearly visible in Fig.~\ref{fig: Non-int}. For  \inline{\bar{\rho}_{\rm{2D}}/k^2\ll 1}, the response is close to the independent-atom Lorentzian, while cooperative recurrent scattering dominates at  \inline{\bar{\rho}_{\rm{2D}}/k^2\sim 1}.

\begin{figure}[!htbp]
    \centering
    \includegraphics[width=0.49\columnwidth]{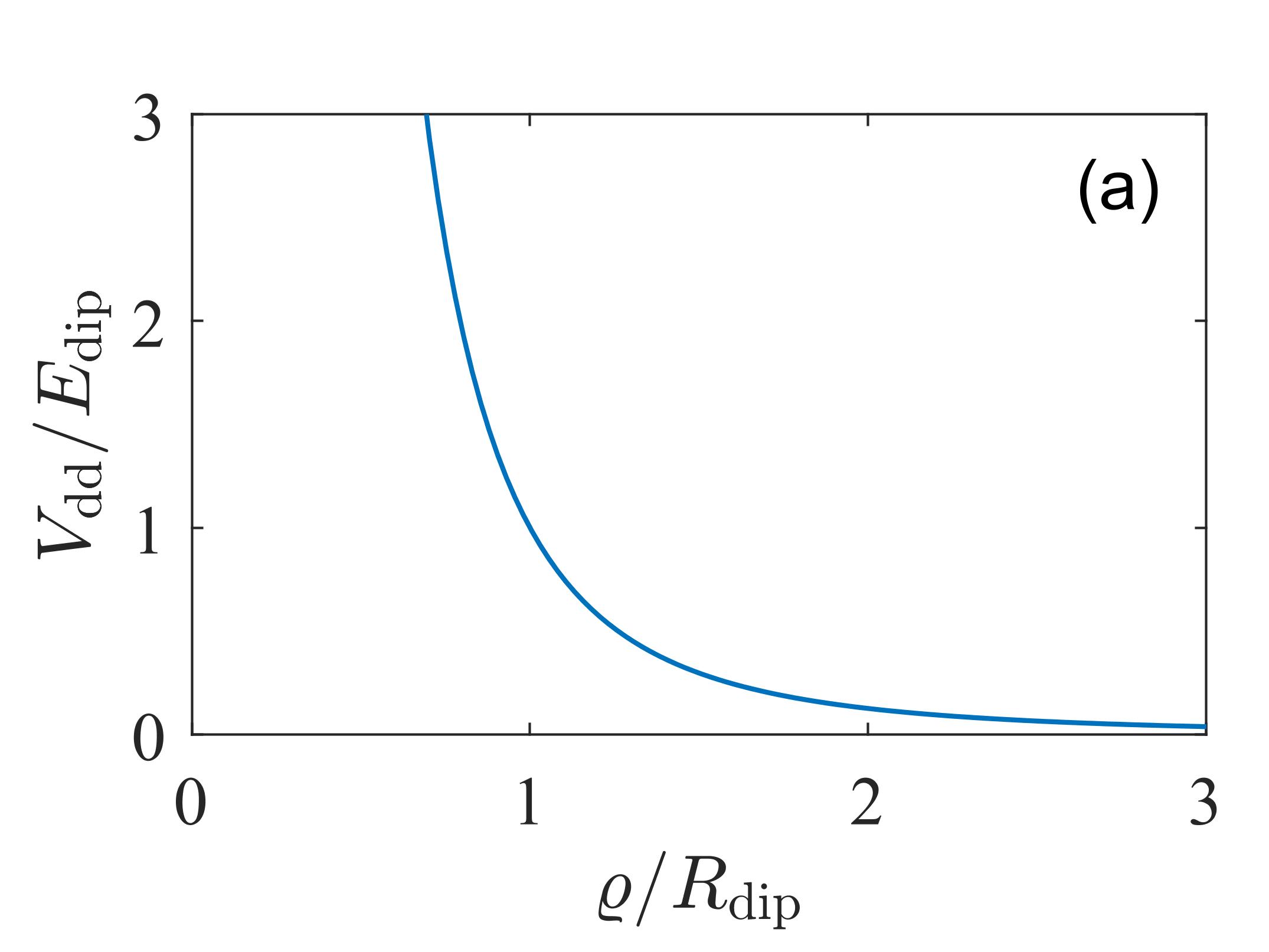}
    \includegraphics[width=0.49\columnwidth]{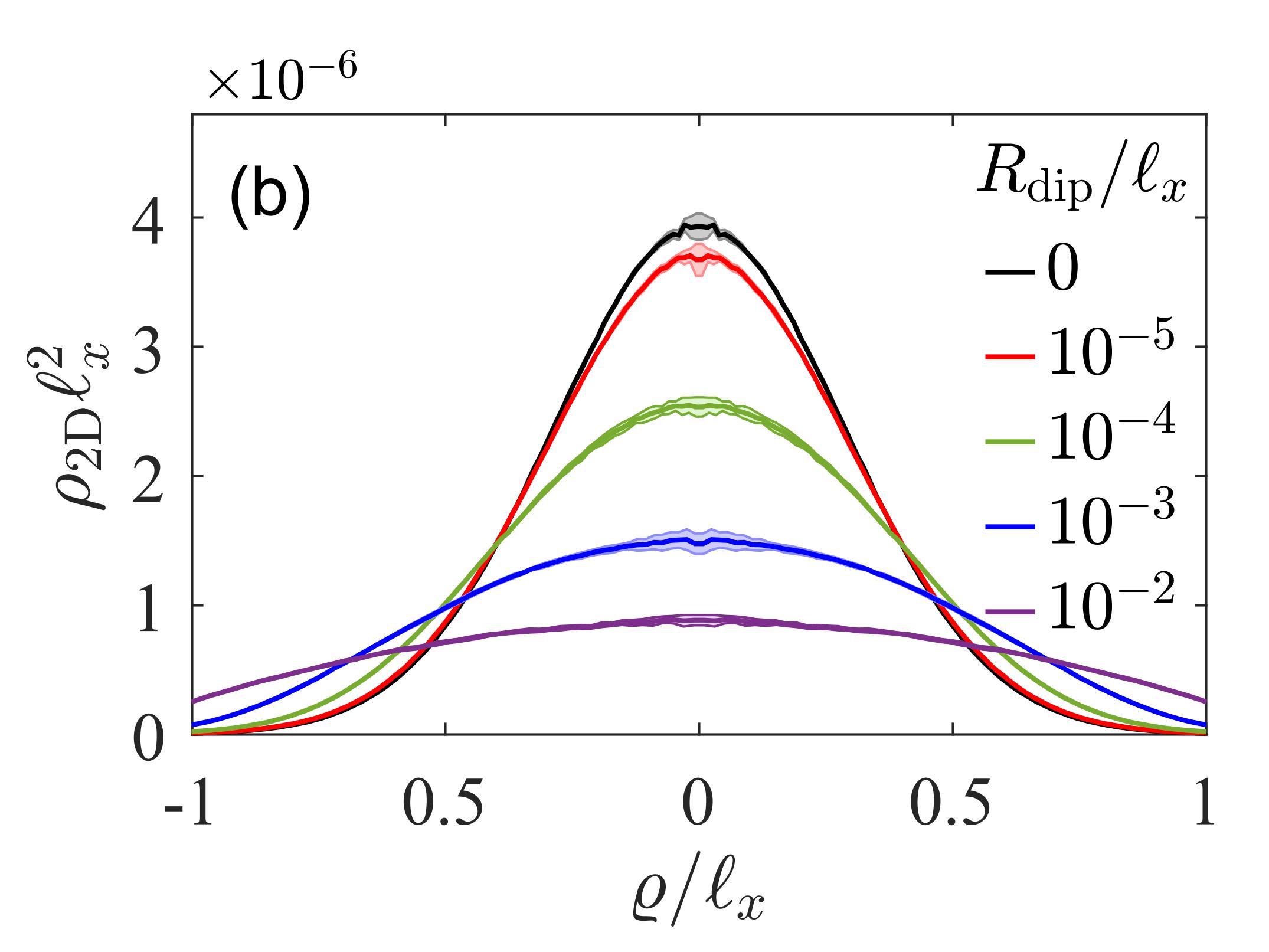}
        \vspace{-0.4cm}
    \caption{Dipolar potential between two dipoles perpendicular to the separation and the 2D density profile as a function of the radius for 100 atoms for different interaction lengths \eqcaption[true]{\inline{{R_{\rm{dip} }}}}{eq:Rdip} for {\inline{\ell_x/\ell_z=100}}.
    }
    \label{fig:shape change}
\end{figure}

The effect of the static DD interactions on the atomic density profile is immediately obvious in Fig.~\ref{fig:shape change} where we show the numerically calculated (using Monte Carlo simulations, Sec.~\ref{sec:qmc}) ground-state density profile of $N=100$ atoms at zero temperature in a symmetric oblate trap. Due to the tight confinement of the atoms along the orientation of the dipoles in the $z$ direction, the lines joining the atoms are almost perpendicular to the dipoles and the interactions are repulsive, as shown in the interatomic dipolar potential in Fig.~\ref{fig:shape change}(a). This results in a well-known effect of increasing cloud radii and flattening density profiles.  In the simulations of the optical response, we accommodate these effects by varying the beam focusing (also to adjust the overlap between the incident and coherently scattered light) and by using a sufficiently large lens to collect the light. Aside from the density profiles, the static DD interactions have a much more substantial effect on the optical response due to the change in atomic correlations that alter the light-mediated interactions between the atoms, as we will show in the next sections.

\subsection{Varying static dipolar interaction in an oblate trap} 
\label{sec:dipolaroblate}

\begin{figure*}
    \centering 
  \includegraphics[width=0.49\columnwidth]{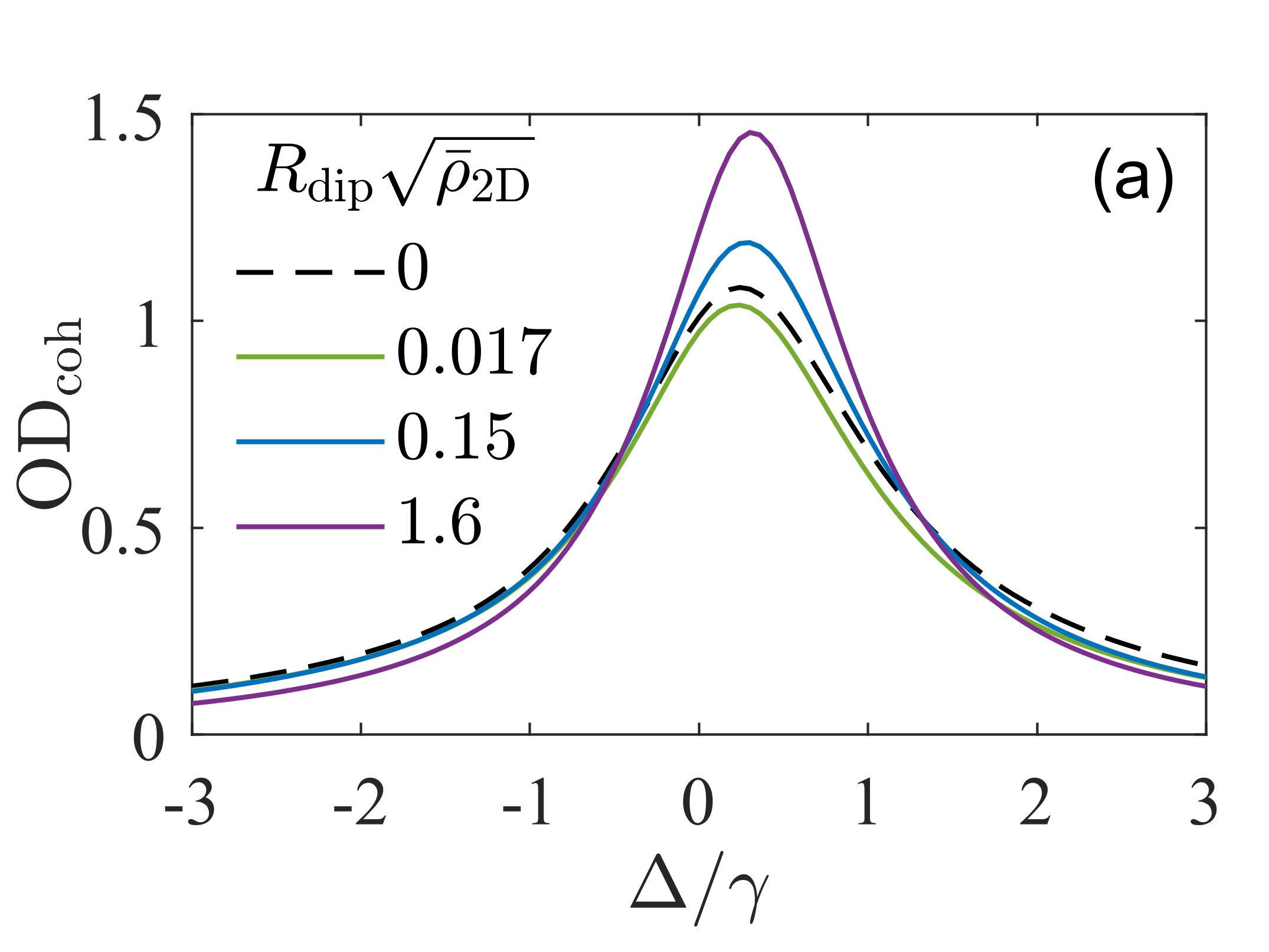}
  \includegraphics[width=0.49\columnwidth]{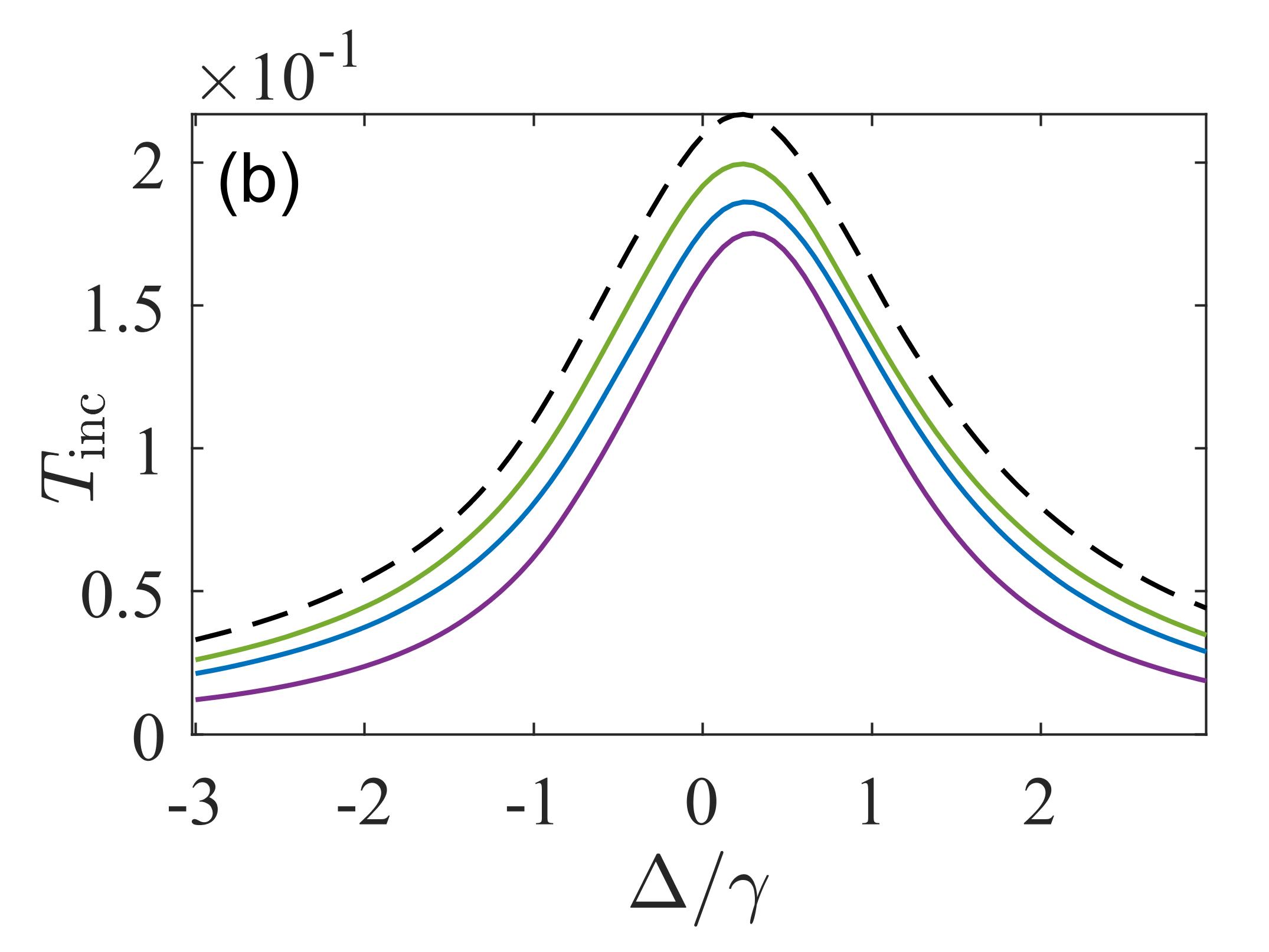}
  \includegraphics[width=0.49\columnwidth]{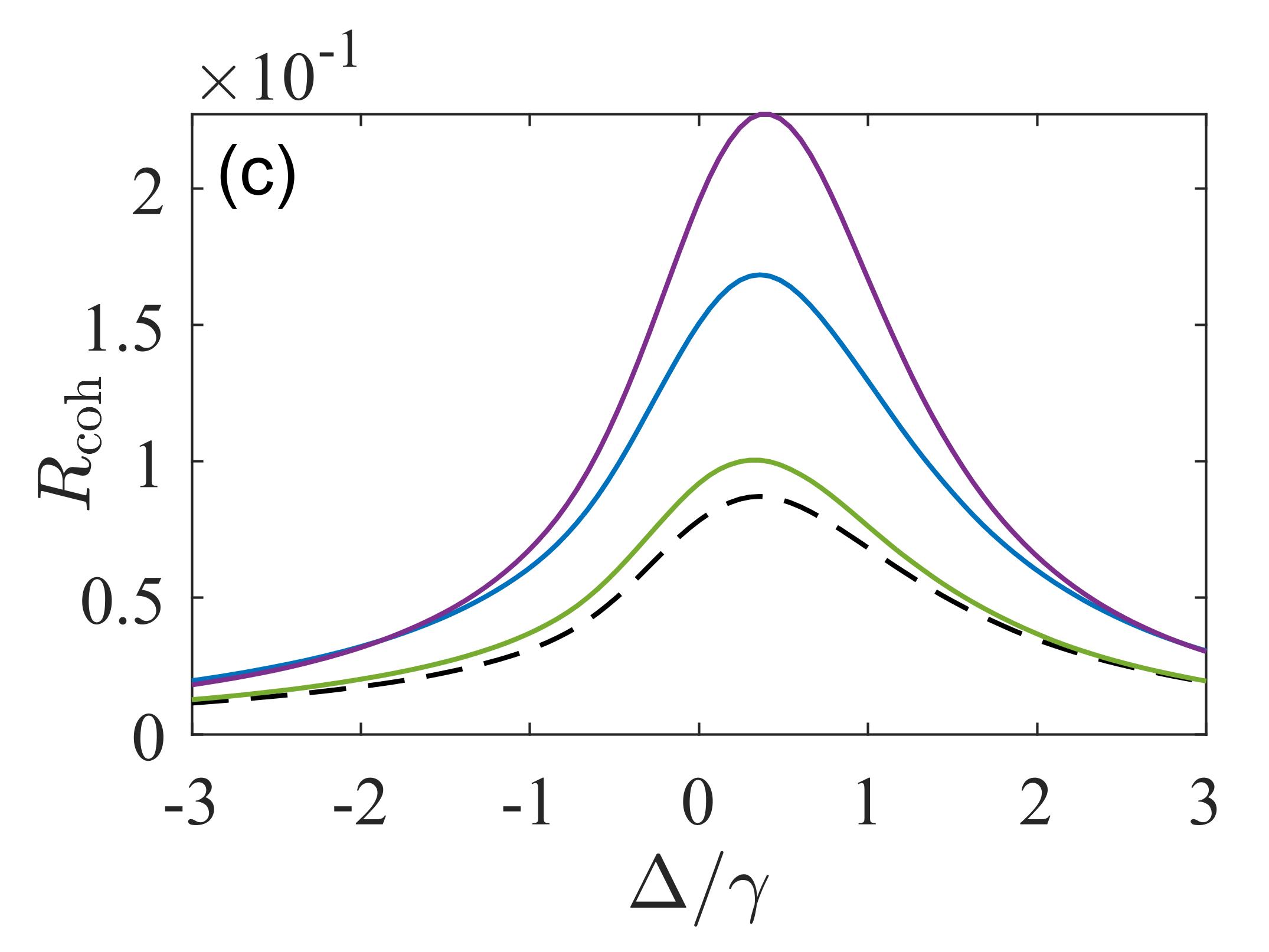}
  \includegraphics[width=0.49\columnwidth]{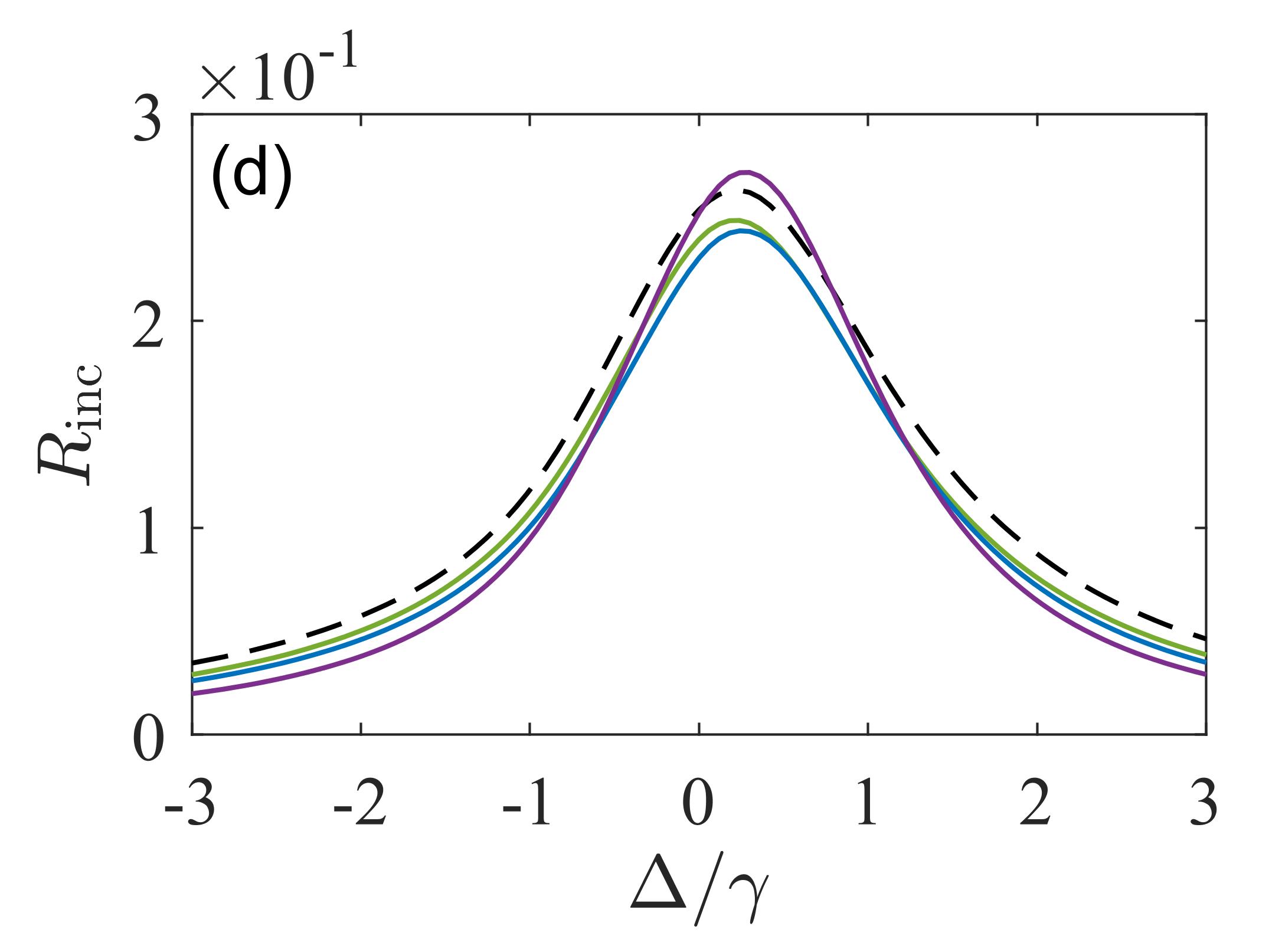}
  \includegraphics[width=0.49\columnwidth]{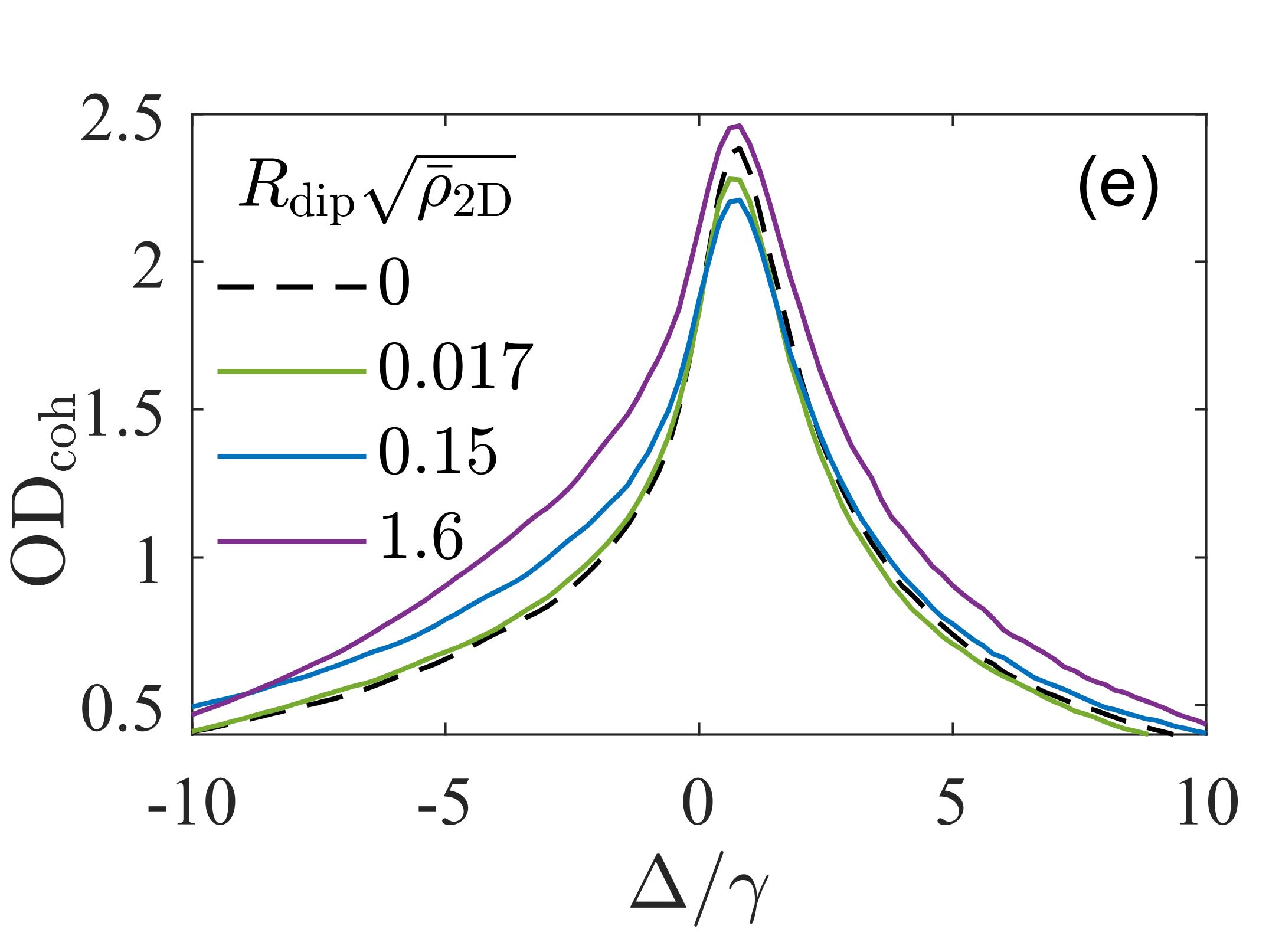}
  \includegraphics[width=0.49\columnwidth]{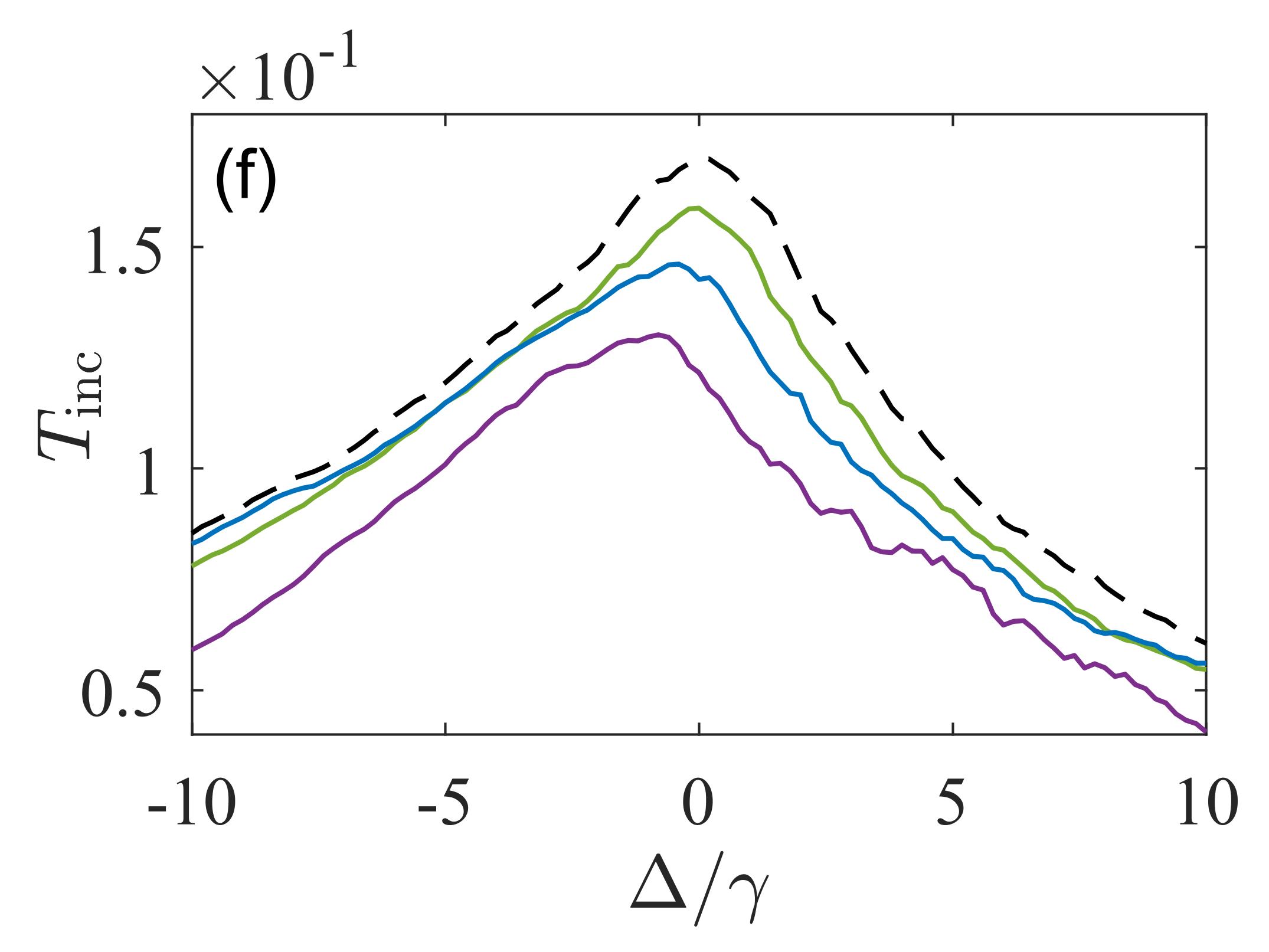}
  \includegraphics[width=0.49\columnwidth]{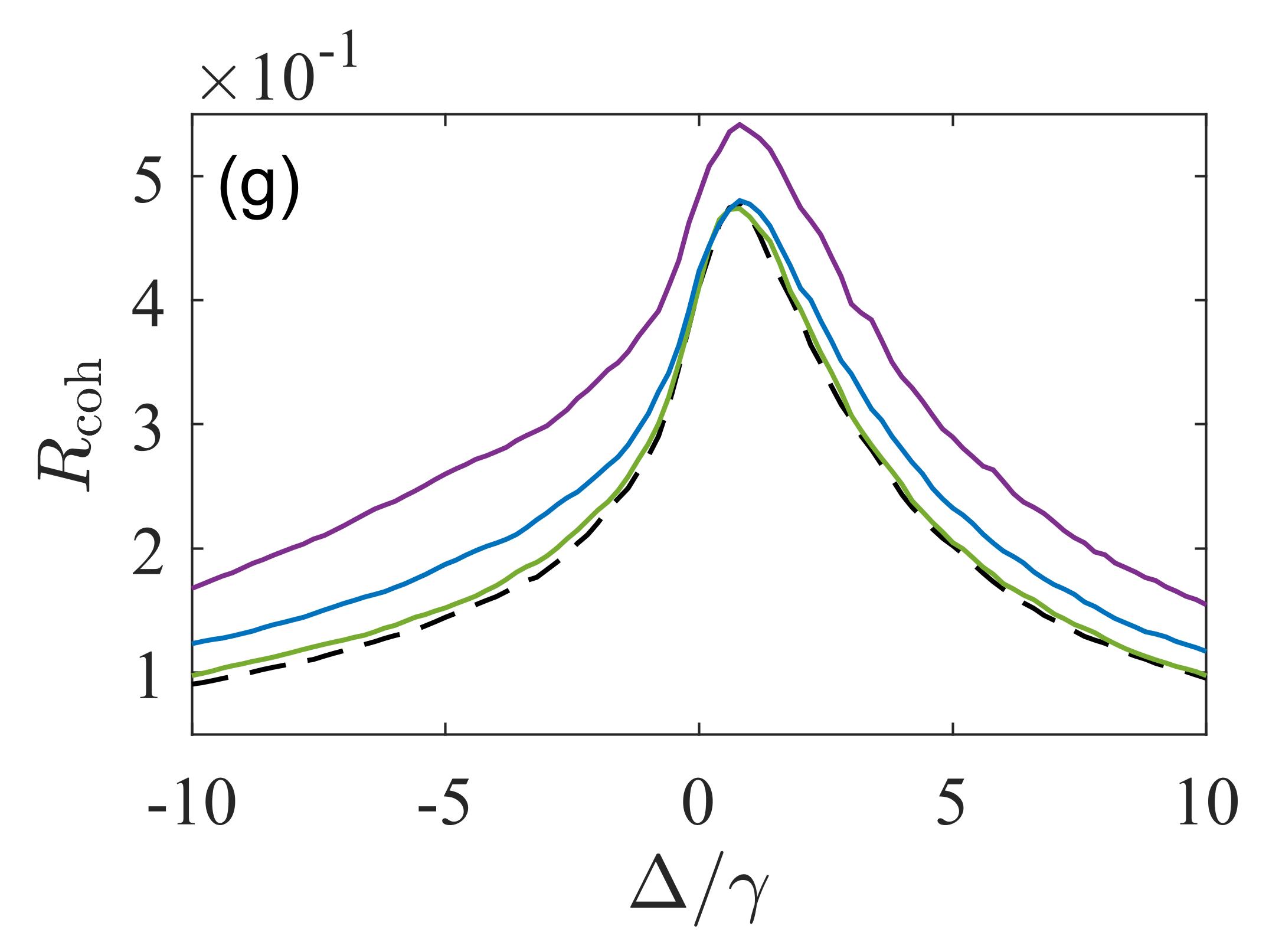}
  \includegraphics[width=0.49\columnwidth]{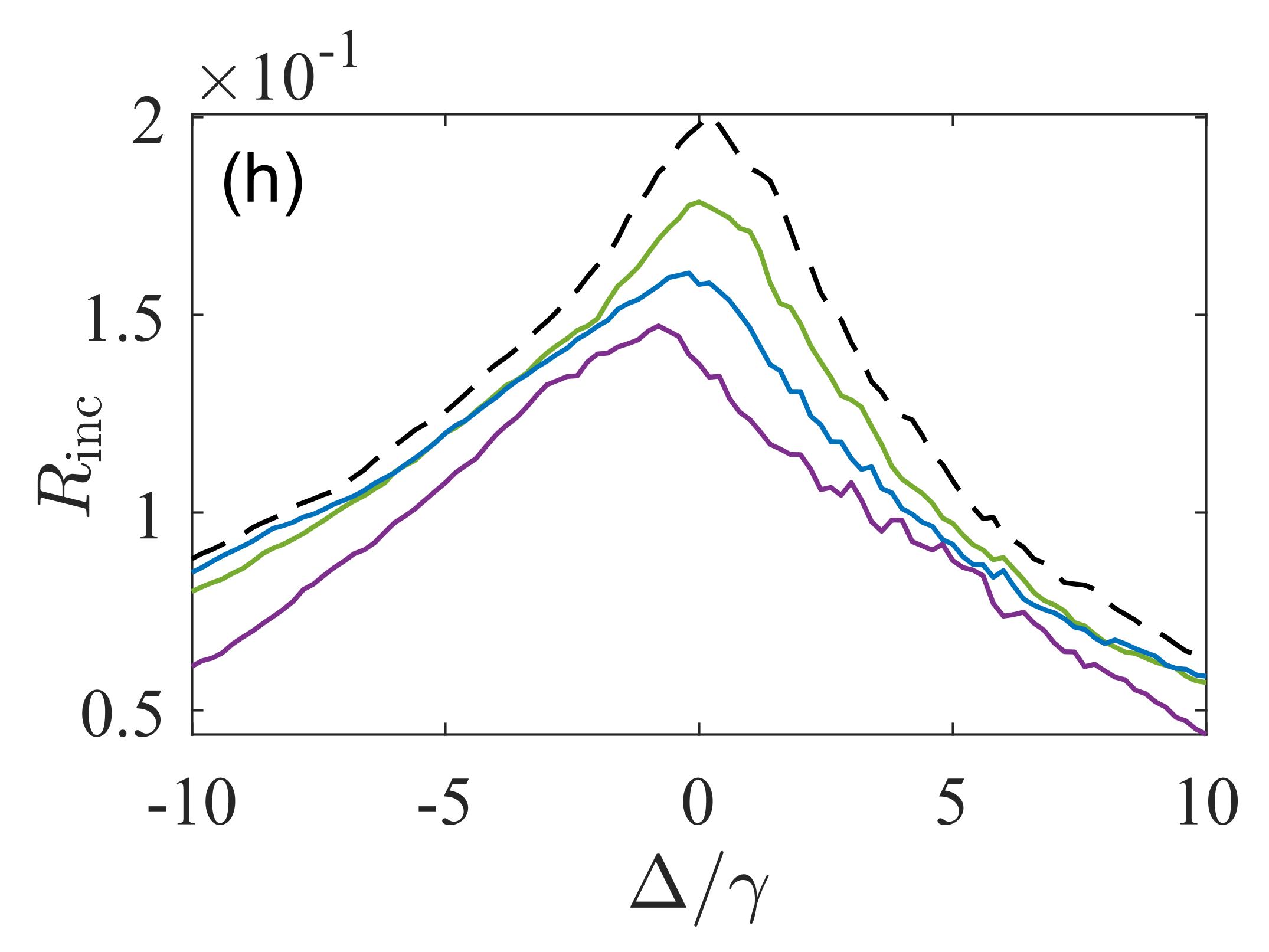}
      \vspace{-0.3cm}
    \caption{Transmission and reflection from 200 two-level atoms in an oblate trap at peak densities \inline{\bar{\rho}_{\rm{2D}}/k^2\simeq0.1} (top row) and 1 (bottom row), illuminated with a Gaussian beam, as a function of the laser detuning for different interaction strengths.  (a), (e) Coherent optical depth \inline{\rm{OD}_{\rm{coh}}};  (b), (f) incoherent transmission \inline{T_{\rm{inc}}}; (c), (g) coherent reflection \inline{R_{\rm{coh}}}; (d), (h) incoherent reflection \inline{R_{\rm{inc}}} [Eqs.~\eqref{eq: Transmission}-\eqref{eq:Es expectation}].  The lens NA 0.2 (top row) 0.25 (bottom row) and 0.8 for coherent and incoherent light, respectively. 
    }
    \label{fig: TransmissionDen004}
\end{figure*}

The crucial parameter in the cooperative response of independent atoms is the atom separation in the units of $1/k$ that determines the strength of recurrent scattering and the emergence of light-induced correlations (Sec.~\ref{sec:sed})~\cite{Morice1995a,JavanainenMFT}. Increasing densities lead to deviations from the continuous media electrodynamics due to these correlations.
To isolate the effect of static DD interactions and the resulting position distributions on the optical response, we therefore consider atomic ensembles with constant \inline{{\bar{\rho}}_{\rm{2D}}/k^2} while varying  the dipolar length \inline{R_{\rm{dip}}} [Eq.~\eqref{eq:Rdip}]. 
We express the  interaction strength as the ratio between \inline{R_{\rm{dip}}} and the average separation between atoms in the $xy$ plane at the trap center \inline{R_{\rm{dip}} \bar{\rho}_{2\rm{D}}^{1/2}} (in a prolate trap we use \inline{R_{\rm{dip}}\bar{\rho}_{1\rm{D}}}).
We take $N=200$ atoms and adjust the peak 2D density  \inline{{\bar{\rho}}_{\rm{2D}}} by changing the trapping frequencies.
The transmission and reflection are shown in  Fig.~\ref{fig: TransmissionDen004} for two-level atoms for linear polarization in the trap plane at the densities \inline{{\bar{\rho}}_{\rm{2D}}/k^2 \simeq0.1} and 1, respectively, by varying \inline{R_{\rm{dip}} \bar{\rho}_{2\rm{D}}^{1/2} \simeq 0, 0.017, 0.15, 1.6}.
\begin{figure*}
    \centering
    \includegraphics[width=0.49\columnwidth]{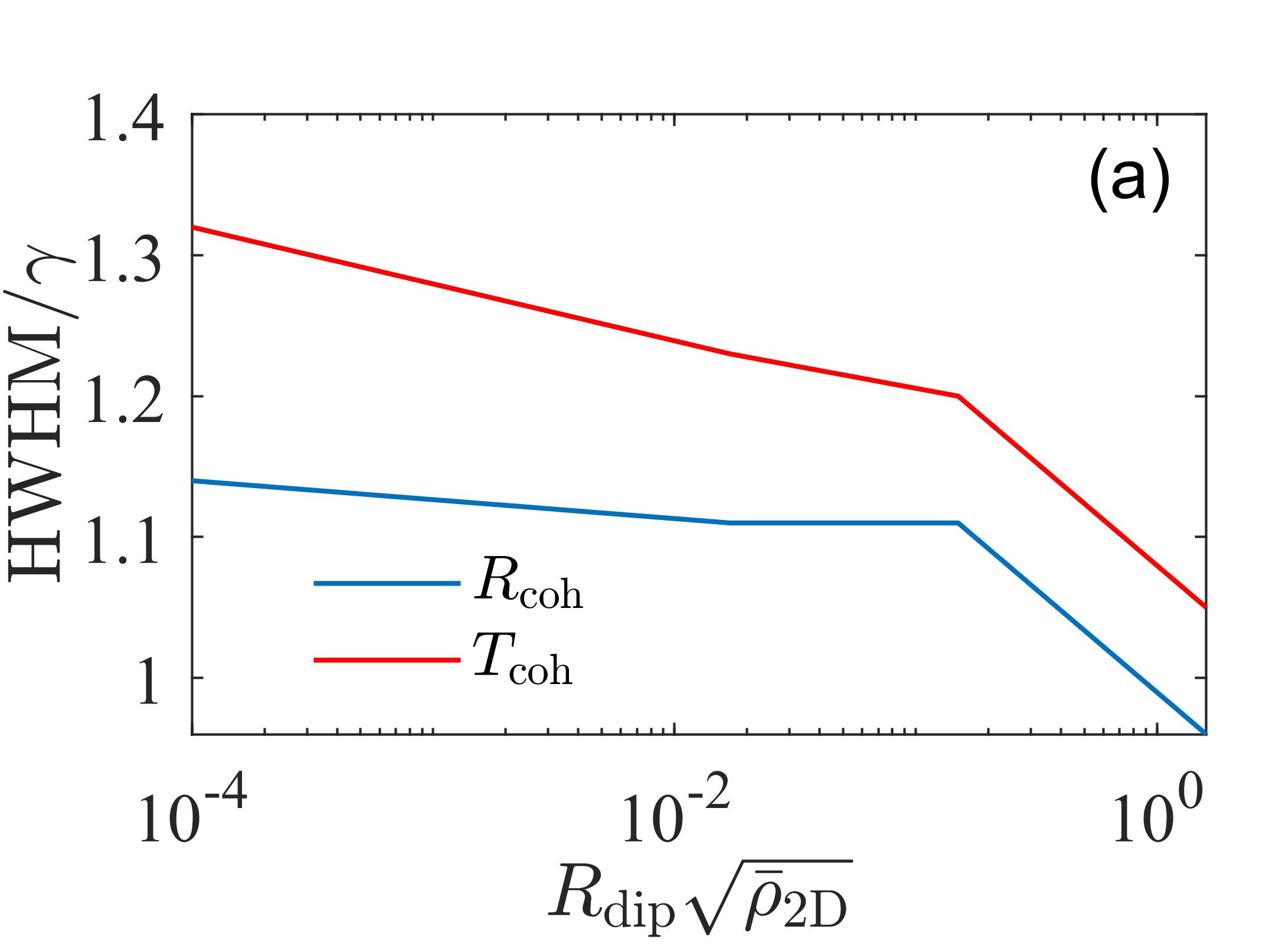}
    \includegraphics[width=0.49\columnwidth]{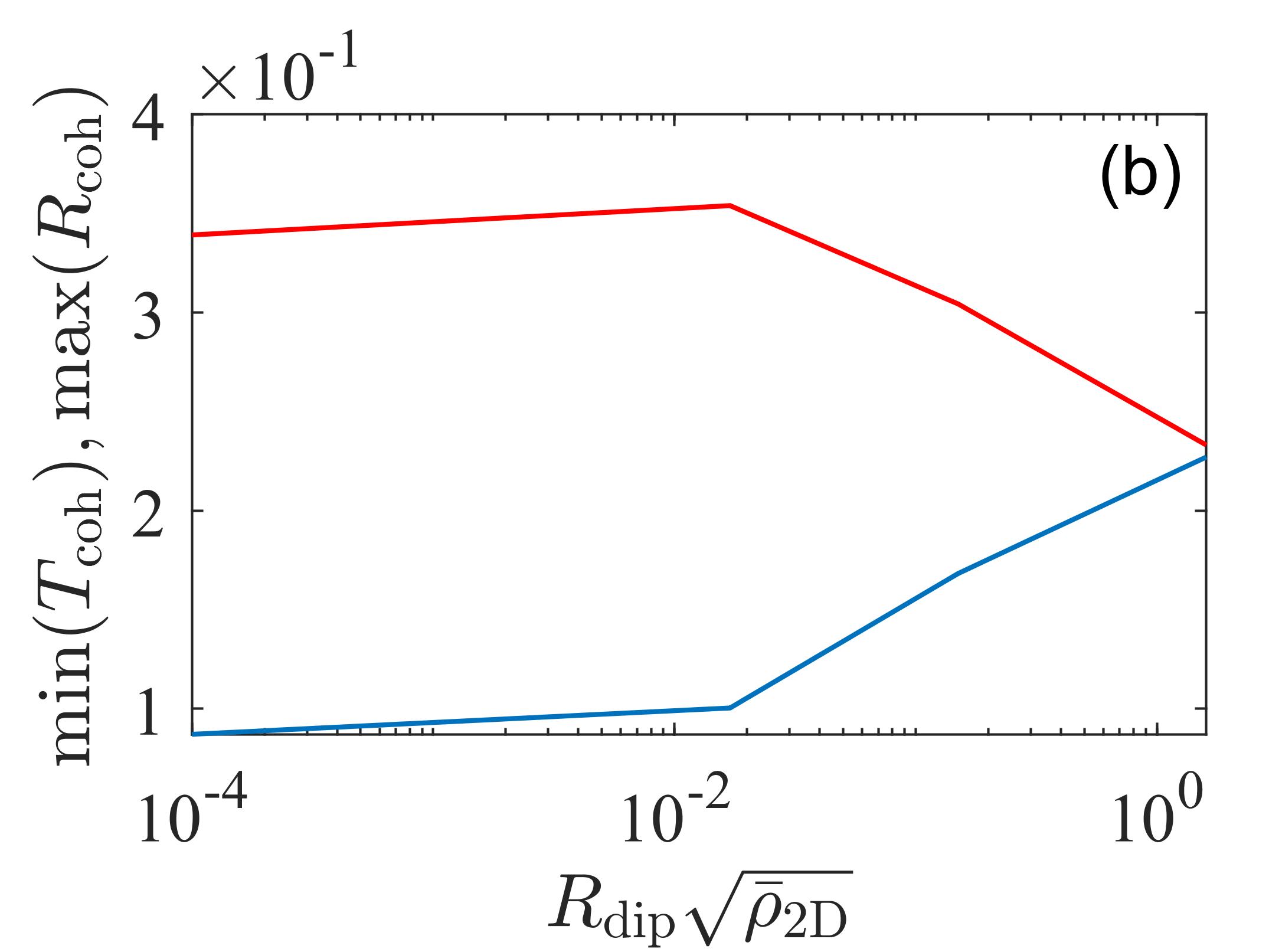}
    \includegraphics[width=0.49\columnwidth]{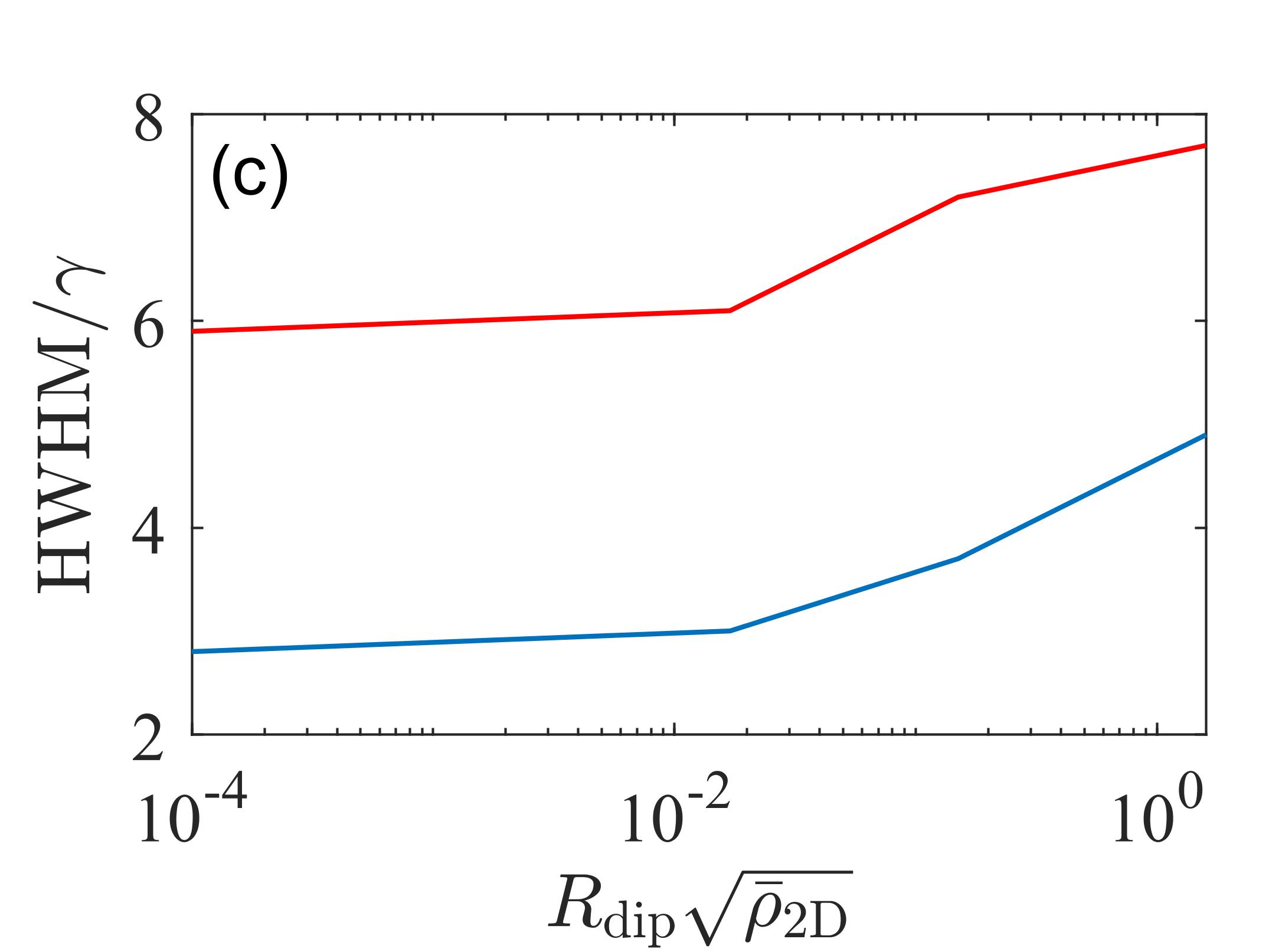}
    \includegraphics[width=0.49\columnwidth]{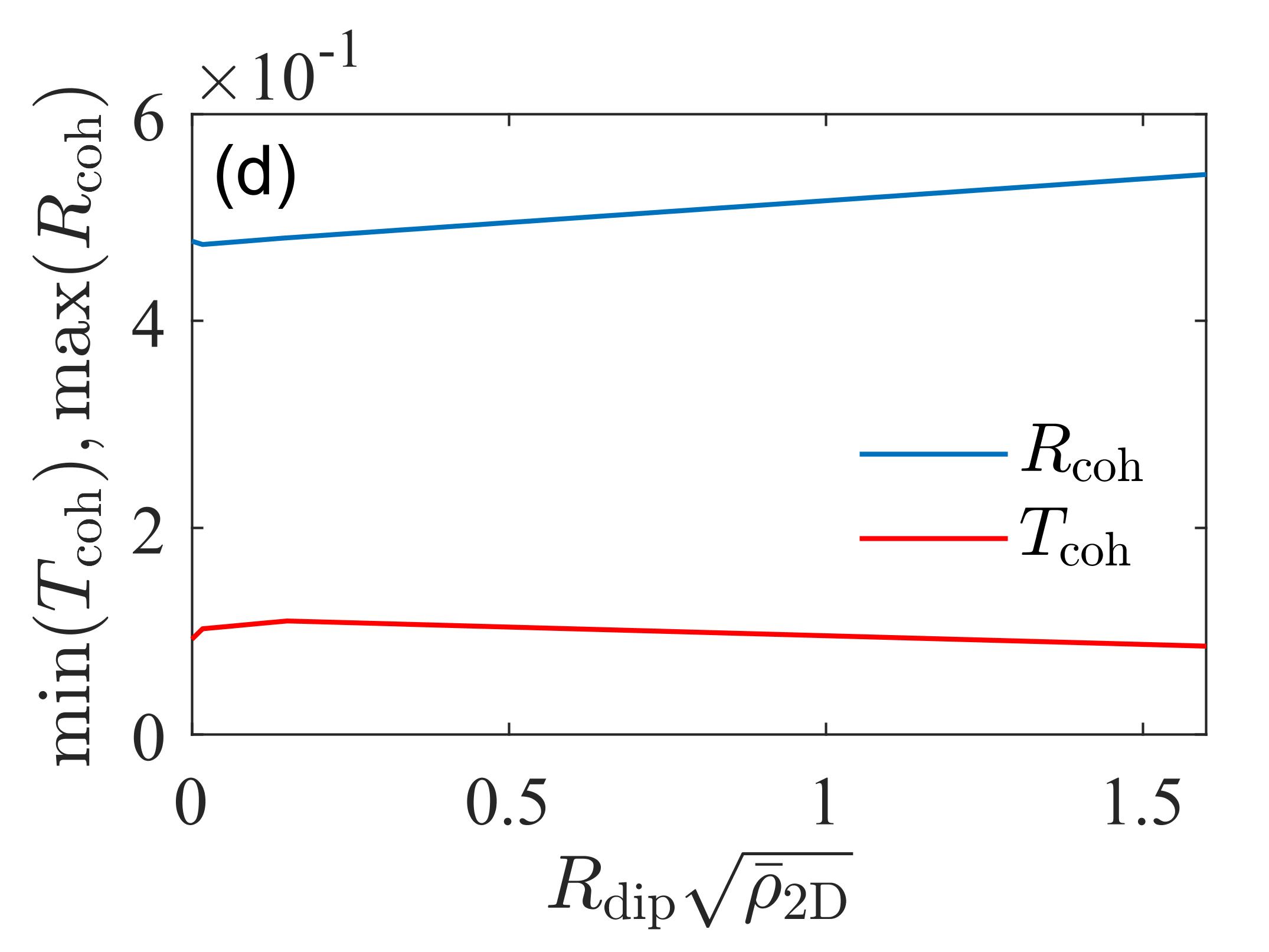}
    \vspace{-0.2cm}
    \caption{Change in the resonance width, shift, and the intensity extrema of coherently reflected and transmitted light for 200 atoms at the peak densities (a), (b) \inline{\bar{\rho}_{\rm{2D}}/k^2\simeq 0.1} and (c), (d) 1 extracted from Fig.~\ref{fig: TransmissionDen004}. (a), (c) Coherent transmission and reflection resonance widths \inline{T_{\rm{coh}}},\inline{R_{\rm{coh}}}; (b), (d) maximum \inline{R_{\rm{coh}}} and minimum \inline{T_{\rm{coh}}}.  }
    \label{fig:Den01 Shifts}
\end{figure*} 
 \begin{figure}
     \includegraphics[width=0.49\columnwidth]{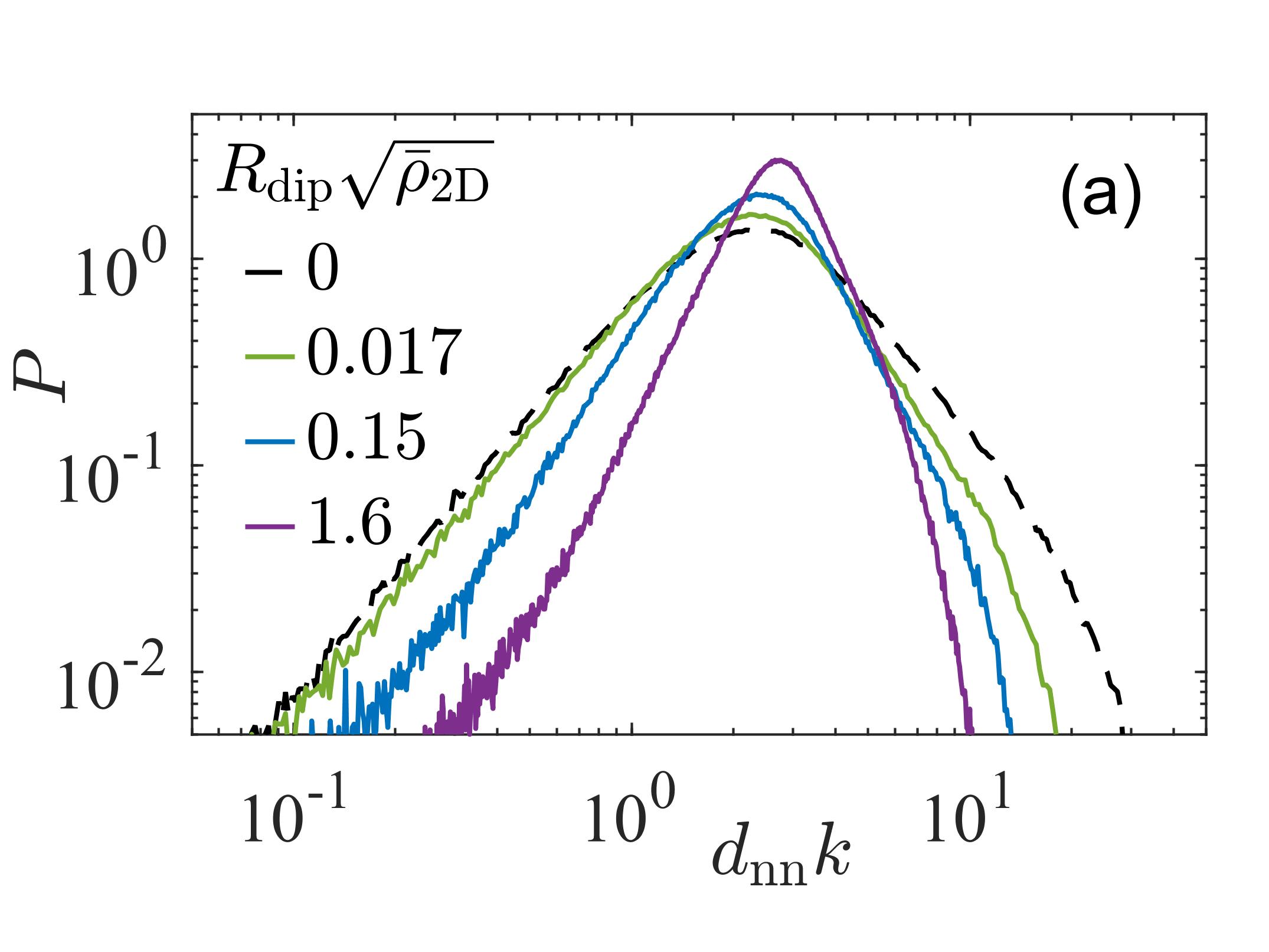}
\includegraphics[width=0.49\columnwidth]{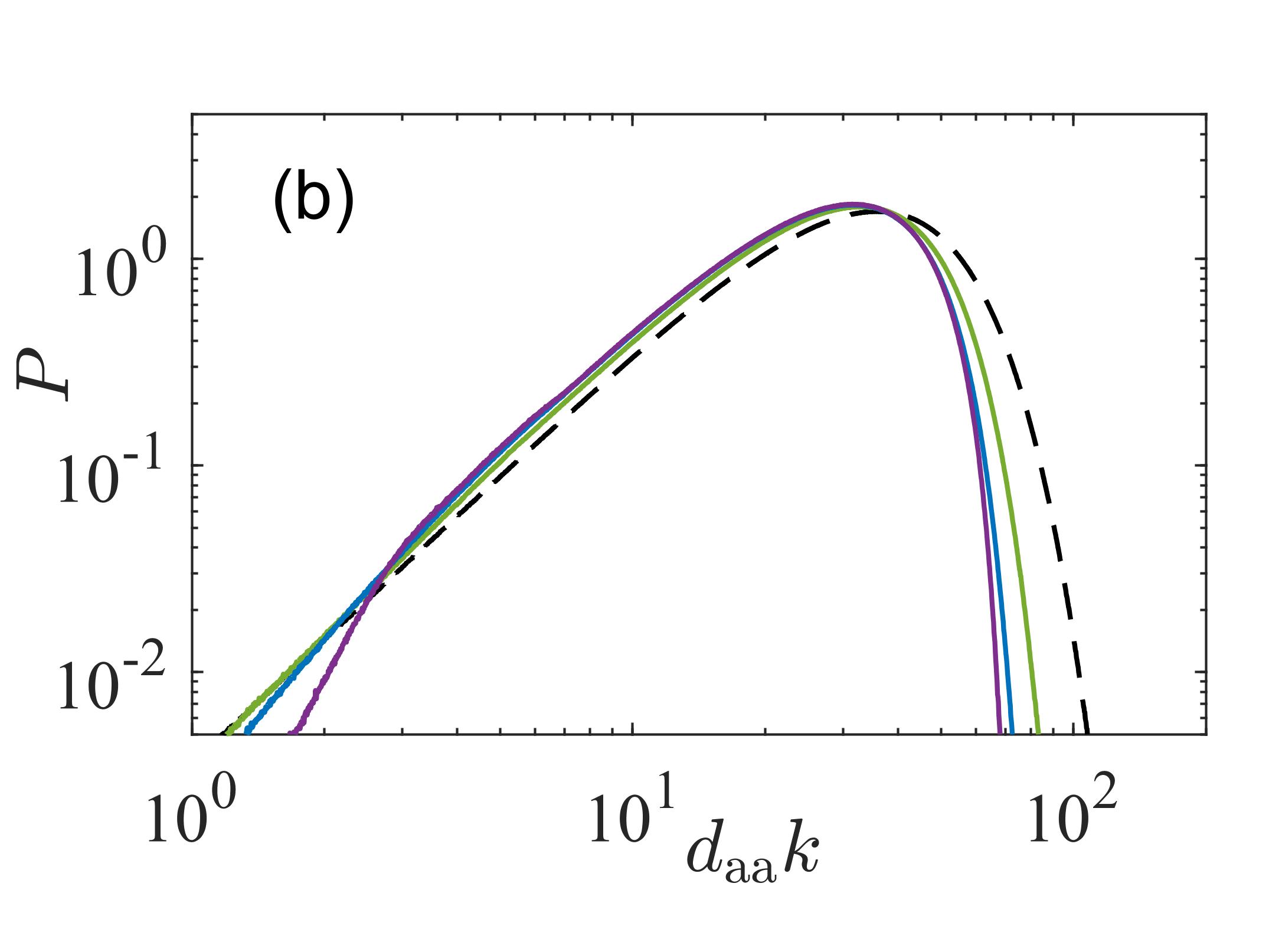}
\includegraphics[width=0.49\columnwidth]{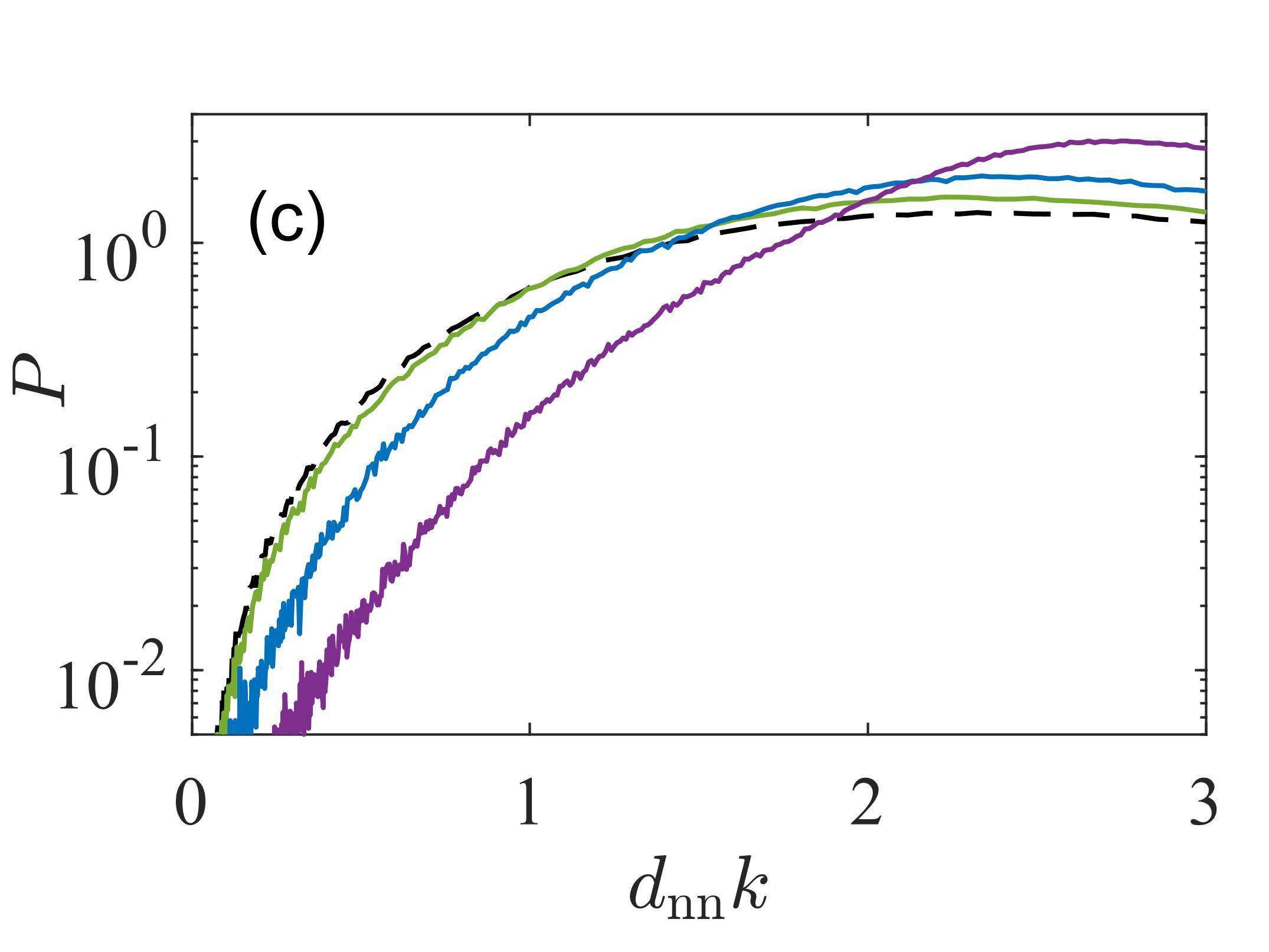}
\includegraphics[width=0.49\columnwidth]{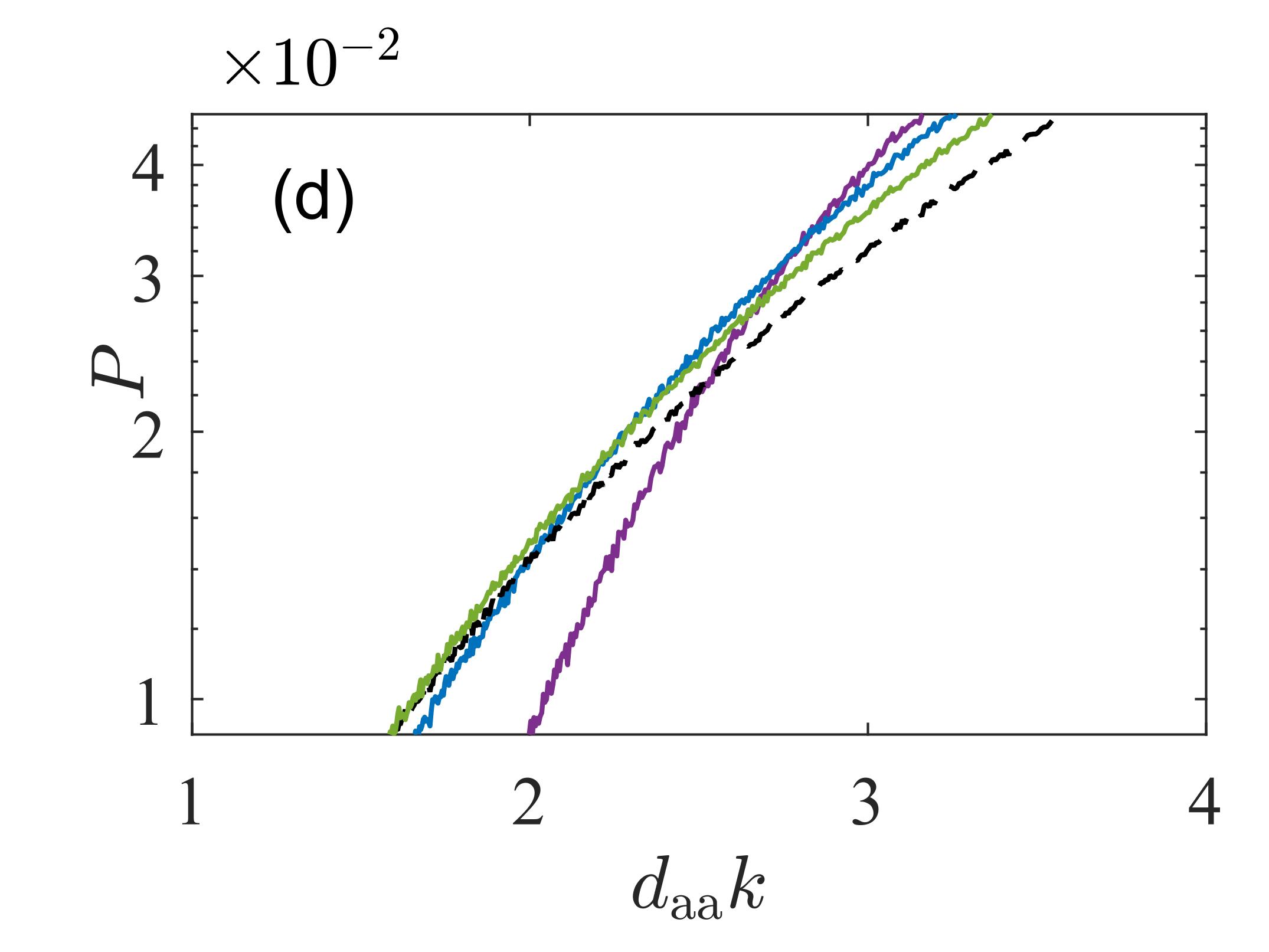}
    \vspace{-0.2cm}
 \caption{ Atom separations for the system of 200 atoms in an oblate trap
at density \inline{\bar{\rho}_{\rm{2D}}/k^2\simeq0.1} with the resonant wavelength and the trap parameters from Fig.~\ref{fig: TransmissionDen004}. Probability distributions of (a), (c)  nearest-neighbor pair \inline{d_{\rm{nn}}} and (b), (d) all-pair separations \inline{d_{\rm{aa}}}.
The short-range distributions are shown on a linear scale while the long-range plots are shown on a logarithmic scale.
    }
    \label{fig: SeparationDen004}
\end{figure}
\begin{figure}
    \centering
\includegraphics[width=0.49\columnwidth]{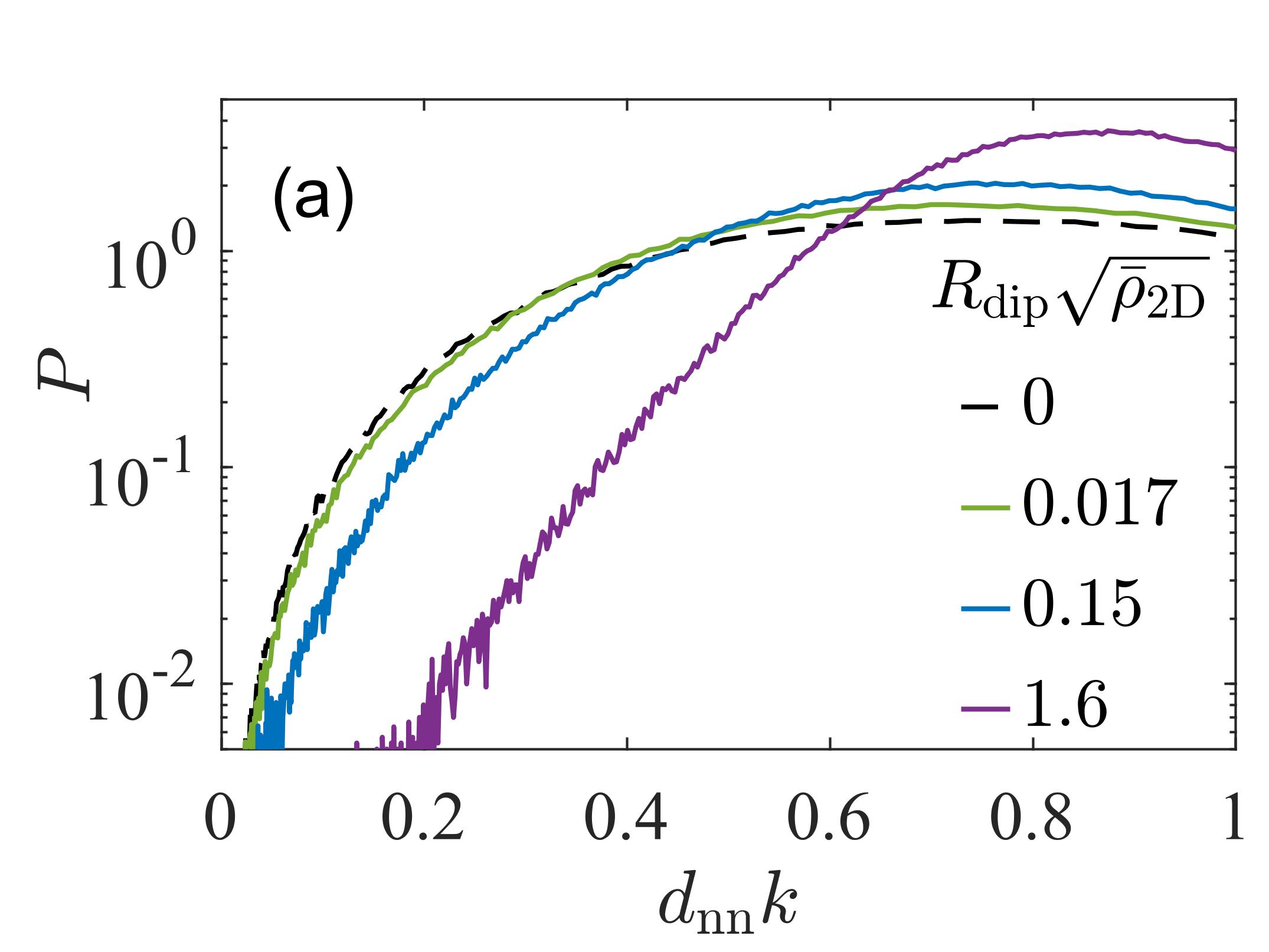}
\includegraphics[width=0.49\columnwidth]{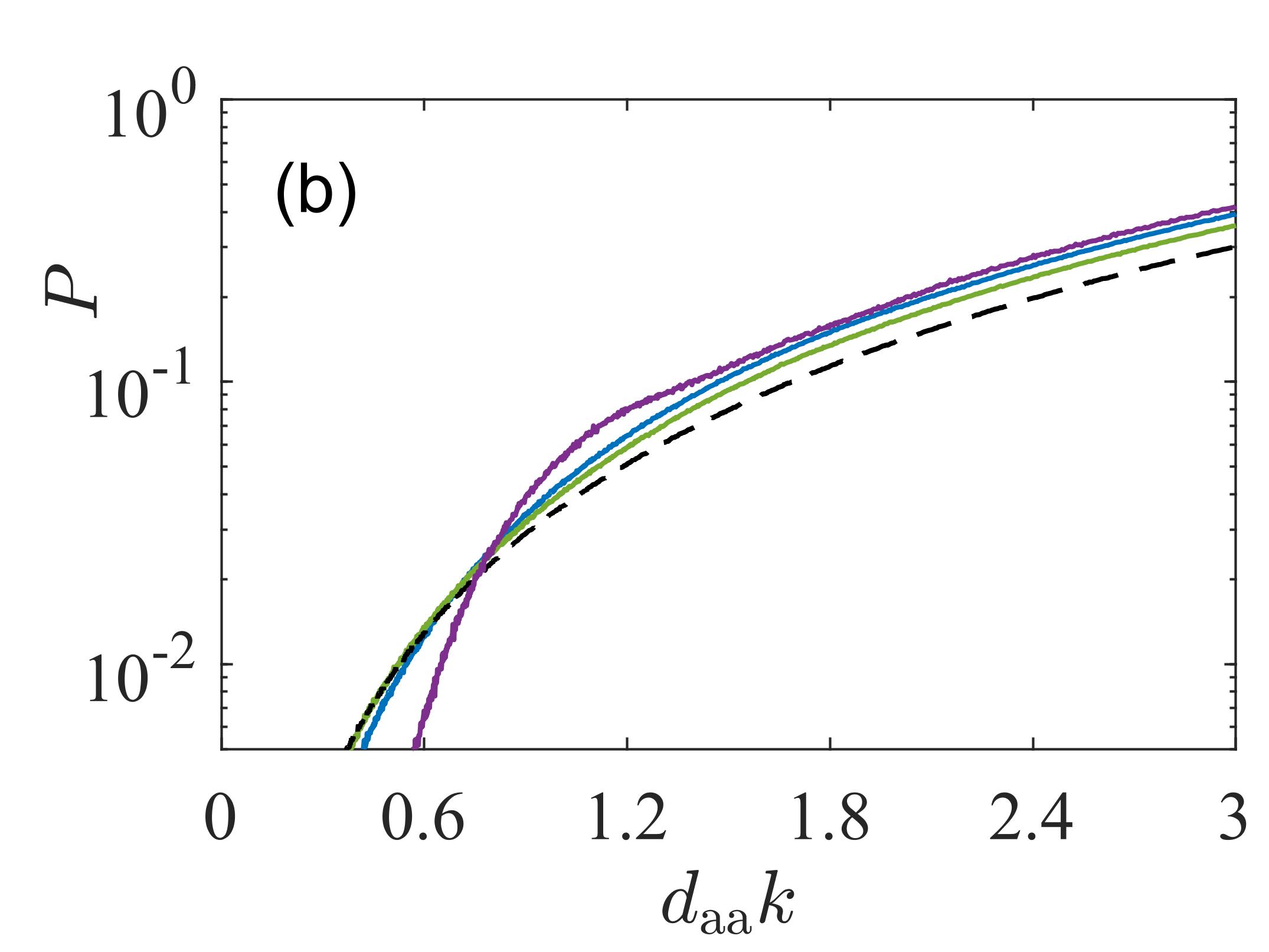}
    \vspace{-0.2cm}
    \caption{Atom separations for the system of 200 atoms in an oblate trap
at density \inline{\bar{\rho}_{\rm{2D}}/k^2\simeq1} with the resonant wavelength and the trap parameters from Fig.~\ref{fig: TransmissionDen004}.   Probability distributions of  (a) nearest-neighbor pair \inline{d_{\rm{nn}}} and (b) all-pair separations \inline{d_{\rm{aa}}}.
 }
    \label{fig: SeparationDen1}
\end{figure}

We find an increase in the resonant coherent reflection and optical depth for increasing \inline{R_{\rm{dip}}}, as also shown in Fig.~\ref{fig:Den01 Shifts}. This is associated with reduced incoherent scattering (with the exception of the incoherent reflection in the strongest interaction case of  \inline{{\bar{\rho}}_{\rm{2D}}/k^2 \simeq0.1} which has particularly high coherent reflection). 
At high densities, the coherent scattering  is already strong for independent atoms  \inline{R_{\rm{dip}}=0}. This is not true for the low density case \inline{{\bar{\rho}}_{\rm{2D}}/k^2\simeq 0.1} when the relative increase in the coherent resonant scattering as a function of \inline{R_{\rm{dip}}} is more noticeable at weak interaction strengths. 
At \inline{{\bar{\rho}}_{\rm{2D}}/k^2 \simeq0.1}, the transmission and reflection resonance linewidths narrow with increasing \inline{R_{\rm{dip}}}, but at the higher density case (\inline{{\bar{\rho}}_{\rm{2D}}/k^2 \simeq1}) there is broadening (Fig.~\ref{fig:Den01 Shifts}). At \inline{{\bar{\rho}}_{\rm{2D}}/k^2 \simeq1}, the resonance is shifted as a function of \inline{R_{\rm{dip}}}, whereas the interaction-dependent shift at the low density case is absent. At \inline{{\bar{\rho}}_{\rm{2D}}/k^2 \simeq1}, the lineshape is substantially deviated from the independent-atom Lorentzian and this effect is magnified by the static interactions. The incoherent scattering exhibits a very broad resonance.

The effect of the static DD interactions on the light-mediated interactions between the atoms at a given density is due to position correlations. Increasing the repulsive interactions modify the distribution of interatomic separations introducing short-range ordering of the atoms, as shown in the probability distributions of the nearest-neighbor and all-atom separations in Figs.~\ref{fig: SeparationDen004} and~\ref{fig: SeparationDen1}.
As \inline{R_{\rm{dip}}} and repulsion increase, the atoms are prevented from coming close to each other. 
This suppression of small interatomic separations prevents the light-mediated DD interactions, which scale as $\propto 1/r^3$ at small separations, becoming very large. In disordered ensembles this reduces the fluctuations of light-induced resonance shifts between any atom pairs, resulting in reduced inhomogeneous broadening of resonance frequencies due to the resonance shifts and reduced incoherent scattering. 
In Fig.~\ref{fig: SeparationDen1}, the most pronounced short-range ordering occurs for  \inline{R_{\rm{dip}}\sqrt{{\bar{\rho}}_{\rm{2D}}} \simeq 1.6} when  the atoms are unable to approach each other, representing the most dramatic lineshape deformation in Fig.~\ref{fig: TransmissionDen004}. 
We additionally find in Figs.~\ref{fig: SeparationDen004} and~\ref{fig: SeparationDen1} that the large separations are suppressed due to increased trap frequencies that maintain the constant peak density value.

\subsubsection{Eigenmodes}
\label{sec:eigenmodes}

\begin{figure}
\includegraphics[width=0.49\columnwidth]{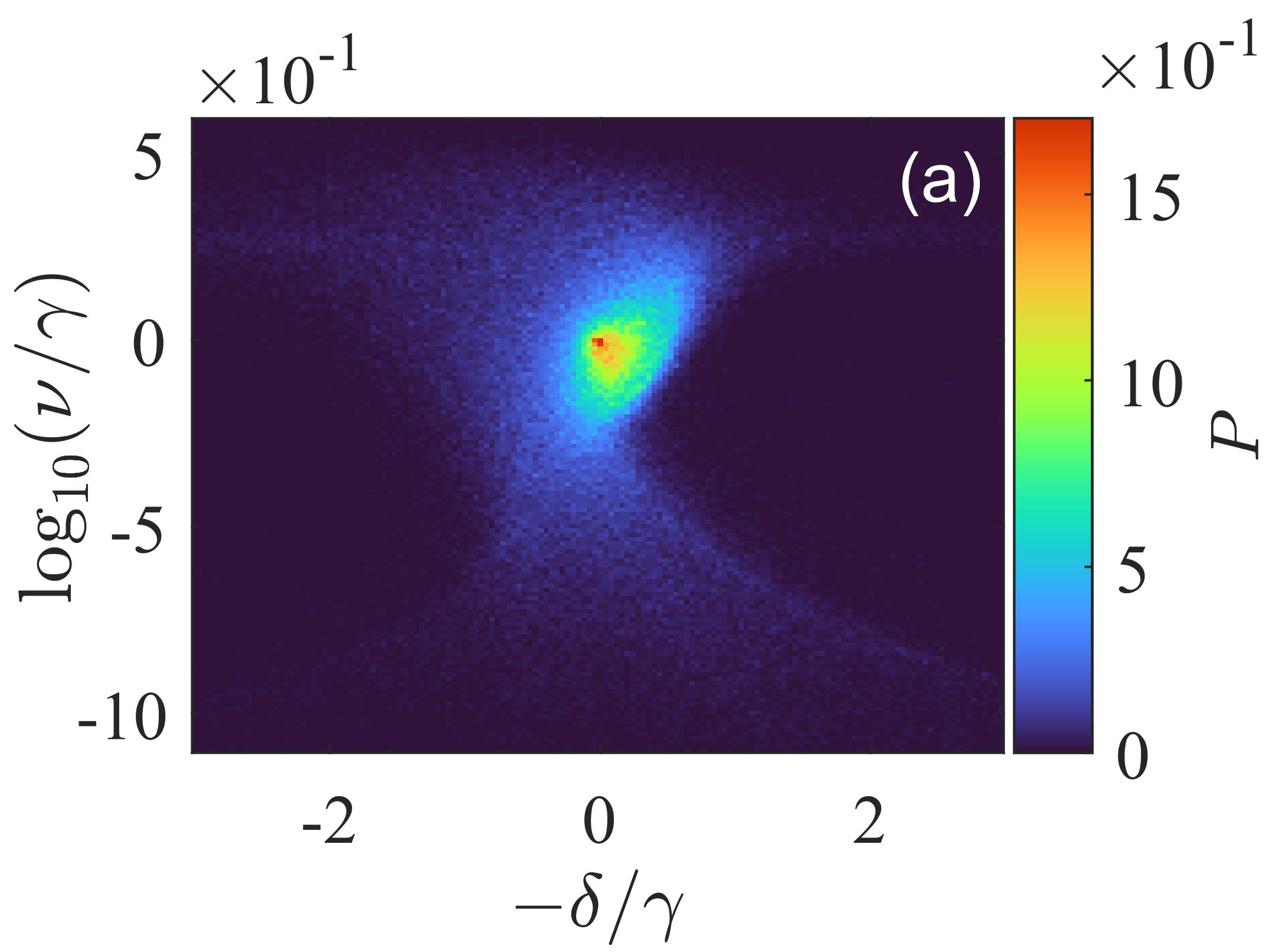}
\includegraphics[width=0.49\columnwidth]{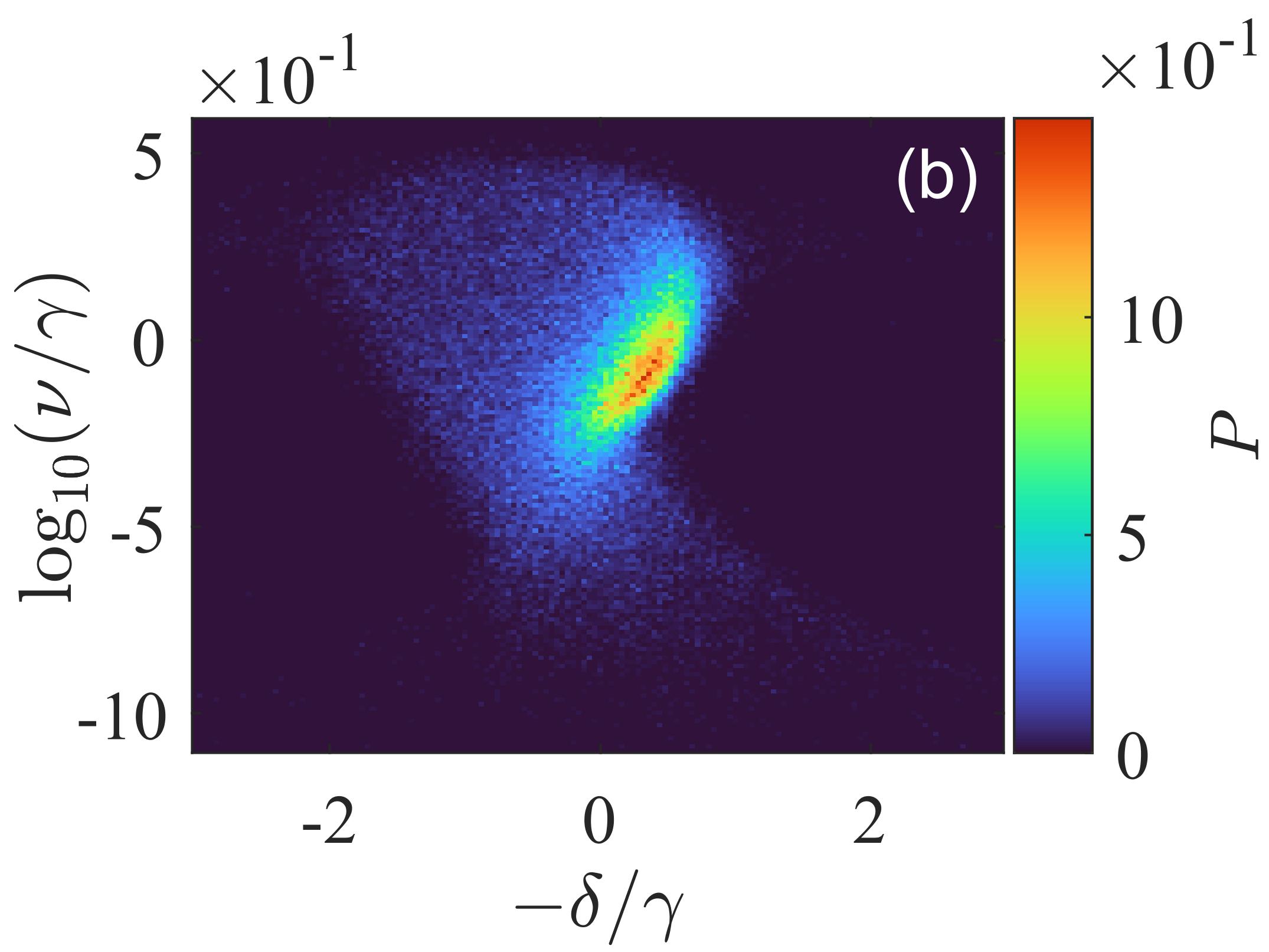}
\includegraphics[width=0.49\columnwidth]{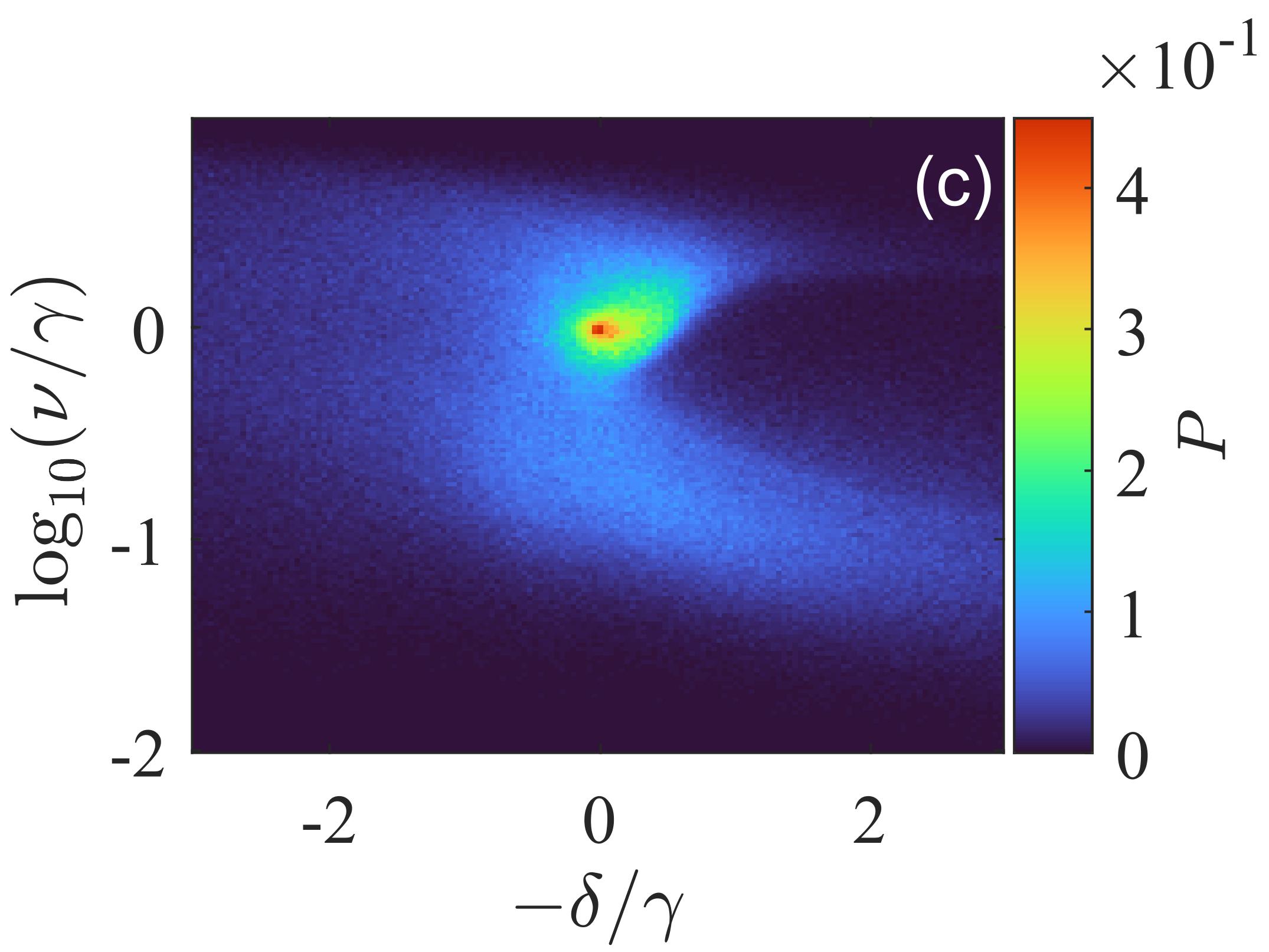}
\includegraphics[width=0.49\columnwidth]{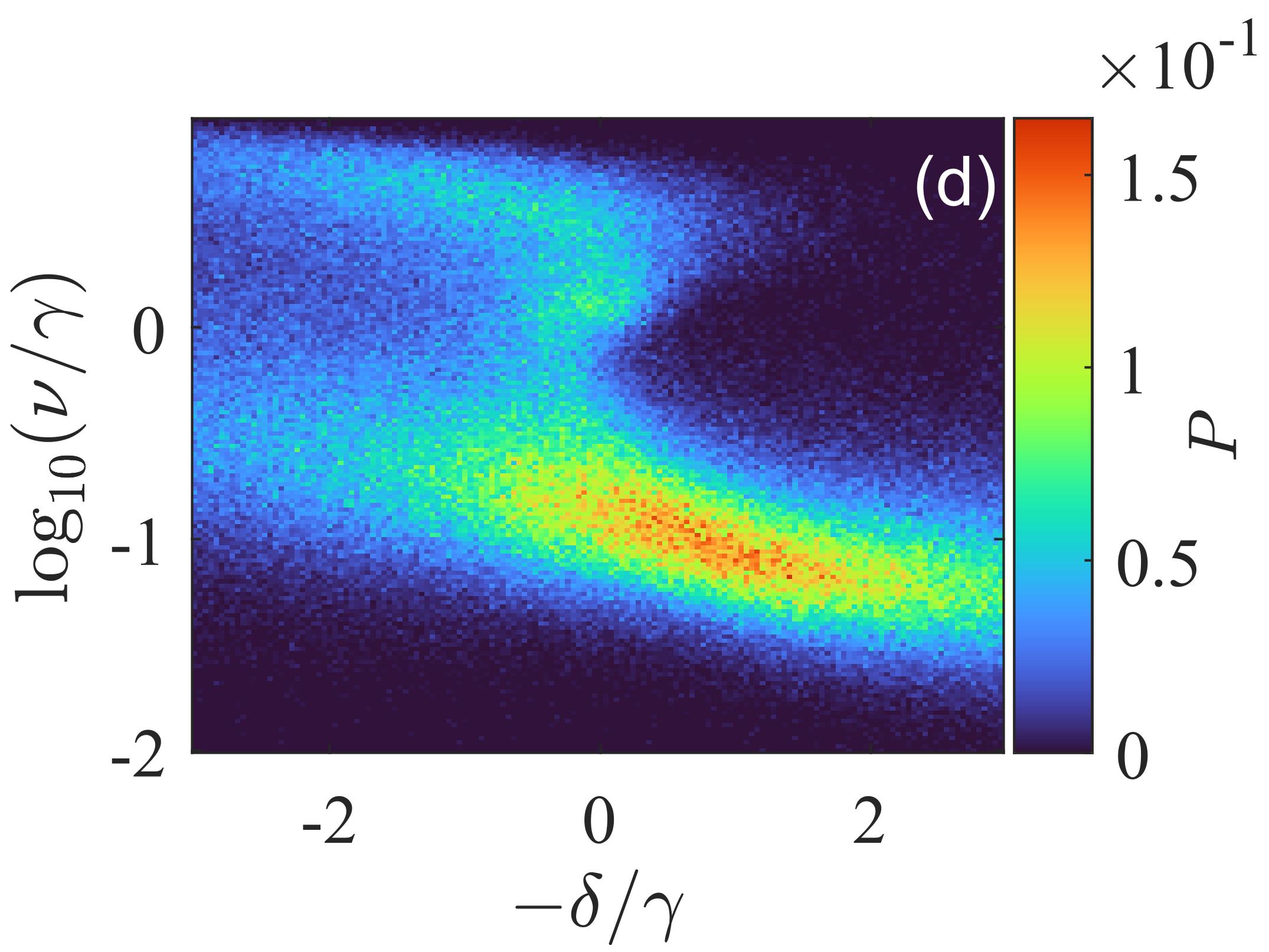}
    \vspace{-0.2cm}
\caption{Distribution of eigenmodes as a function of the collective resonance linewidth and line shift for two cases of Fig.~\ref{fig: TransmissionDen004}  with  (a), (b) \inline{\bar{\rho}_{\rm{2D}}/k^2\simeq0.1} and (c), (d) 1 for (a), (c) independent atoms \inline{R_{\rm{dip}} = 0} and (b), (d) static DD interactions with \inline{R_{\rm{dip}} \sqrt{\bar{\rho}_{\rm{2D}}} \simeq 1.6}.
The bin size \inline{\Delta \log_{10} (\nu/\gamma) \times \Delta \delta/\gamma = [0.017 \times 0.04]}.
\label{fig: Eigenmode low Den Prob}
}
\end{figure}

The optical response of the atoms can be analyzed in terms of the collective excitation eigenmodes. In Fig.~\ref{fig: Eigenmode low Den Prob}, we show the normalized histogram distribution of the eigenmodes as a function of their collective resonance linewidths \inline{\nu} and line shifts \inline{-\delta} for \inline{\bar{\rho}_{\rm{2D}}/k^2\simeq0.1} and 1 (the minus sign is used to align the plots with the laser detuning in lineshapes).
This corresponds to the optical responses shown in Fig.~\ref{fig: TransmissionDen004} for  the independent-atom case  \inline{R_{\rm{dip}}=0} and for the most strongly interacting case \inline{R_{\rm{dip}} \sqrt{\bar{\rho}_{\rm{2D}}} \simeq 1.6}. 
Owing to the relatively low density of the \inline{\bar{\rho}_{\rm{2D}}/k^2\simeq0.1} case, the eigenmodes for \inline{R_{\rm{dip}}=0} are strongly peaked around the single atom resonance and linewidth. The static DD interactions cause the highly occupied region to spread, extending towards blue-detuned super-radiance and red-detuned subradiance. This peak region is also shifted consistently with the collective resonance shift in Fig.~\ref{fig: TransmissionDen004}. 

With increasing density, the size of the central region generally increases both in width in resonance and in linewidth. The distribution also forms long arms that extend into regions of super-radiant and subradiant modes. 
At \inline{\bar{\rho}_{\rm{2D}}/k^2\simeq1} in Fig.~\ref{fig: Eigenmode low Den Prob}, the mode density is high far from the single-atom resonance even for independent atoms. The distribution becomes highly asymmetric at high density with pronounced blue-detuned subradiant eigenmodes. The static interactions further magnify these modes and also produce very super-radiant red-detuned modes. These changes in eigenmode distributions correspond to resonance broadening of the lineshapes with increasing density in Fig.~\ref{fig: TransmissionDen004}. 

\begin{figure}
    \centering
\includegraphics[width=0.49\columnwidth]{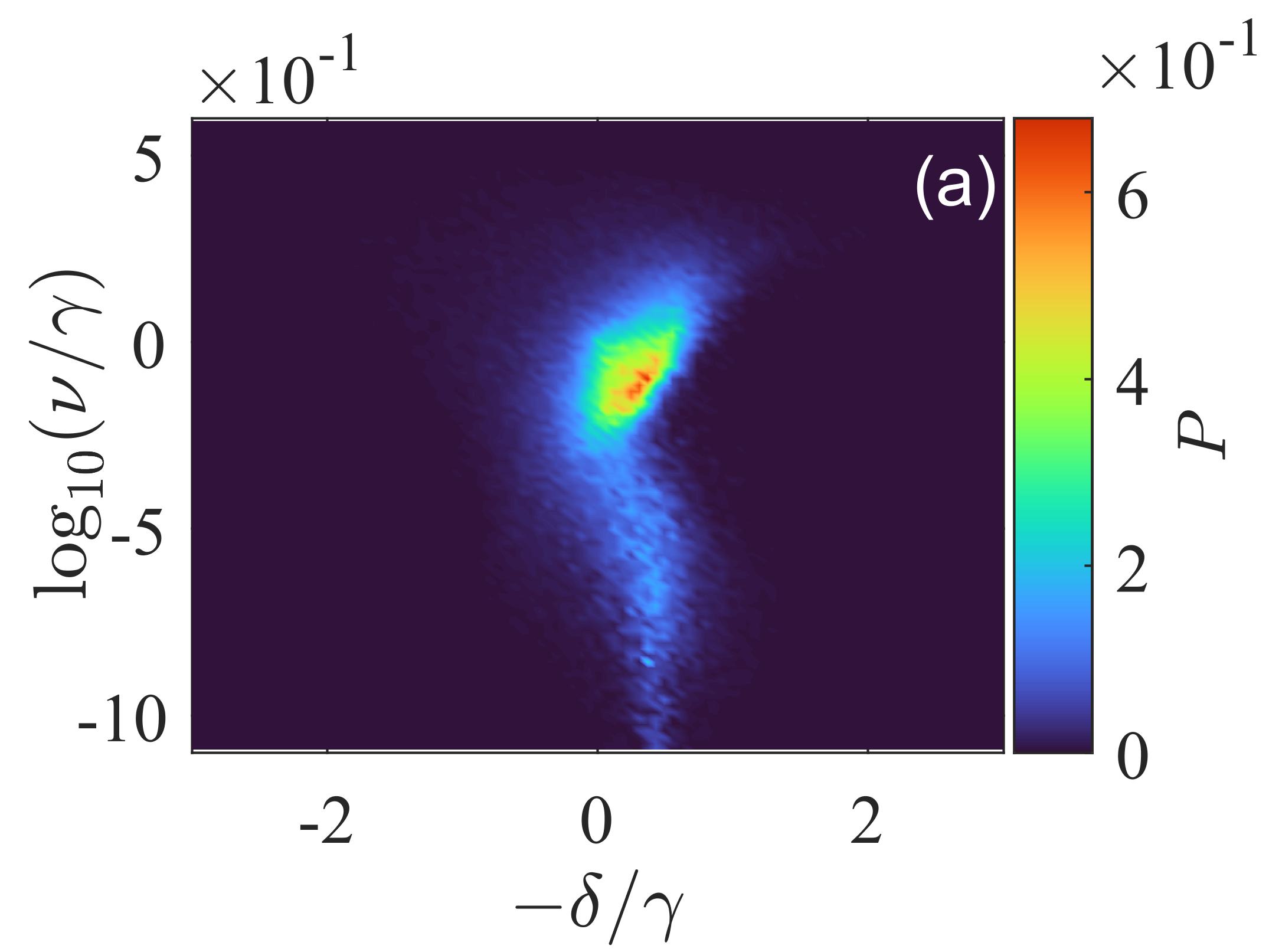}
\includegraphics[width=0.49\columnwidth]{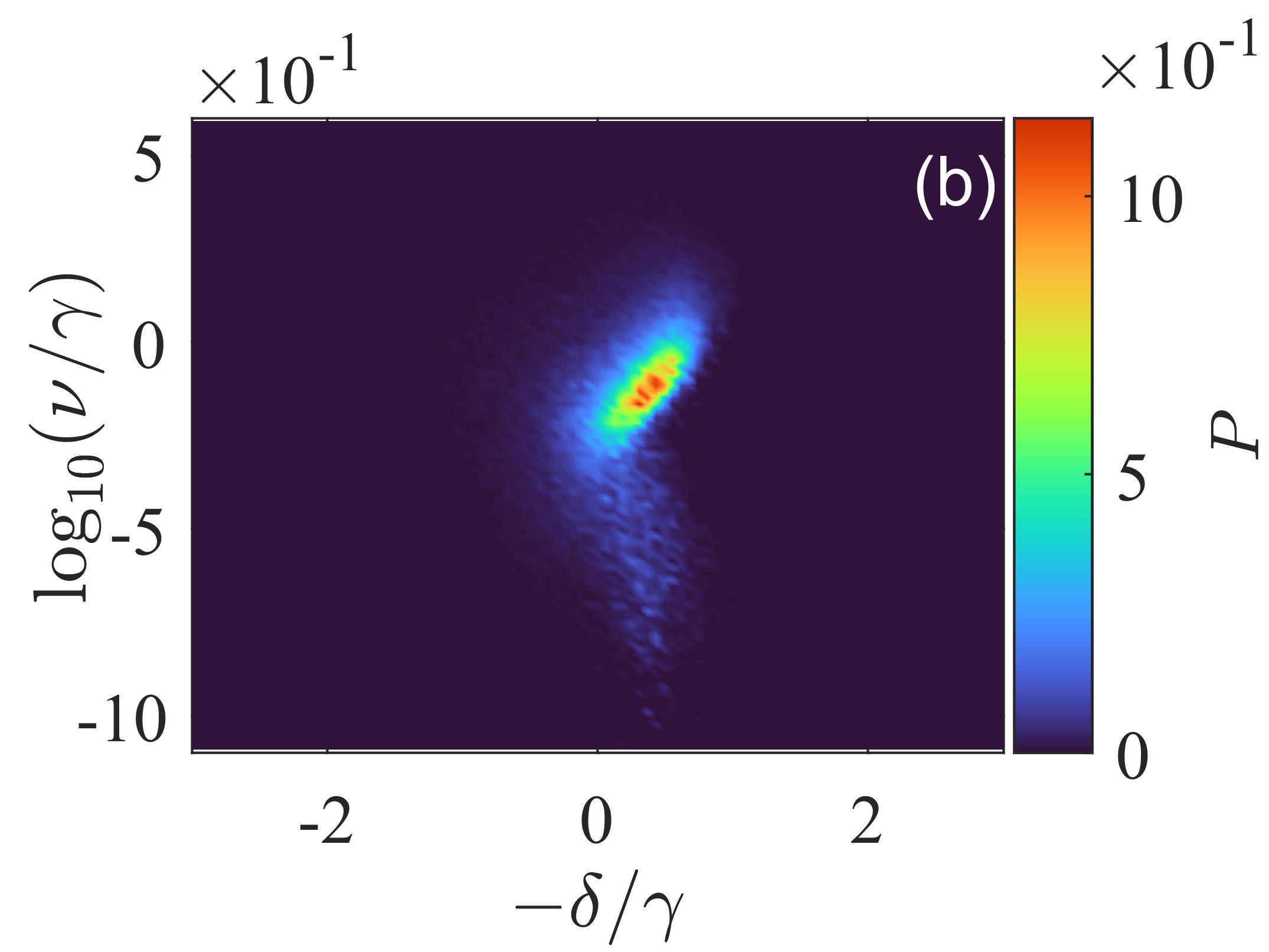}
\includegraphics[width=0.49\columnwidth]{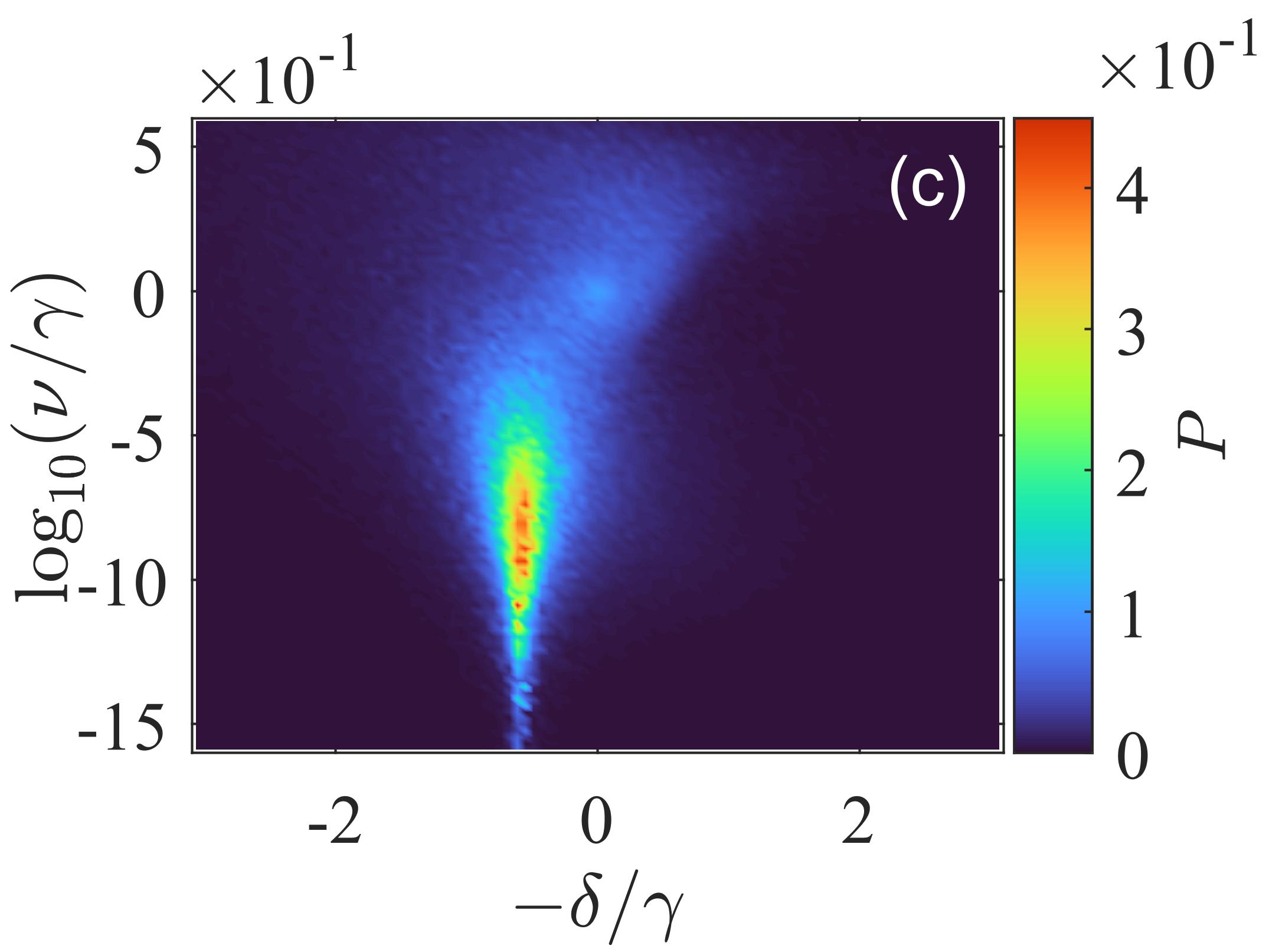}
\includegraphics[width=0.49\columnwidth]{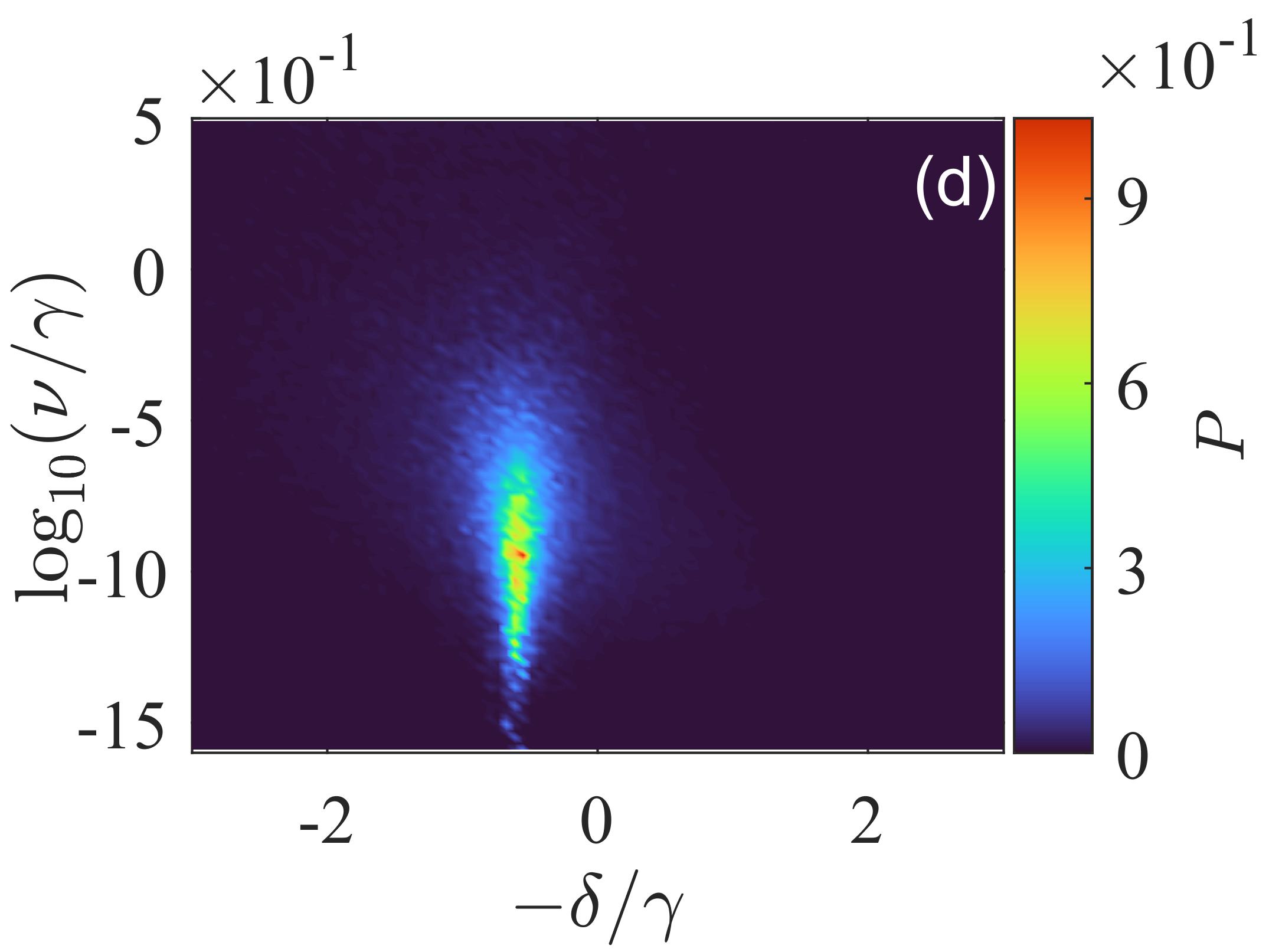}
    \vspace{-0.2cm}
    \caption{Occupation of collective excitation eigenmodes in the steady-state response of Fig.~\ref{fig: TransmissionDen004} at peak densities  \inline{\bar{\rho}_{\rm{2D}}/k^2 \simeq 0.1} (top row) and 1 (bottom row), at detunings $\Delta/\gamma = 0.4 $ and $-0.6$, respectively, as a function of the collective linewidth \inline{\nu} and line shift \inline{\delta}. (a), (c) Independent atoms; (b), (d)  \inline{R_{\rm{dip}}\sqrt{{\bar{\rho}}_{\rm{2D}}} \simeq  1.6} [Eq.~\eqref{eq:Rdip}].}
    \label{fig:Eigenmodes low den}
\end{figure}

We also calculate which eigenmodes are occupied at specific laser frequencies in Fig.~\ref{fig: TransmissionDen004}. Here Eq.~\eqref{eq: occupation} is used to produce a histogram plot of the occupied mode probability distribution in Fig.~\ref{fig:Eigenmodes low den}. 
Intriguingly, the presence of the static DD interactions allows for better targeted excitation of subradiant eigenmodes at high atom densities, as shown in terms of increased and more localized occupations in Fig.~\ref{fig:Eigenmodes low den}(c), (d)
for a subradiant eigenmode resonance at the detuning  \inline{\Delta\simeq-0.6\gamma}. 
Although the incident light strongly couples to a single eigenmode already for the case of independently distributed atoms, the coupling is even more selective in the presence of static DD interactions.
Similarly, due to the static coupling, it is more difficult to excite the modes off resonance. Although the modes are only excited over a narrow range of frequencies, they still extend over a wide range of linewidths due to the position fluctuations even with strong static interactions. 

Selectively exciting subradiant eigenmodes of atoms in oblate trapping geometries, as well as storing photons in such modes, has been actively investigated when the atoms form periodic arrays and individual atoms are trapped at specific spatial locations, see, e.g., Refs.~\cite{Facchinetti16,Ruostekoski23,Ballantine21quantum,Rubies22,Ballantine22bilayer,Manzoni18}. Here the driving of subradiant modes is studied instead in disordered ensembles with short-range ordering due to the static DD interactions. 
In Sec.~\ref{sec:dipolarprolate}, we show how these effects are even more pronounced in a prolate trap where small atom numbers also permit better targeting of individual eigenmodes. 


\subsection{Isotropic vs two-level transition}
\label{sec:iso}

\begin{figure}
    \centering
\includegraphics[width=0.49\columnwidth]{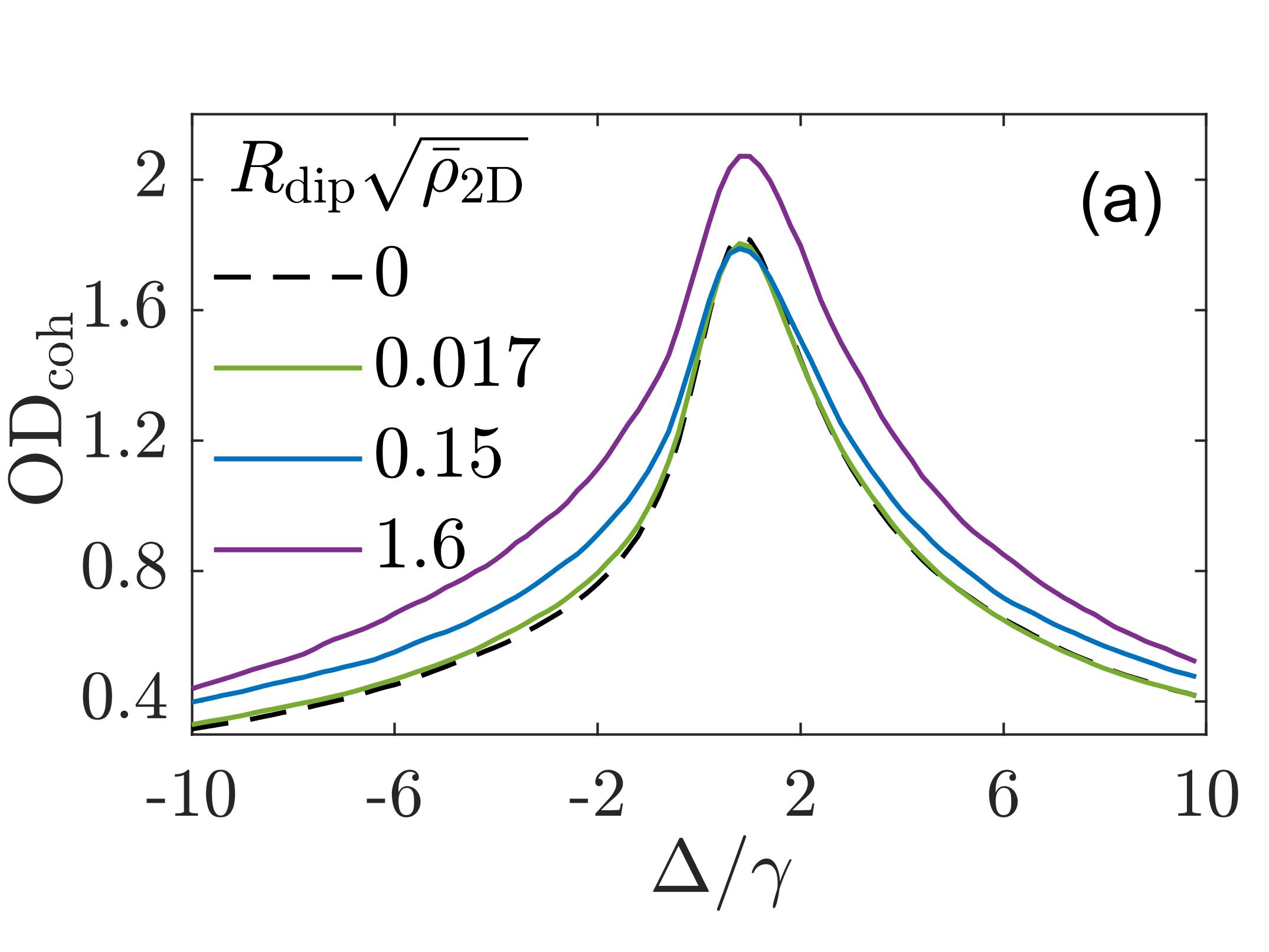}
\includegraphics[width=0.49\columnwidth]{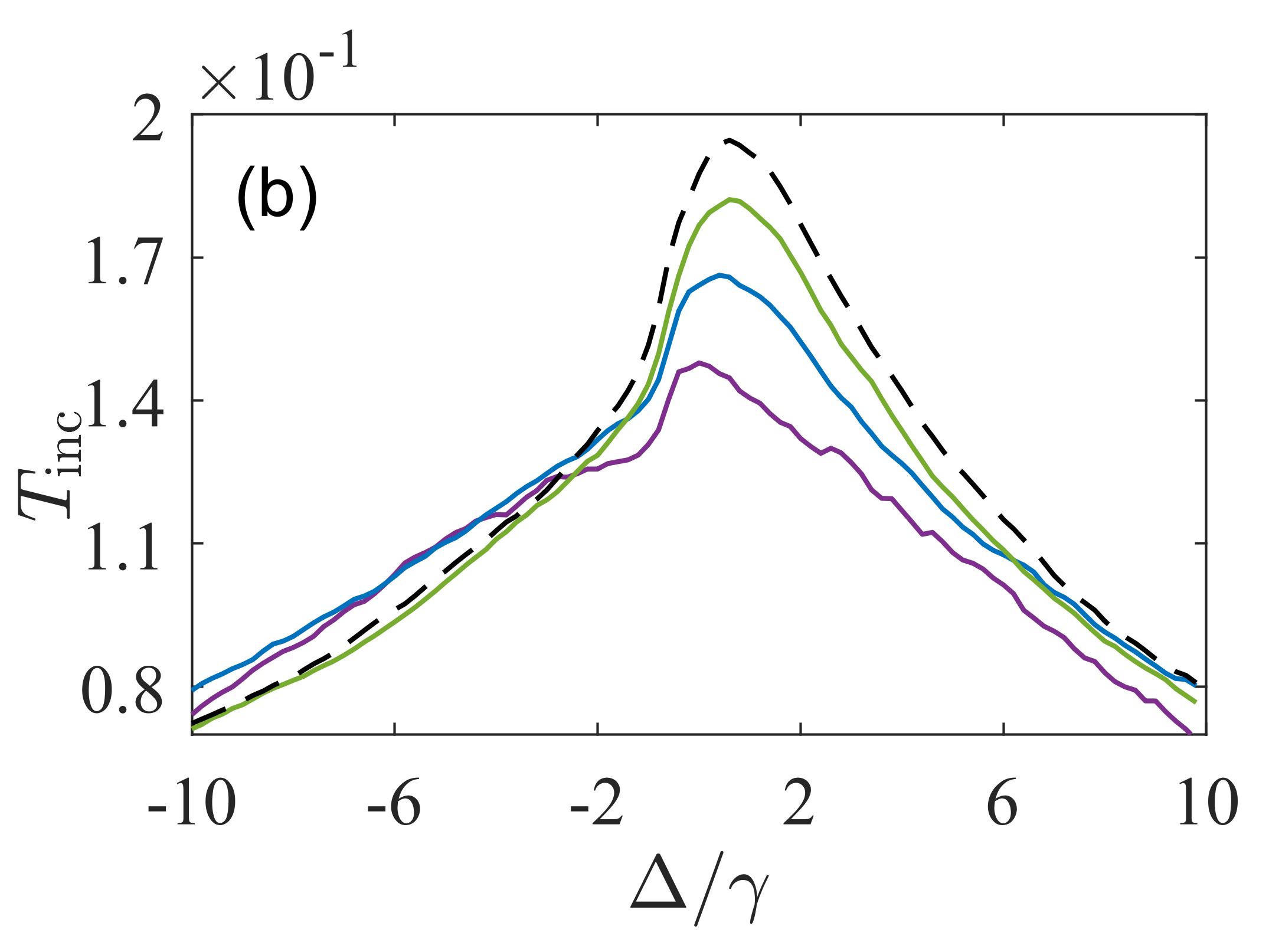}
    \includegraphics[width=0.49\columnwidth]{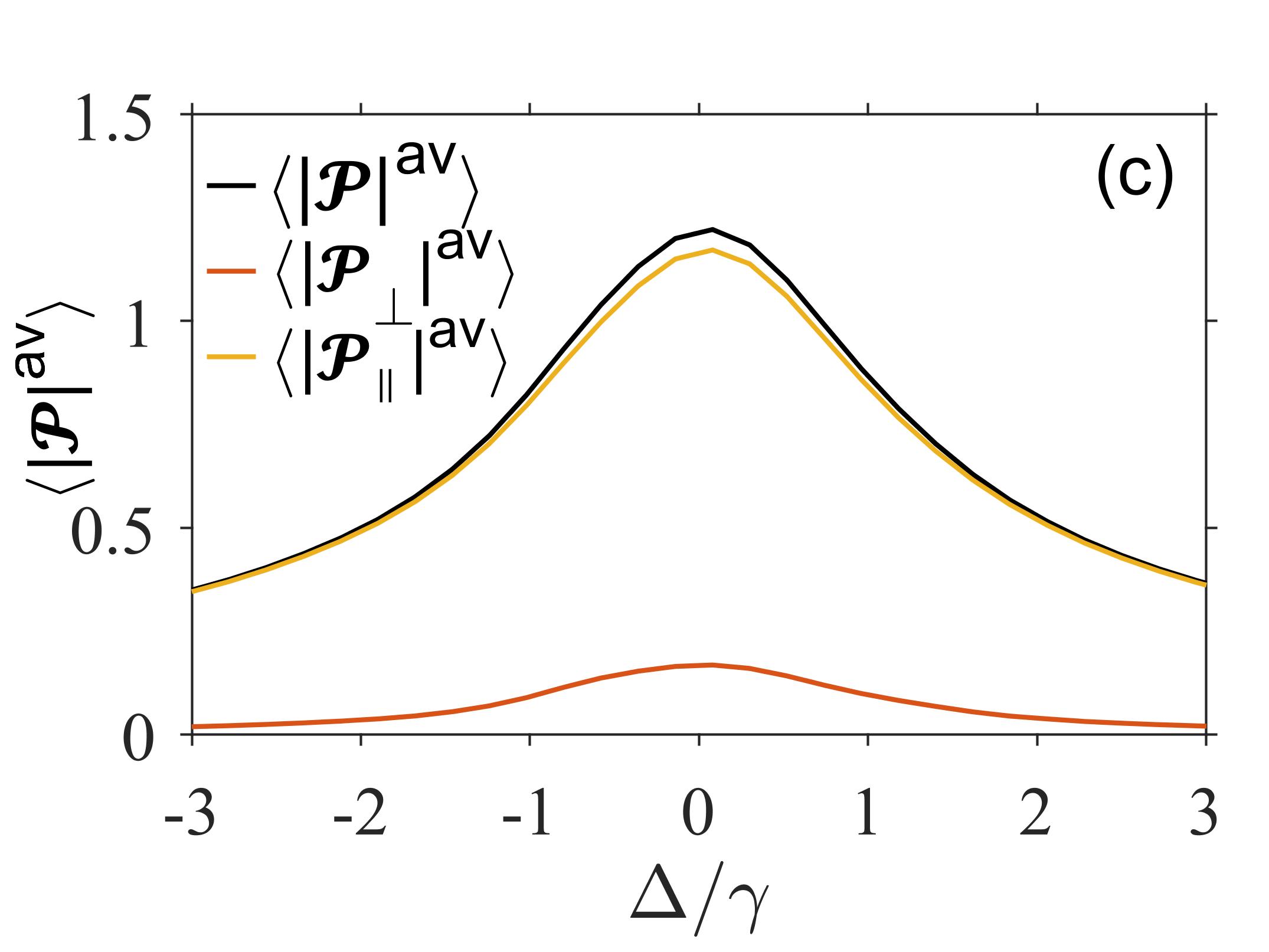}
   \includegraphics[width=0.49\columnwidth]{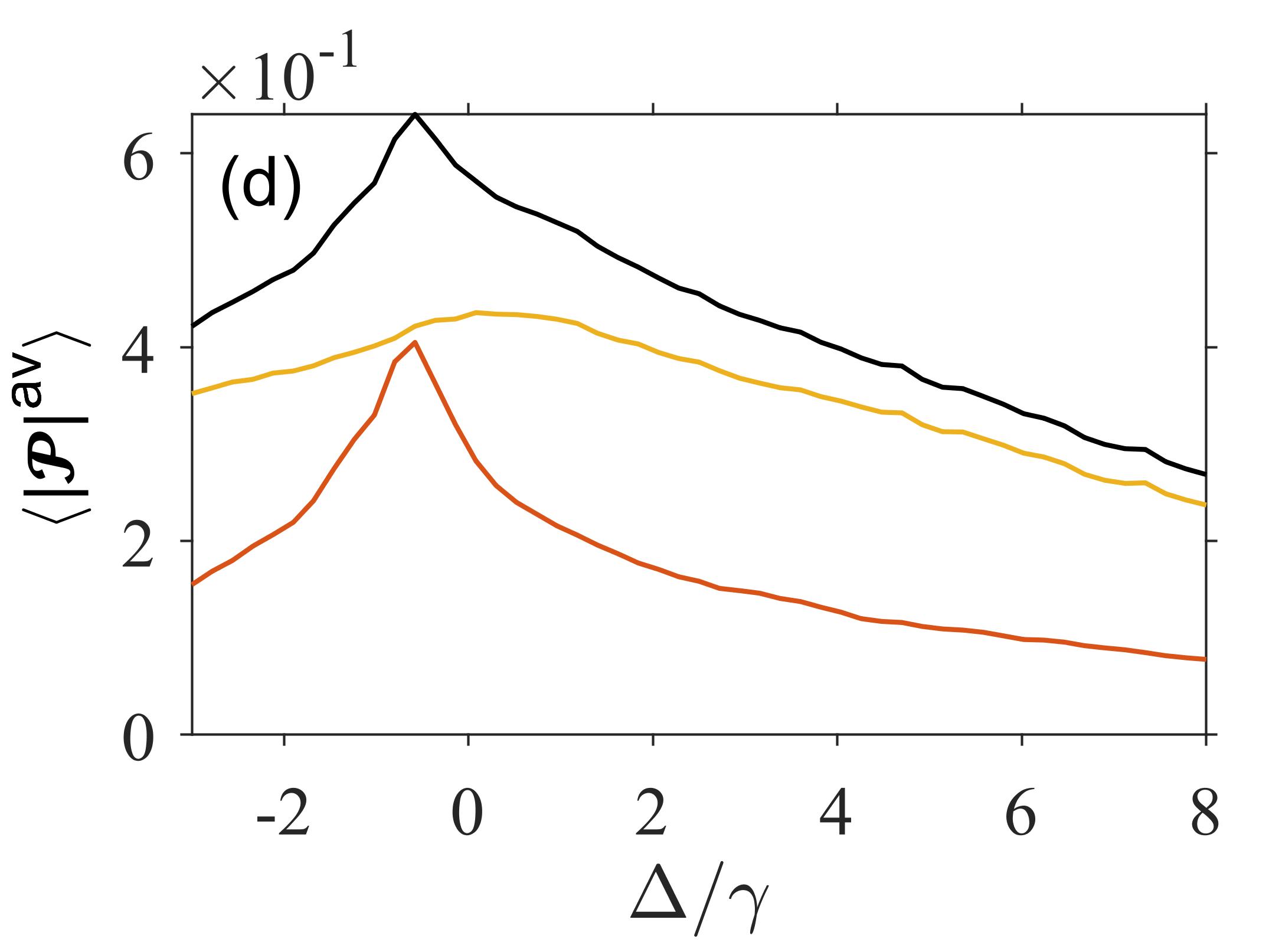}
\caption{(a) Coherent optical depth \inline{\rm{OD}_{\rm{coh}}}, (b) incoherent transmission \inline{T_{\rm{inc}}} as in Fig.~\ref{fig: TransmissionDen004} at \inline{\bar{\rho}_{\rm{2D}}/k^2\simeq1}, but for 
an isotropic $J=0\rightarrow J'=1$ transition.The expectation values of the magnitudes of the average excitation \inline{\< |\boldvec{\Pc}|^{\rm{av}}\>}, and its in-plane $\< |\boldvec{\Pc}_\parallel|^{\rm{av}}\>$, and out-of-plane $\< |\boldvec{\Pc}_\perp|^{\rm{av}}\>$ components for \inline{R_{\rm{dip}}\sqrt{{\bar{\rho}}_{\rm{2D}}} \simeq 0.15} for (c) \inline{\bar{\rho}_{\rm{2D}}/k^2\simeq0.1} and (d) 1.
}
    \label{fig: Transmissioniso}
\end{figure}

The optical response of two-level atoms in Fig.~\ref{fig: TransmissionDen004} is qualitatively similar to the response of atoms with an isotropic $J=0\rightarrow J'=1$ transition to positive circular polarization in the trap plane. At low densities, the transmission and reflection lineshapes are very similar (Fig.~\ref{fig: Transmissioniso}). At the high density case with \inline{\bar{\rho}_{\rm{2D}}/k^2\simeq1}, the coherent scattering is still only slightly modified, but more notable deviations in the lineshape appear in the incoherent transmission.
In the case of the $J=0\rightarrow J'=1$ transition, the incoherent scattering has a dominant peak at $\Delta\simeq 0$ and a smaller peak at $\Delta\simeq-2\gamma$, while in the two-level case
 the $\Delta\simeq0$ peak is less pronounced.
 
In the isotropic case, the dipoles can be excited in the direction normal to the trap plane, despite these components not being directly driven by the incident light. This can result in scattering that is not captured by the lenses. To calculate this out-of-plane excitation, we define the expectation values of the magnitudes of the atomic polarization components,
\begin{equation}
   \langle{|\boldvec{\Pc}^{\rm{av}}|}\rangle= \sum_{q=1}^Q \frac{1}{Q} \left|\sum_{j=1}^N \frac{\boldvec{\Pc}^{(j,q)}}{N}  \right|,
   \label{eq: Expected Polarisation}
\end{equation}
where  $q$ denotes the stochastic realization and \inline{Q} is the total number of realizations.  The in-plane \inline{\langle|\boldvec{\Pc}_\parallel|^{\rm{av}}\rangle} and out-of-plane  \inline{\langle|\boldvec{\Pc}_\perp|^{\rm{av}}\rangle} excitation magnitudes are shown in Fig.~\ref{fig: Transmissioniso}. While the out-of-plane component is weak 
at low density \inline{\bar{\rho}_{\rm{2D}}/k^2\simeq0.1}, it becomes substantial at \inline{\bar{\rho}_{\rm{2D}}/k^2\simeq1}. The resonance of the out-of-plane component is narrower and shifted to the negative detuning, dominating the resonance of the total excitation.

\subsection{Inhomogeneous broadening due to static dipoles}
\label{sec:Zeeman}

So far we have studied the effect of the static DD interactions on optical responses through the change of the electronic ground-state atom distributions and correlations.
Depending on the physical system considered, the static DD coupling can also influence the optical transition frequencies. Since the static DD field experienced by each atom depends on the relative positions of the other atoms, the effect is not uniform throughout the sample. For example, in the case of static magnetic DD interactions where the electronic ground and excited levels of each atom exhibit a magnetic dipole, the atoms can experience level shifts due to ground state -- ground state, ground state -- excited state, and excited state -- excited state interactions which depend on the specific level structure and atom.\footnote{For a discussion of level shifts in such a scenario due to two-body interactions in the LLI limit, see, e.g, Ref.~\cite{bcslinelong}.} For simplicity, we consider a general model to demonstrate the effect of inhomogeneous broadening of resonance frequencies that results from the nonuniformly experienced static DD interactions between the atoms. We introduce position dependent transition frequency shifts in individual atoms
\begin{equation}
       \delta^{(j)} =D  \sum_\ell \left(\frac{3 z_{j\ell}^2 }{|\boldvec{r}_{j\ell}| ^5}- \frac{1}{|\boldvec{r}_{j\ell}| ^3}\right), 
       \label{Eq: Zeeman Coupling}
\end{equation}
where the shift in the atom $j$ at $\rv_j$ is caused by the static DD interaction of all the other atoms $\ell$ at positions $\rv_\ell$ and $D$ denotes the effect of the static field on the transition frequency.

Introducing Eq.~\eqref{Eq: Zeeman Coupling} in the simulations broadens the distribution of the atomic transition frequencies in the ensemble and shifts the peak value of the atomic transition frequency as a function of $D$. In optical responses, the broadening is analogous to inhomogeneous broadening in thermal and other ensembles~\cite{Javanainen2014a,Jenkins_thermshift,Palffy21}. The transition frequencies are most strongly shifted for any atom pairs that are very close to each other [see the ground state -- ground state coupling in Fig.~\ref{fig:shape change}(a)].
However, the repulsive force between the atoms reduces the likelihood of very strong shifts; the DD interaction between the electronic ground level atoms inhibits atomic distributions with short interatomic separations.

\begin{figure}
    \centering
\includegraphics[width=0.49\columnwidth]{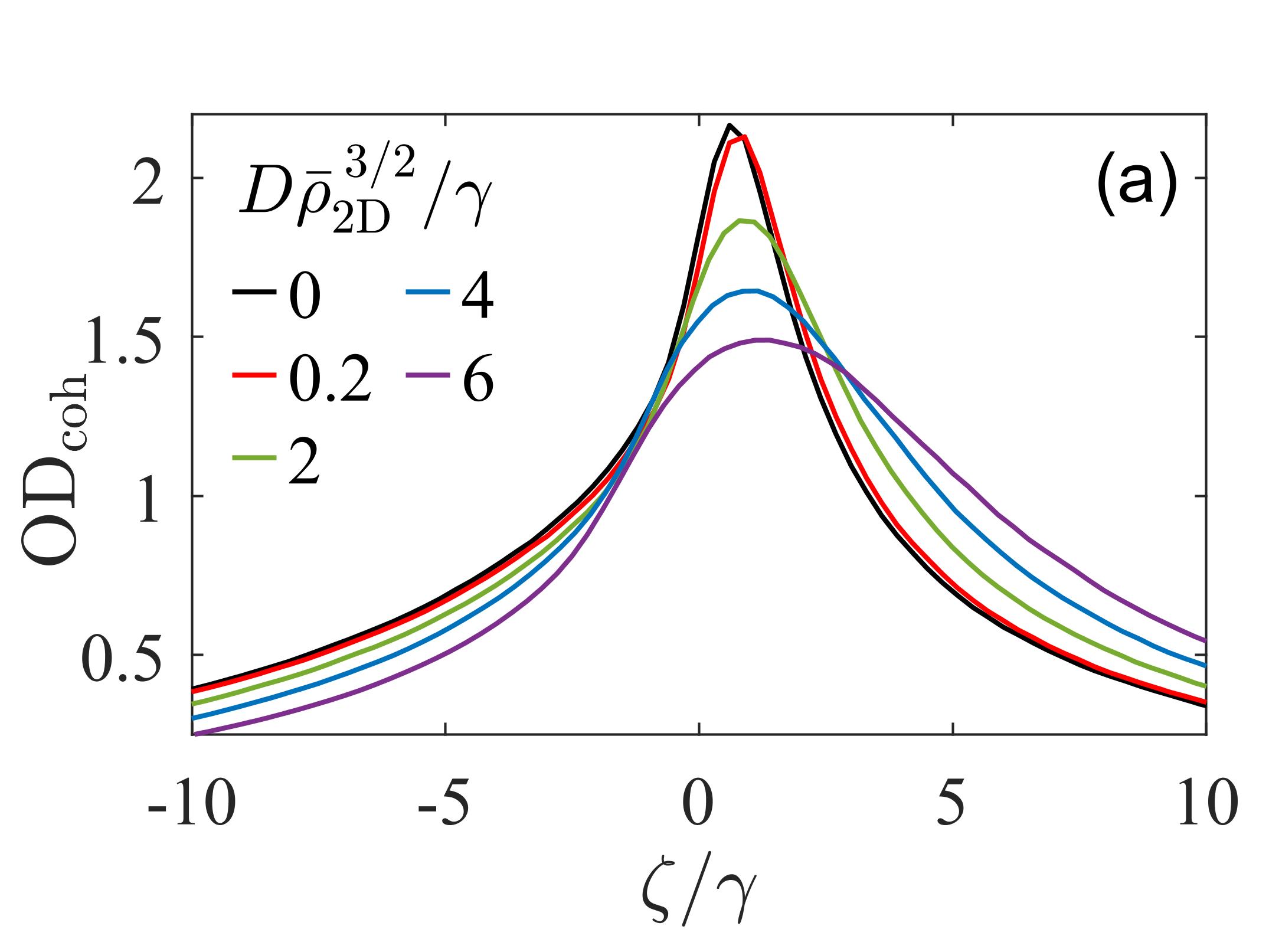}
    \includegraphics[width=0.49\columnwidth]{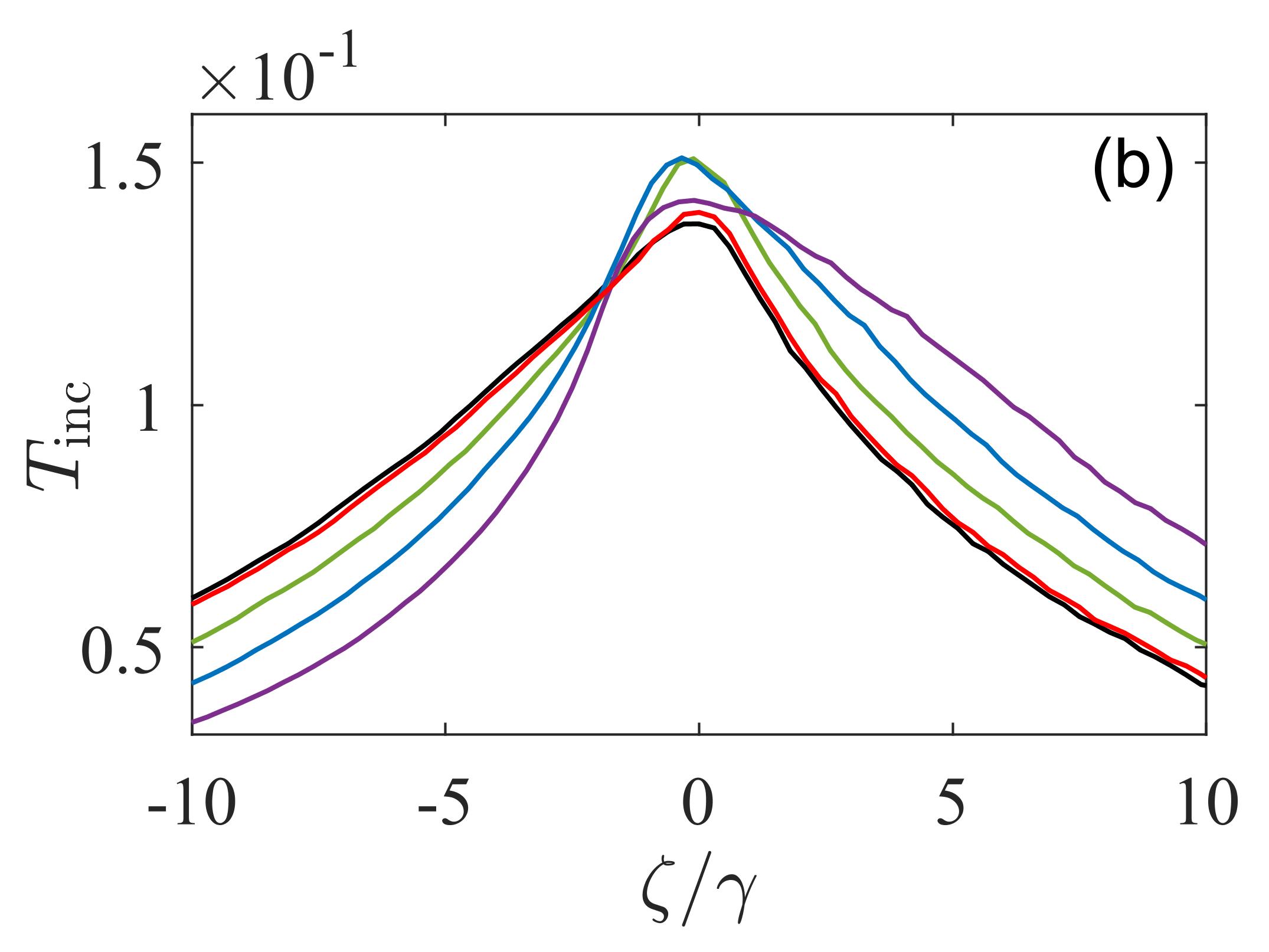}
     \includegraphics[width=0.49\columnwidth]{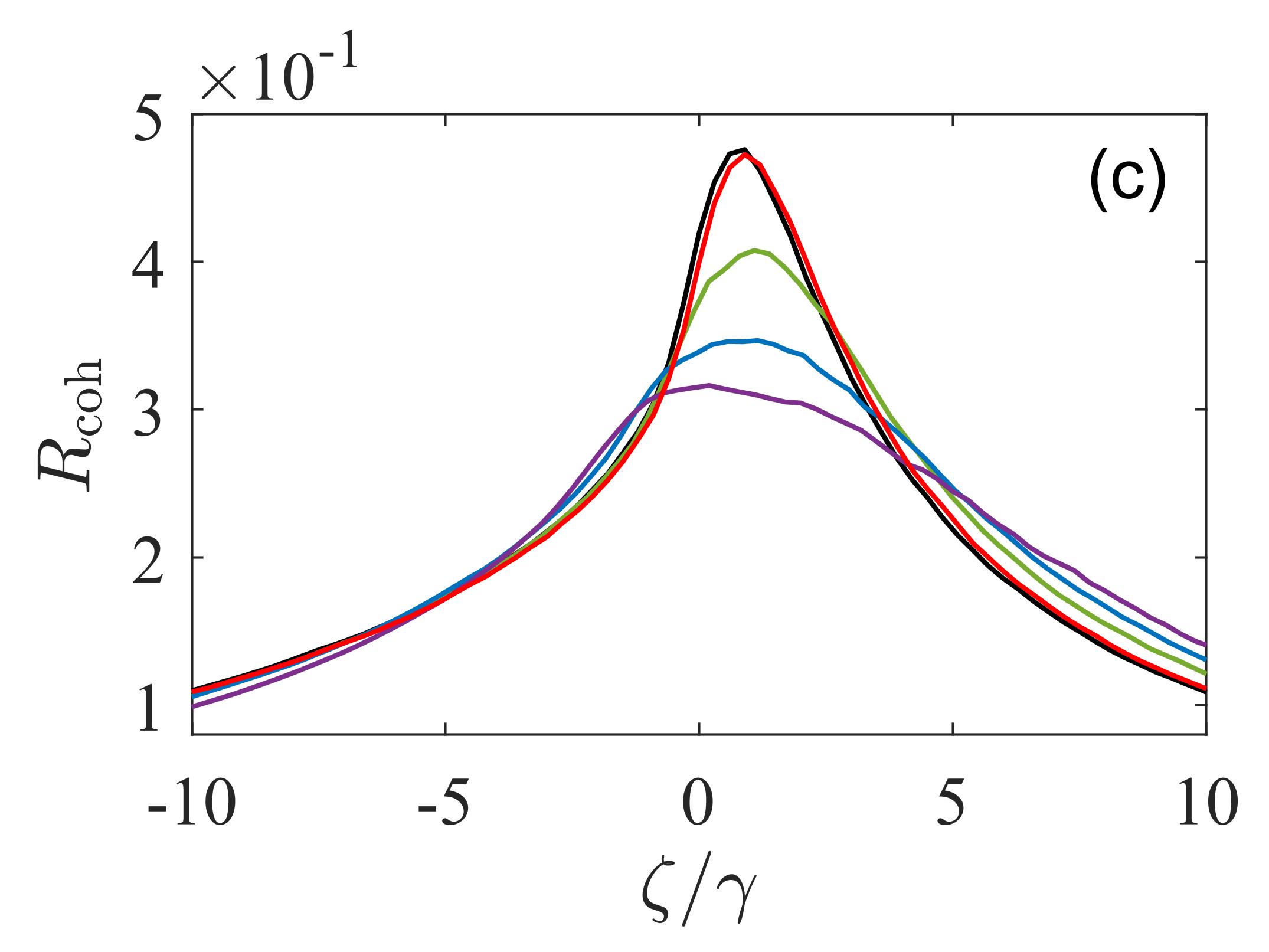}
      \includegraphics[width=0.49\columnwidth]{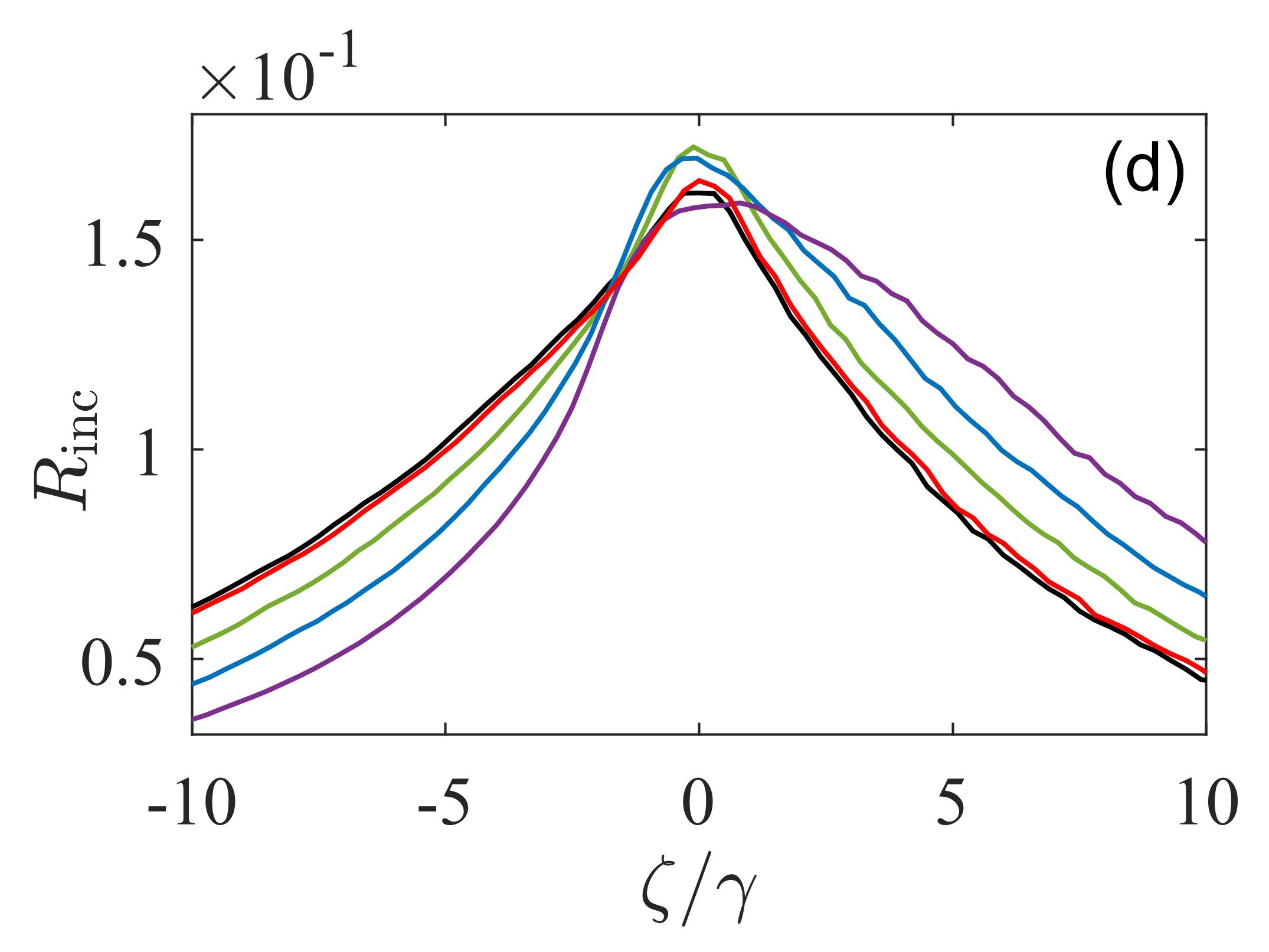}
          \vspace{-0.2cm}
      \caption{Inhomogeneously broadened transition frequencies induced by static dipolar interactions for two-level atoms in an oblate trap for $N=200$, \inline{\ell_x k= 6}, \inline{\; \ell_z k= 0.15 }, \inline{R_{{\rm dip}}k = 0.15}  at  \inline{\bar{\rho}_{\rm{2D}}/k^2\simeq1}.  (a) Coherent optical depth \inline{\rm{OD}_{\rm{coh}}}, 
     (b) incoherent transmission \inline{T_{\rm{inc}}}, (c) coherent reflection \inline{R_{\rm{coh}}}  and (d) incoherent reflection \inline{R_{\rm{inc}}} [Eqs.~\eqref{eq: Transmission}-\eqref{eq:Es expectation}]  with coupling strength \inline{D}~[Eq.~\eqref{Eq: Zeeman Coupling}] in terms of detuning \inline{\zeta} from the most likely atomic transition frequency $\bar\delta$ in each case.   }
     \label{fig: Weak zeeman trans}
\end{figure}

The nonuniform change in the resonance frequencies broadens the transmission and reflection resonances at different coupling strengths in Fig.~\ref{fig: Weak zeeman trans}. To make the different cases comparable, the resonances are shown with respect to the most likely shift in the transition frequency \inline{\bar{\delta}} in each case. 
The consequences of these level shifts on Dy atoms are discussed in Sec.~\ref {sec:conc} where we argue that their effect on the optical response is negligible.

\subsection{Varying static dipolar interaction in a prolate trap}
\label{sec:dipolarprolate}

\begin{figure}
    \centering
      \includegraphics[width=0.49\columnwidth]{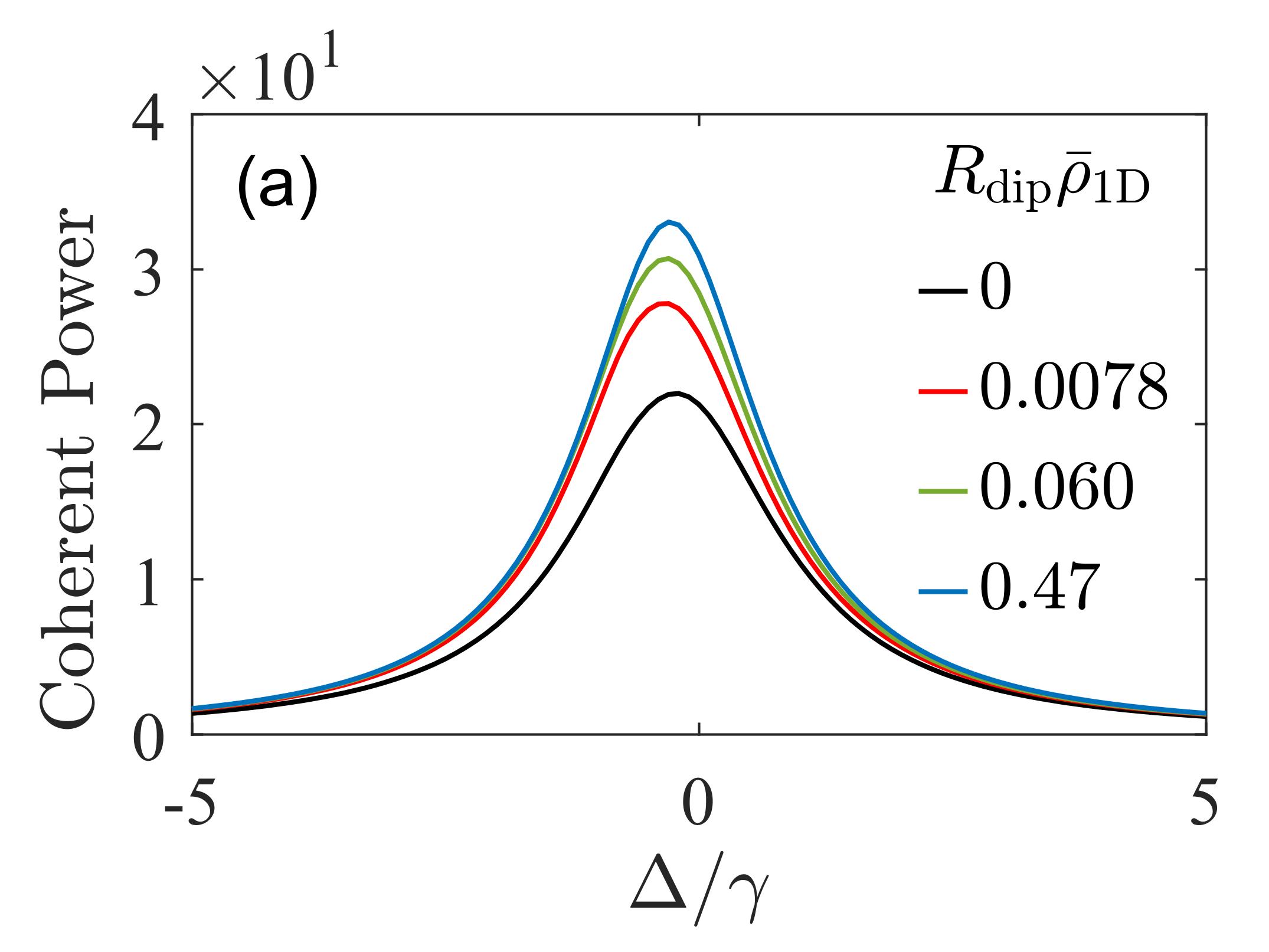} 
      \includegraphics[width=0.49\columnwidth]{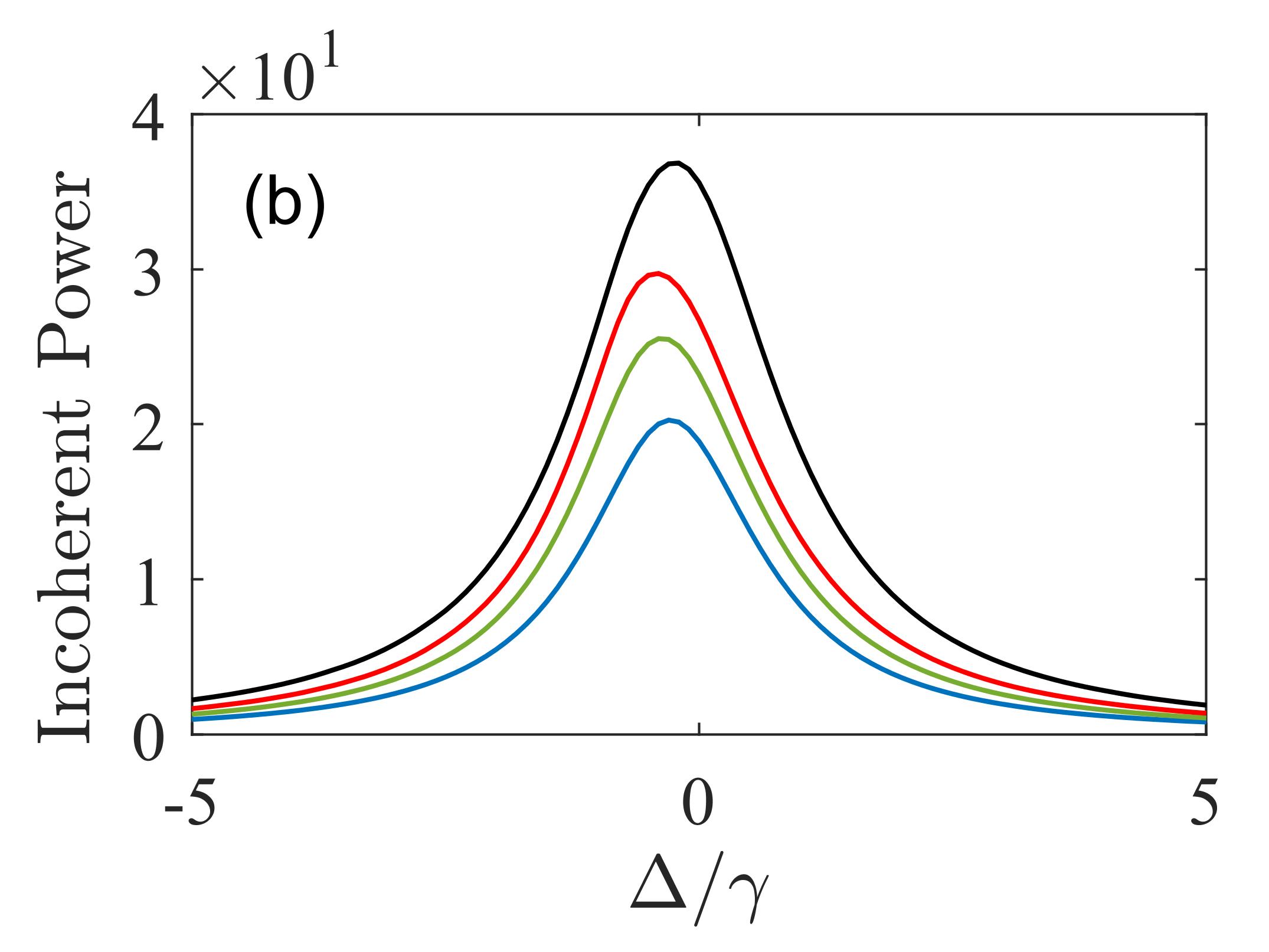}
          \includegraphics[width=0.49\columnwidth]{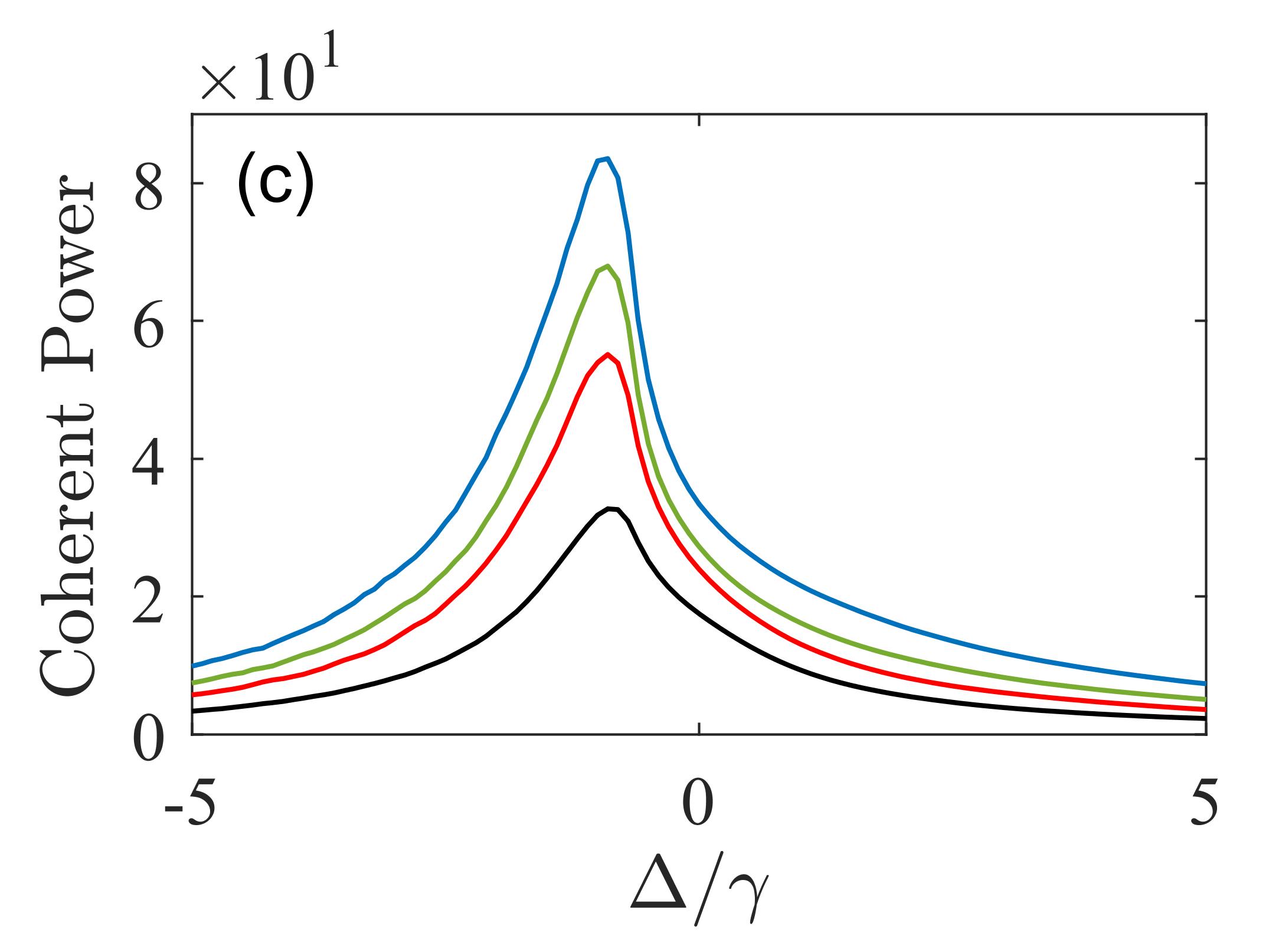}
     \includegraphics[width=0.49\columnwidth]{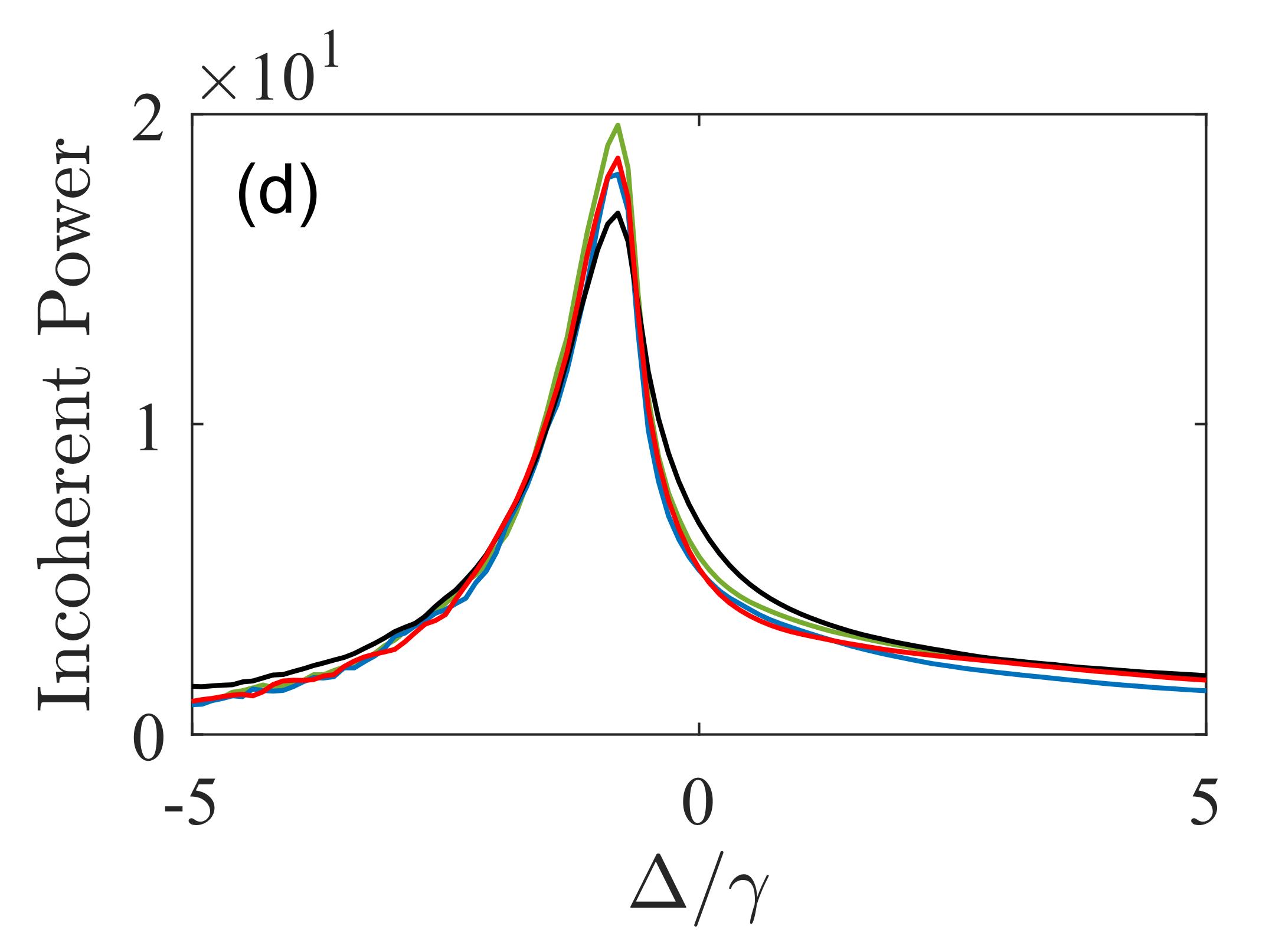}
     \includegraphics[width=0.49\columnwidth]{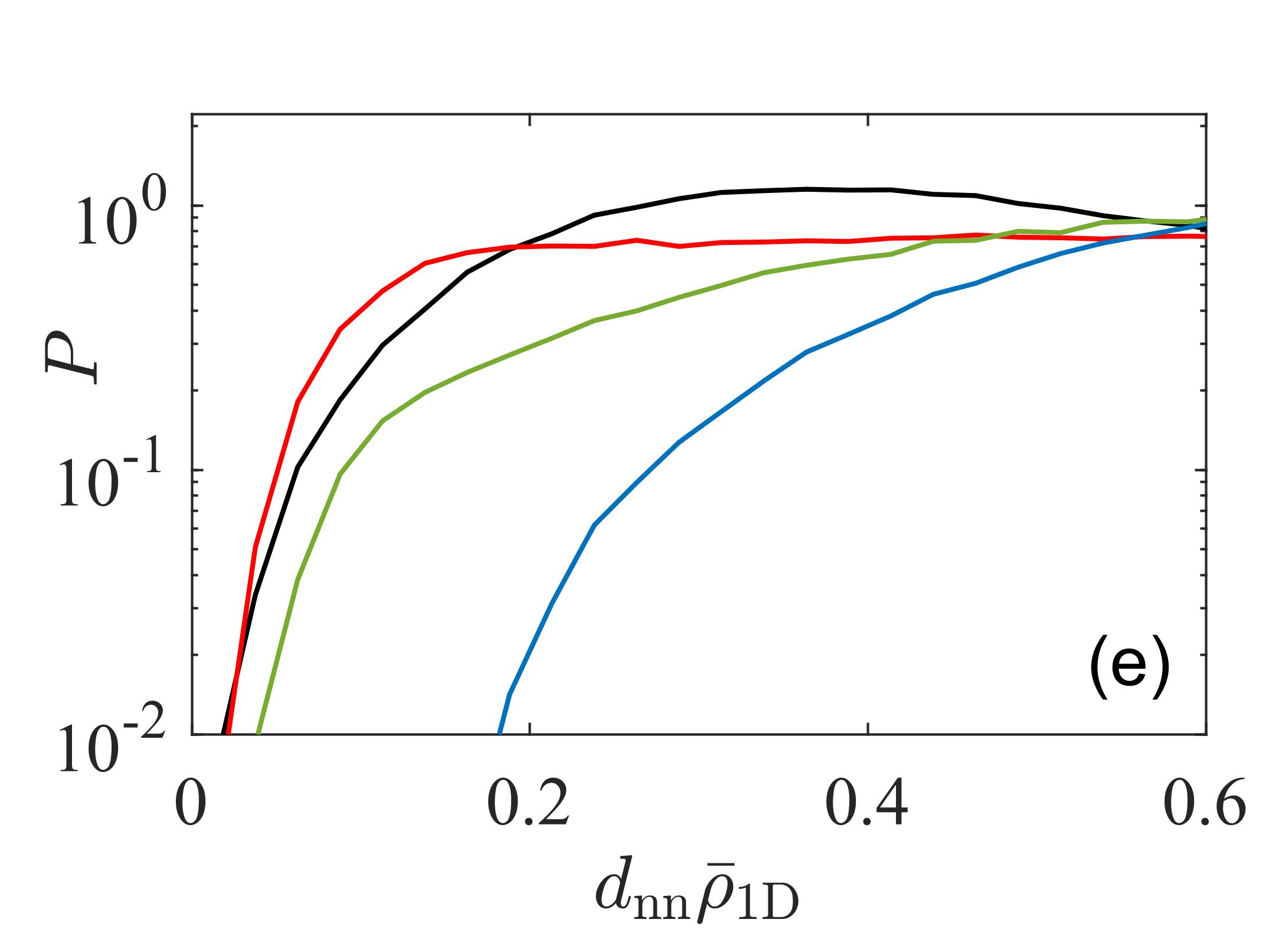}
  \includegraphics[width=0.49\columnwidth]{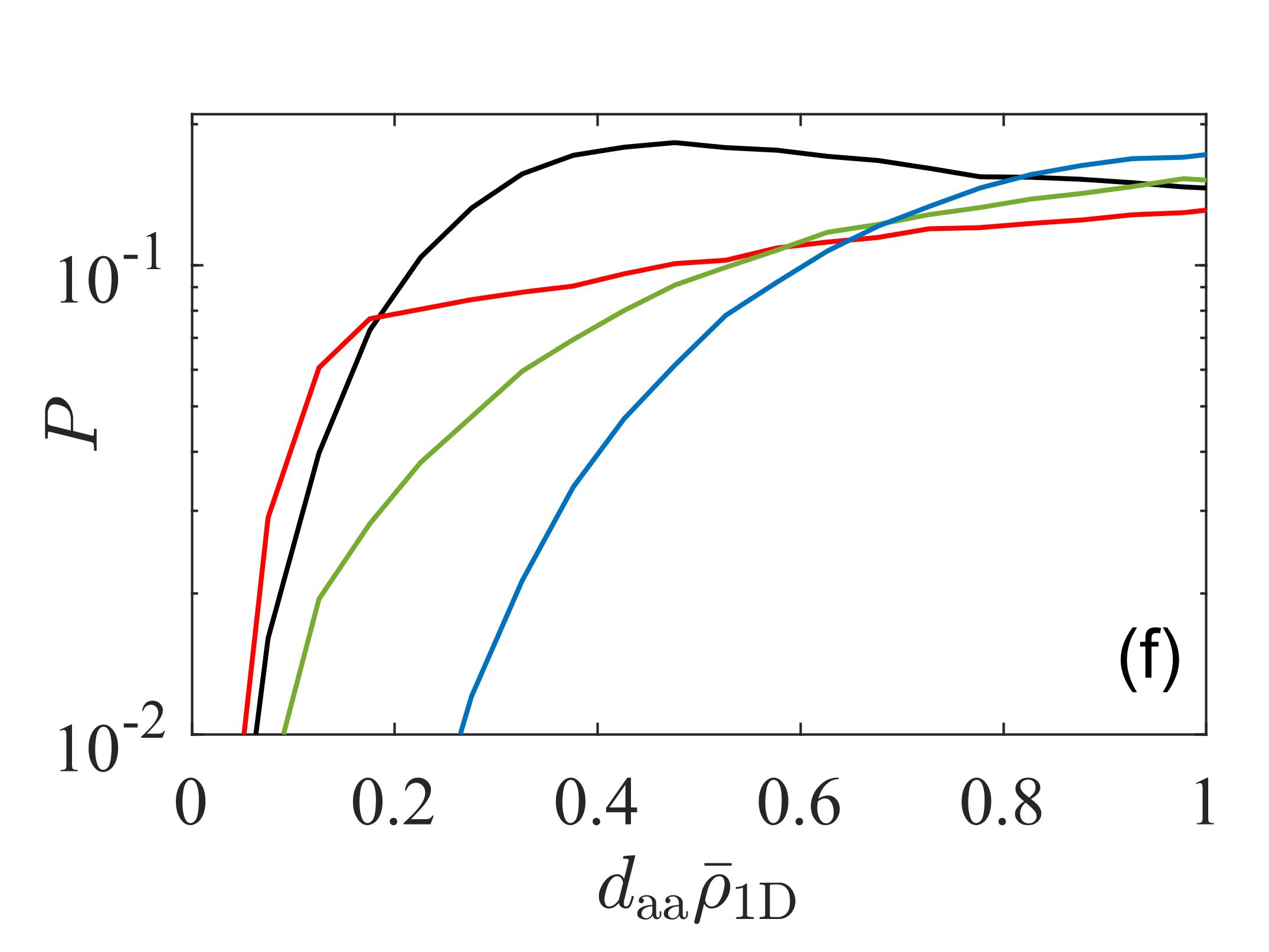}
      \vspace{-0.2cm}
    \caption{Coherently and incoherently forward-scattered power (in units of incident light intensity/$k^2$) from 10 atoms in a prolate trap (\inline{\ell_z/ \ell_x=25}) for different static dipole interaction strengths at the peak densities (a), (b) \inline{\bar{\rho}_{\rm{1D}}/k\simeq0.1} and (c), (d) \inline{\bar{\rho}_{\rm{1D}}/k\simeq1}, and (e), (f) the atomic pair distributions. The light propagates parallel to the long trap axis.  Normalized (e) nearest-neighbor and  (f) all-pair distributions between the atoms. The lens NA 0.8.}
    \label{fig: on-axis transmission}
\end{figure}

\begin{figure}
\includegraphics[width=0.49\columnwidth]{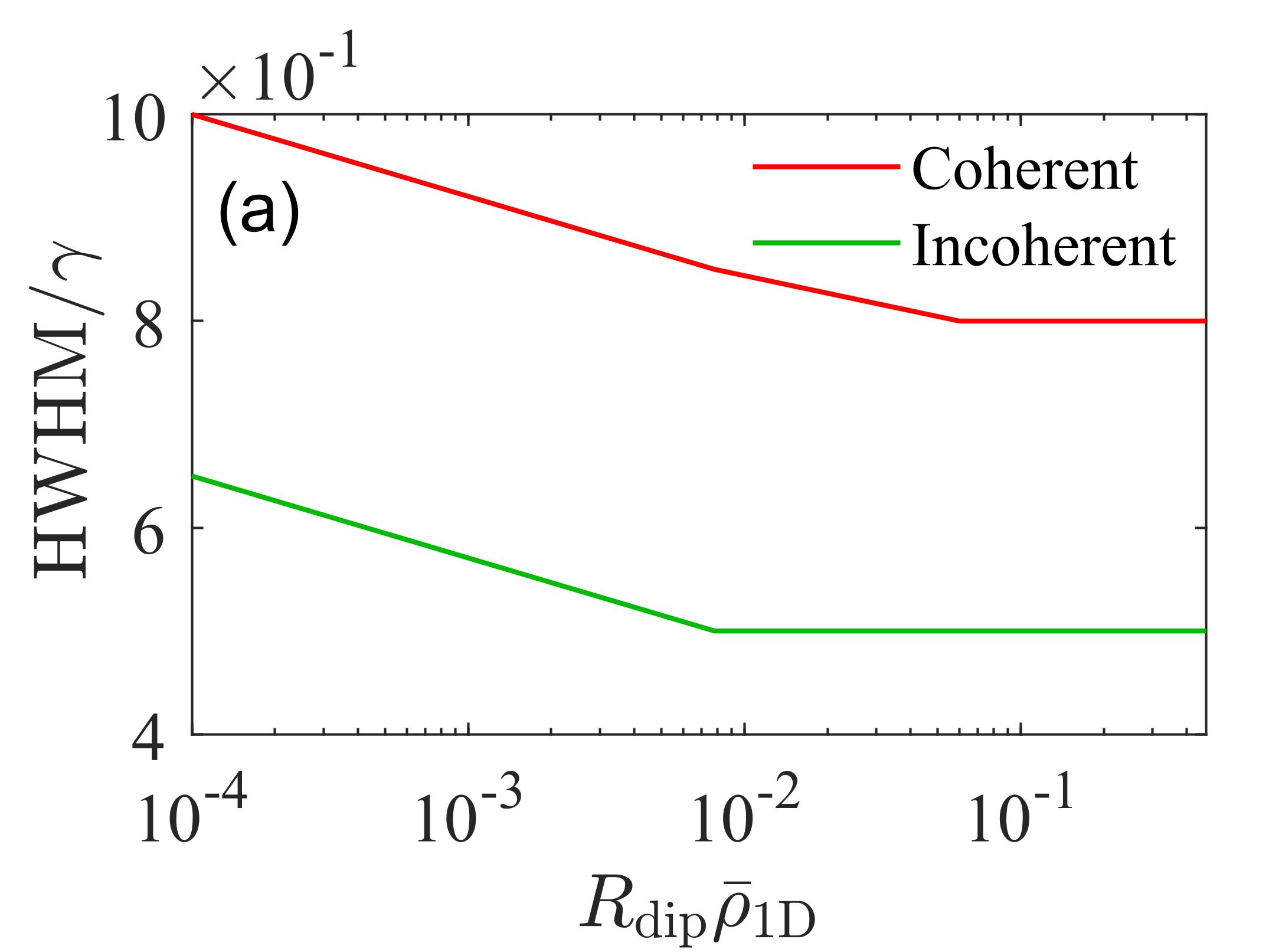}
\includegraphics[width=0.49\columnwidth]{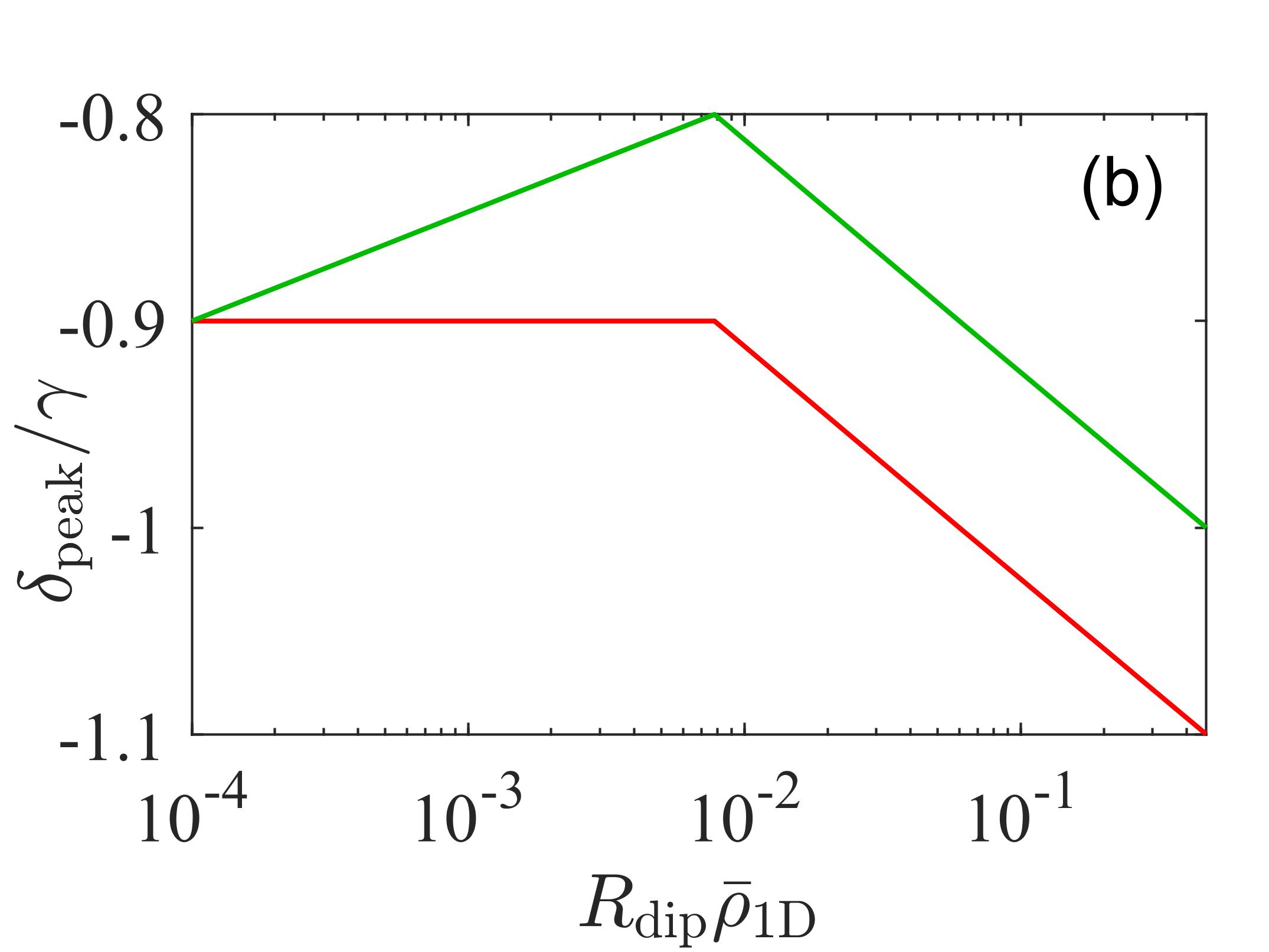}
    \vspace{-0.2cm}
\caption{Scattered light resonance (a) widths and (b) shifts of maxima from Fig.~\ref{fig: on-axis transmission} for  \inline{\bar{\rho}_{\rm{1D}}/k\simeq1}. The HWHM is calculated by averaging the contributions of the left and right sides of the peak.
}
\label{fig: Lineshape Params on-axis}
\end{figure}

We consider $N=10$ atoms trapped in a prolate trap at peak densities \inline{\bar{\rho}_{\rm{1D}}/k \simeq 0.1} and 1. The static dipoles again are all parallel and repulsive, oriented perpendicular to the long trap axis. The atoms, with the $J=0\rightarrow J'=1$ transition, are illuminated by a Gaussian beam, with positive circular polarization, propagating either along the long axis of the trap (`on-axis') or perpendicular to it (`off-axis'). The on-axis scattered light power and the atomic pair distributions are shown in  Fig.~\ref{fig: on-axis transmission}. At low atom densities, the lineshape is close to the Lorentzian with the resonance near the single atom resonance, while at high densities it is asymmetric.
The  coherent scattering resonance is shifted towards negative detuning. Similarly to the oblate case in Sec.~\ref{sec:dipolaroblate}, we find the resonance narrowing with increasing \inline{R_{{\rm dip}}} (Fig.~\ref{fig: Lineshape Params on-axis}), increasing coherent scattering, and decreasing incoherent scattering, but now the HWHM resonances are very narrow and below the linewidth of an isolated atom. These changes due to \inline{R_{\rm{dip}}} originate from the short-range ordering of the atoms  that suppresses the fluctuations of the light-mediated DD interactions.

\begin{figure}
\includegraphics[width=0.49\columnwidth]{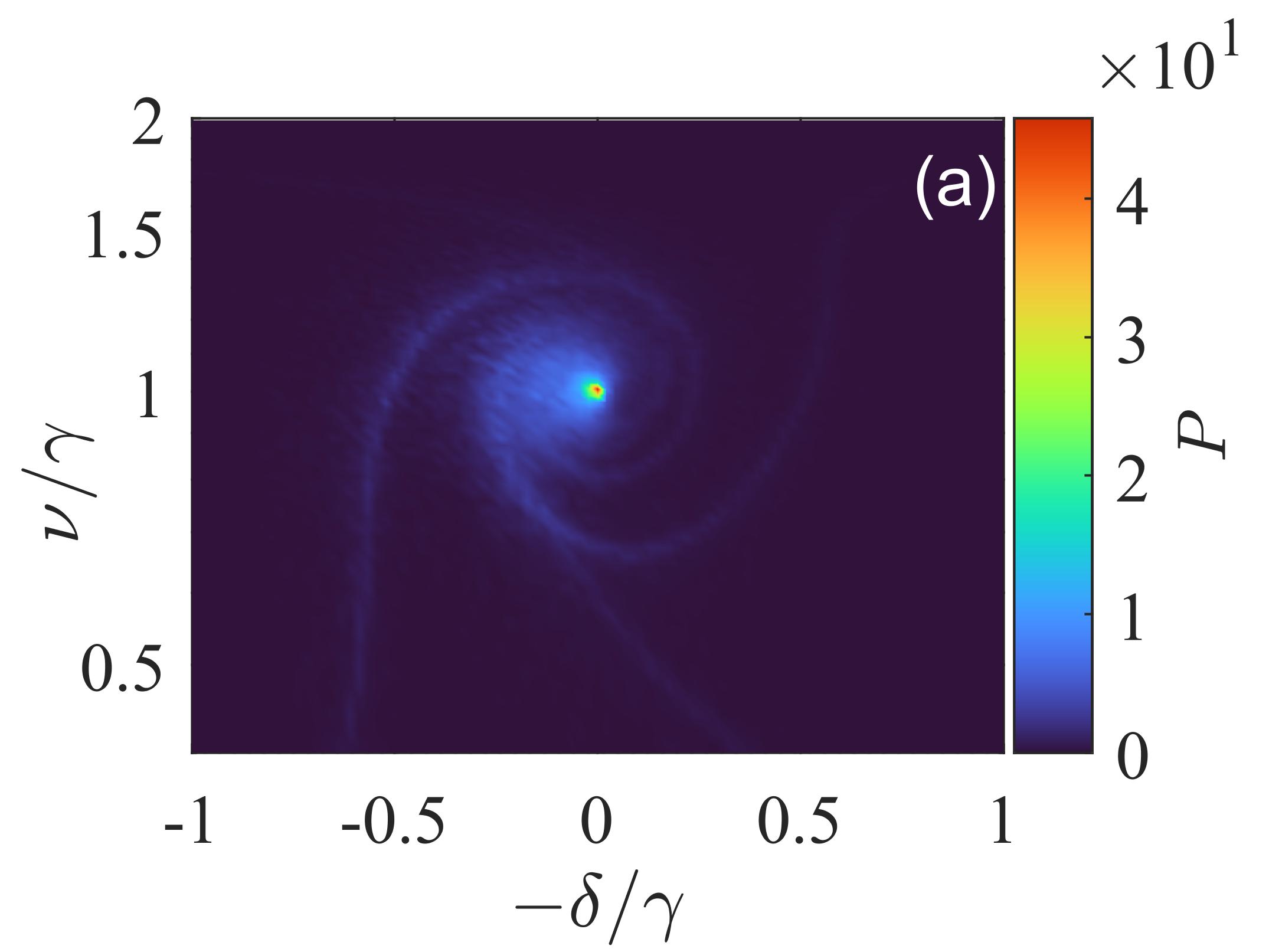}
\includegraphics[width=0.49\columnwidth]{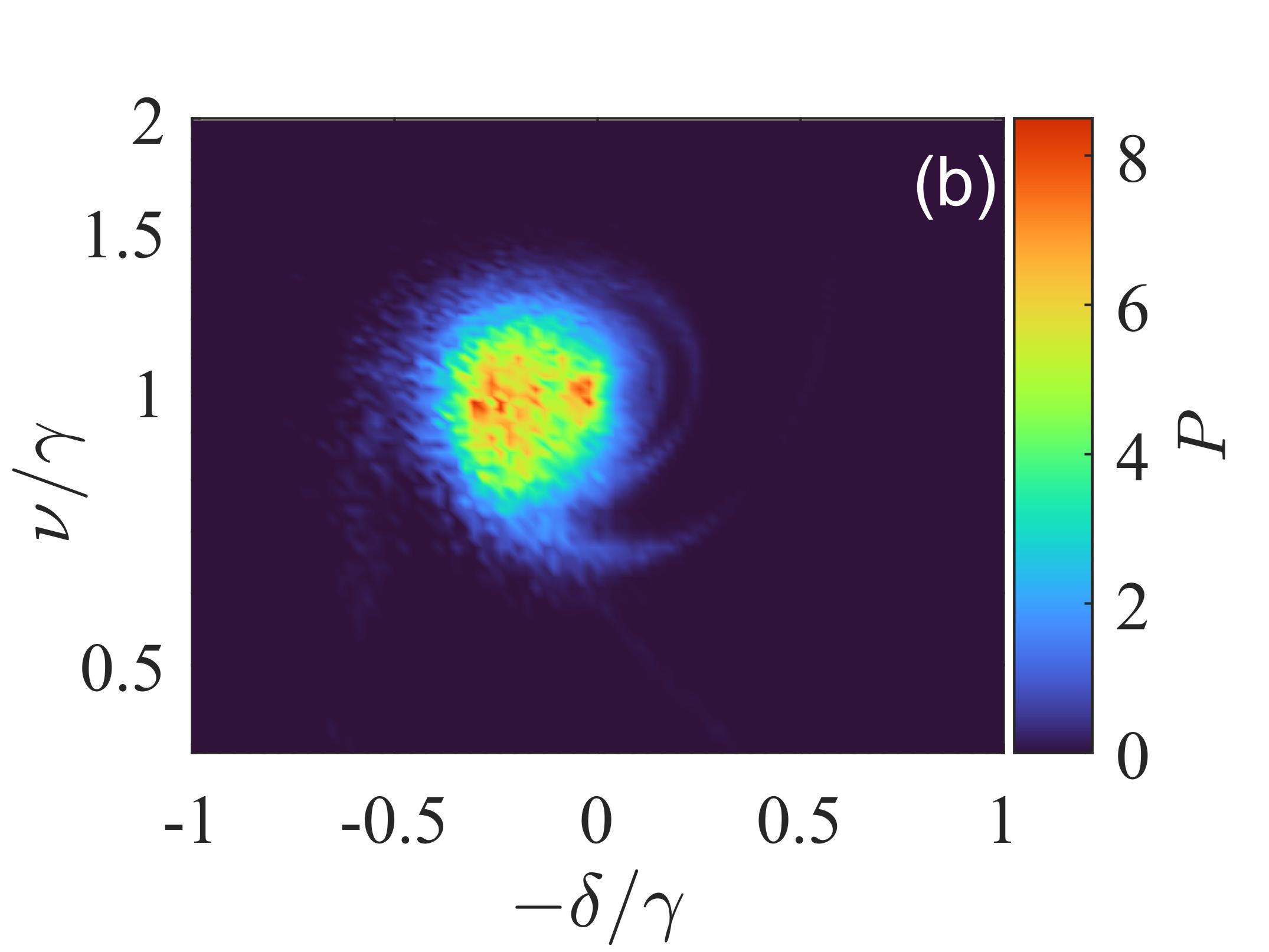}
\includegraphics[width=0.49\columnwidth]{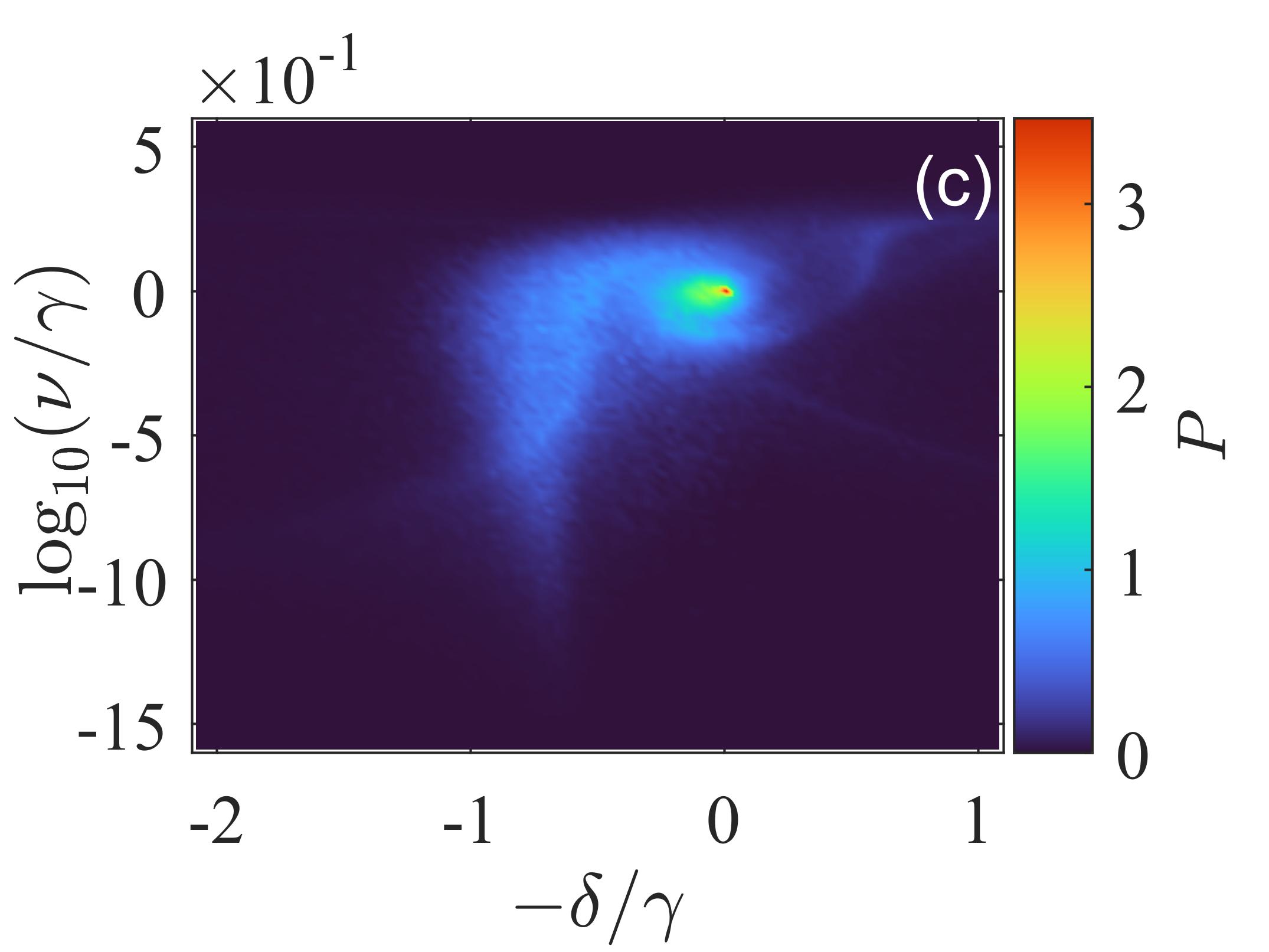}
\includegraphics[width=0.49\columnwidth]{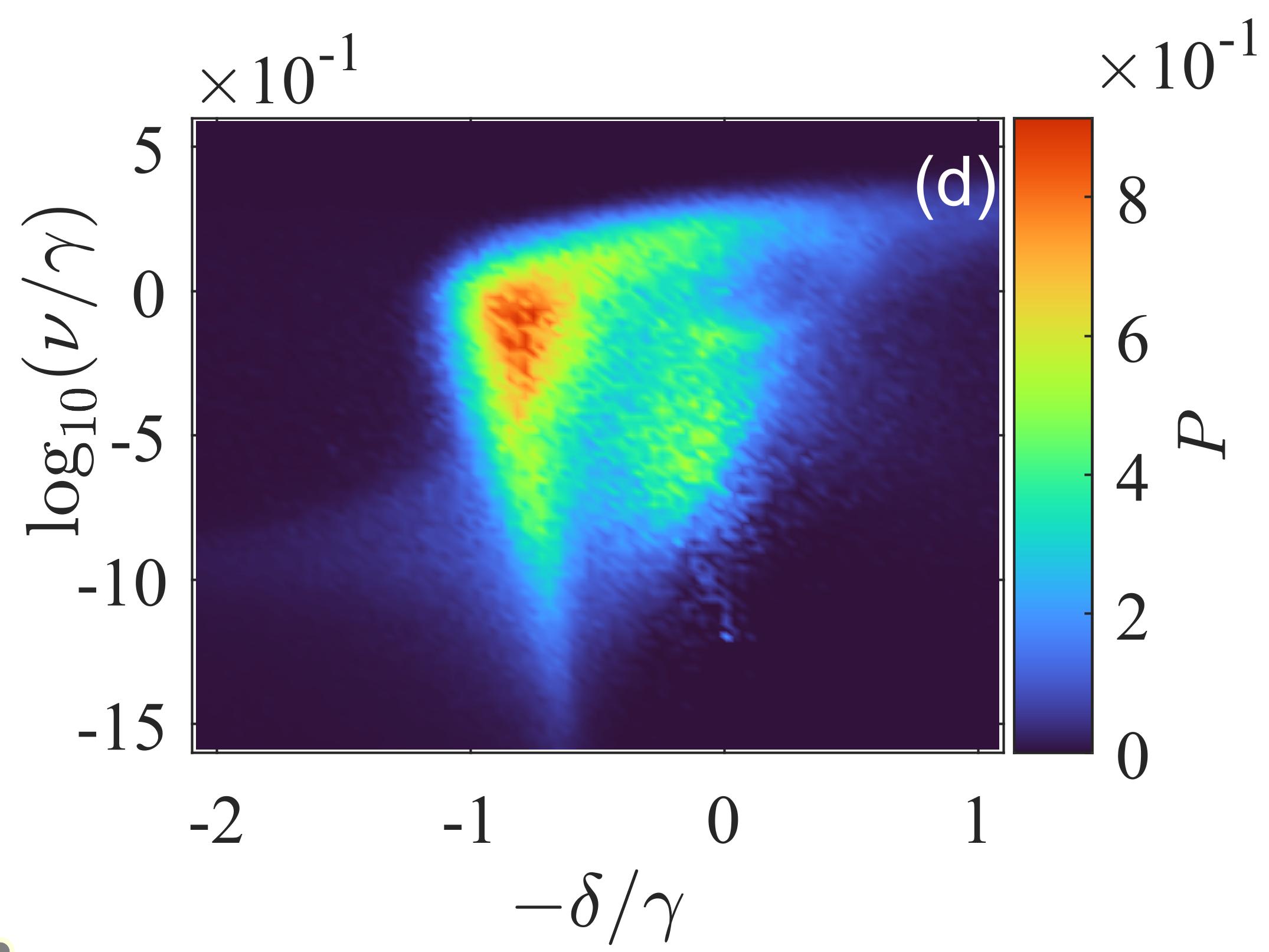}
\includegraphics[width=0.49\columnwidth]{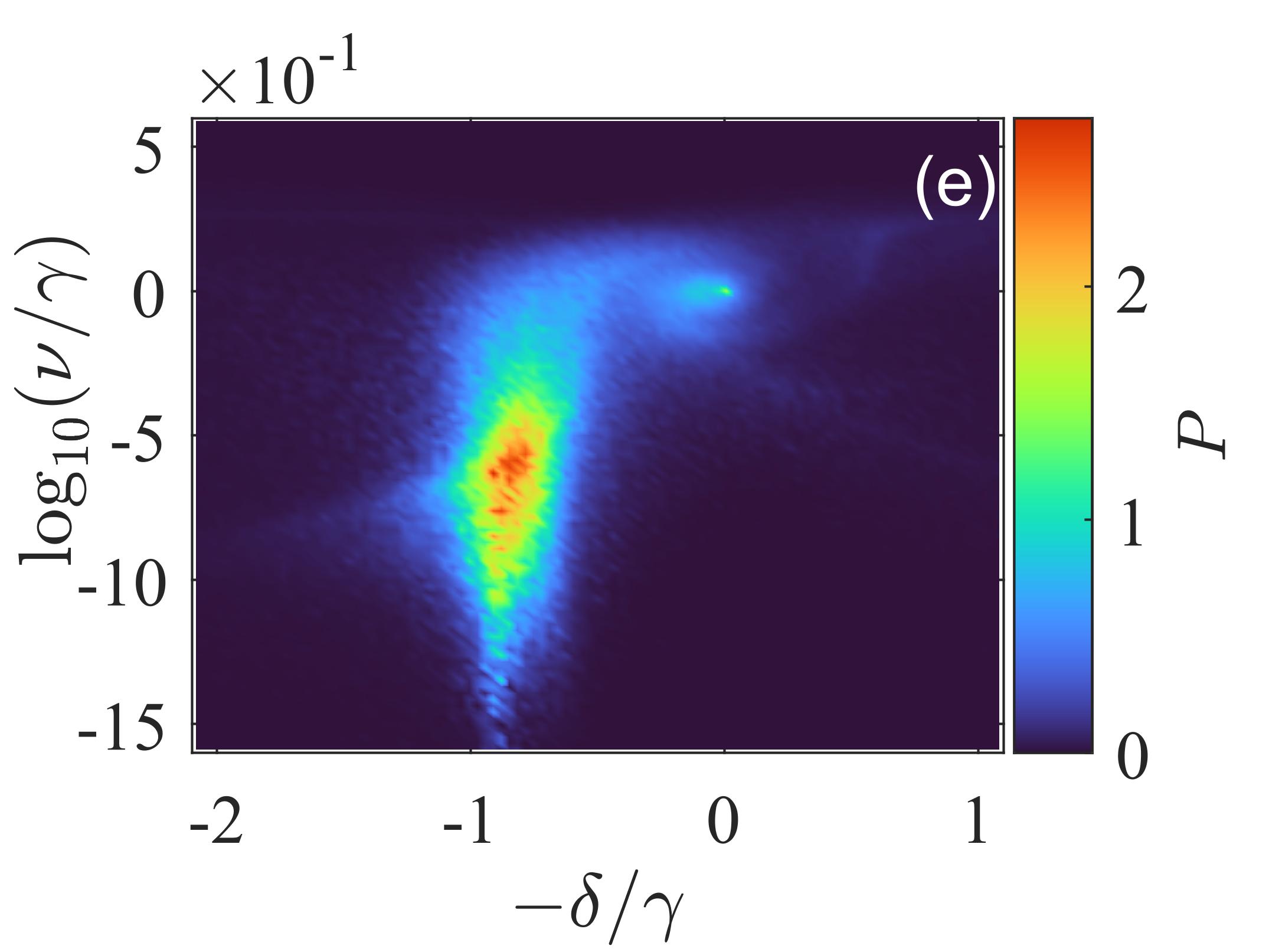}
\includegraphics[width=0.49\columnwidth]{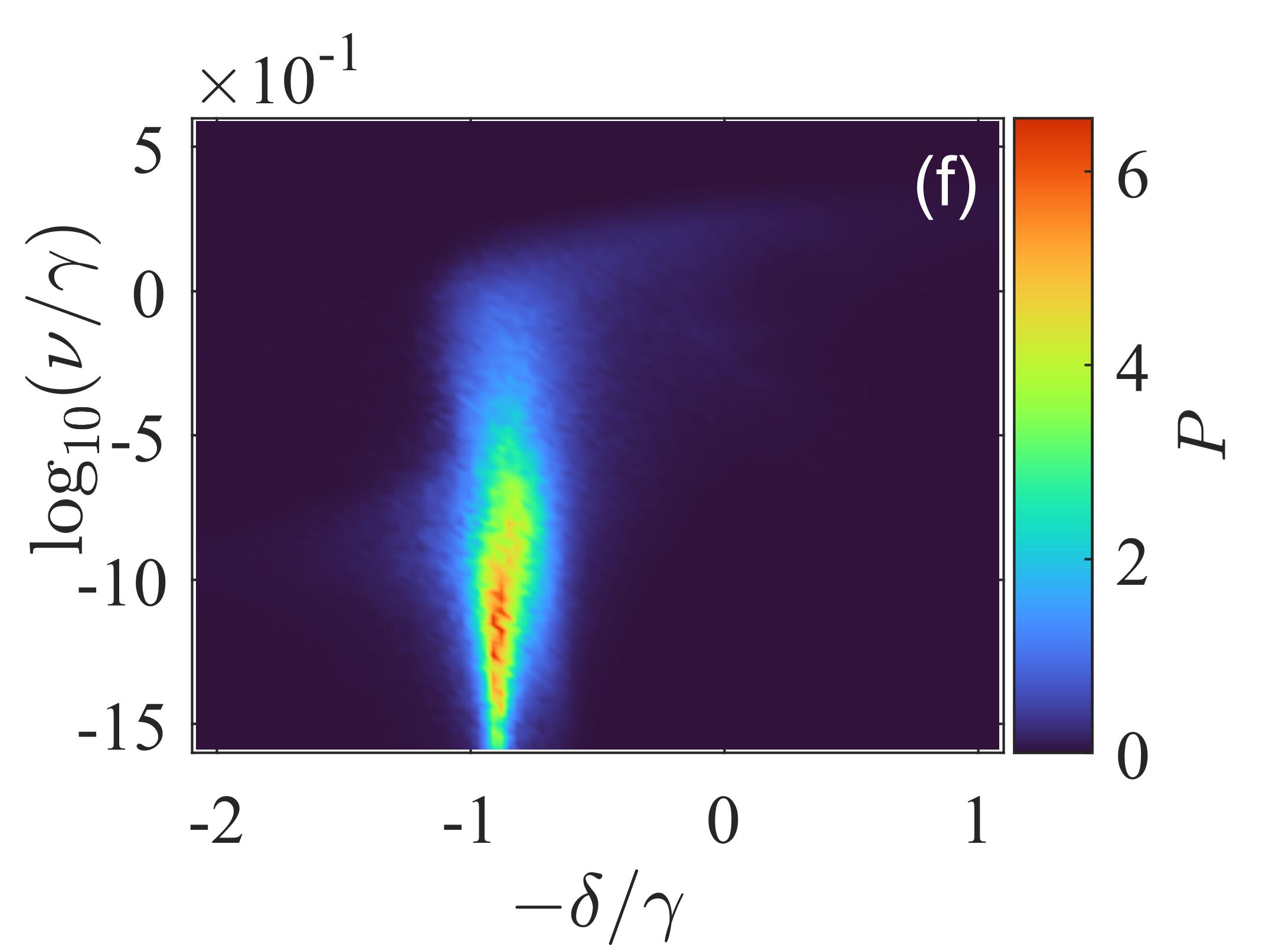}
    \vspace{-0.2cm}
\caption{Eigenmode occupations at the steady-state response in Fig.~\ref{fig: on-axis transmission} at the detuning (a)-(d) $\Delta=0$ and (e), (f) $-0.9\gamma$ for  (a), (b) \inline{\bar{\rho}_{\rm{1D}}/k\simeq0.1} and (c)-(f) 1.  The static interaction (a), (c), (e) \inline{R_{\rm{dip}}=0}  
and (b), (d), (f) \inline{R_{\rm{dip}}\bar{\rho}_{1D}\simeq 0.47} }
\label{fig: on-axis Eigenmodes}
\end{figure}

Due to small atom numbers and eigenmodes, it is easier to identify individual modes at high densities in a prolate trap than in an oblate one. The possibility of highly selected targeting of the eigenmodes as a result of static DD interactions, highlighted in Sec.~\ref{sec:dipolaroblate}, becomes even more pronounced in a prolate trap, as shown in Fig.~\ref{fig: on-axis Eigenmodes}(e), (f) for  \inline{\Delta=-0.9\gamma}. 
We also find that the most occupied eigenmode in most stochastic realizations for the system of Fig.~\ref{fig: on-axis Eigenmodes}(f) has close to 0.9 normalized occupation. The average occupation of the most occupied eigenmode over many realizations is about 0.66 (with the standard deviation of 0.16), as a result of some realizations having two eigenmodes more highly occupied. 
The linewidth of the excited subradiant mode for $N=10$ atoms in this case considerably varies between stochastic realizations, but some subradiant modes with a broader resonance in Fig.~\ref{fig: on-axis Eigenmodes}(f) have a smaller wavenumber than those with a narrower resonance.  

\begin{figure}
    \centering
      \includegraphics[width=0.49\columnwidth]{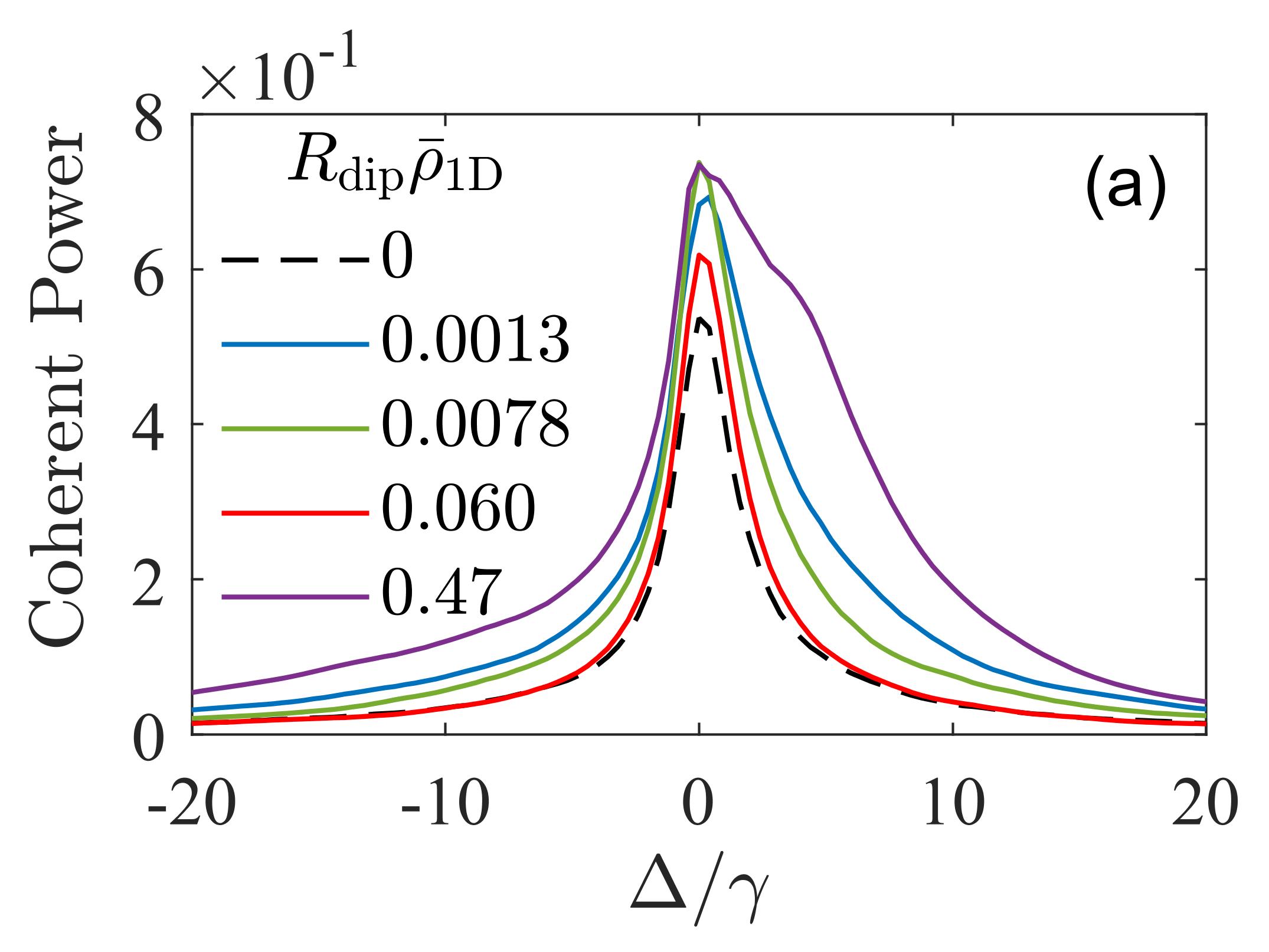}
 \includegraphics[width=0.49\columnwidth]{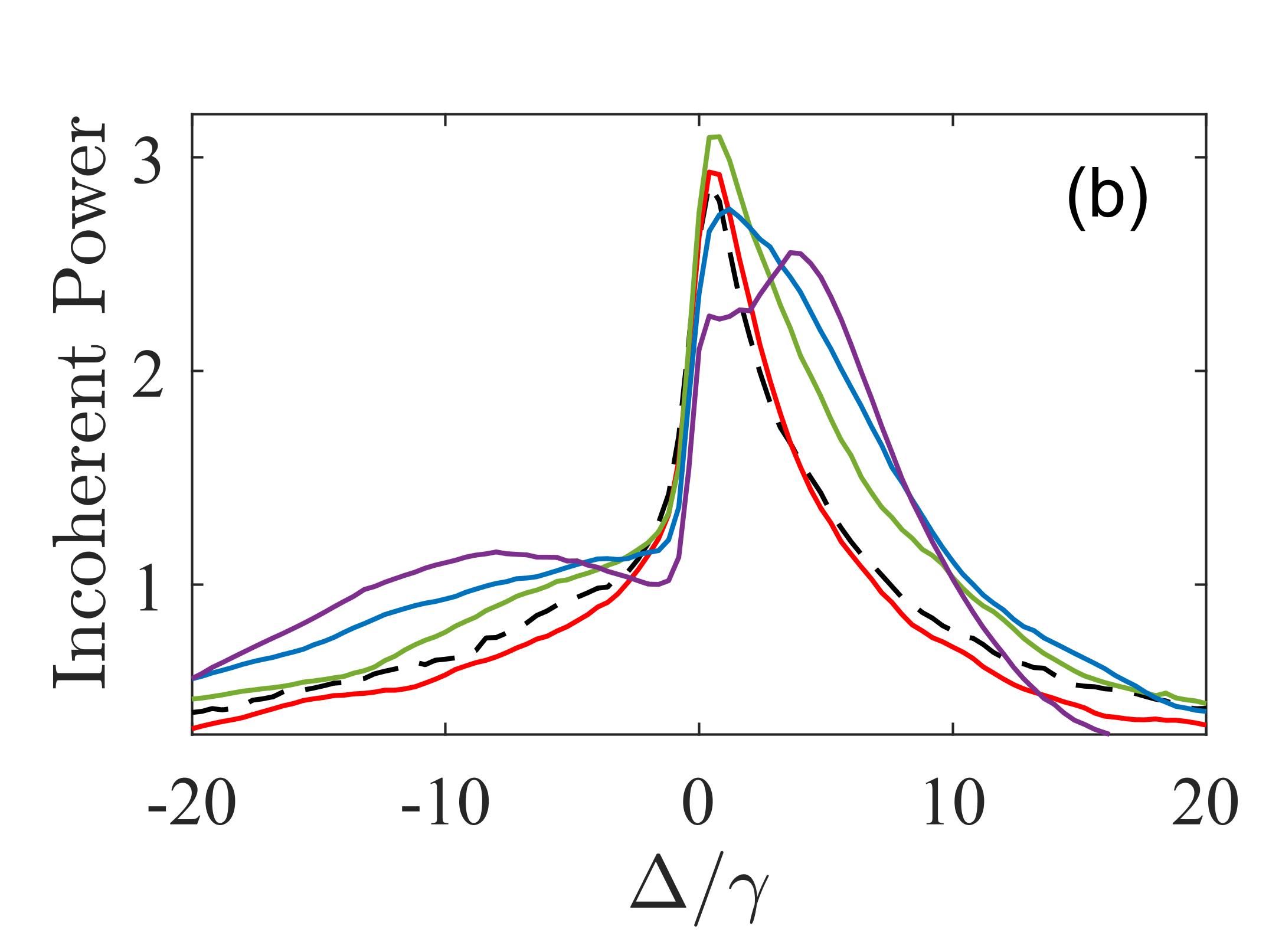}
    \vspace{-0.2cm}
    \caption{(a) Coherently and (b) incoherently forward-scattered power (in units of incident light intensity/$k^2$)  from 10 atoms in a prolate trap for different static dipole interaction strengths at the peak density \inline{\bar{\rho}_{\rm{1D}}/k\simeq1} and \inline{\ell_z/ \ell_x=25}. The light propagates perpendicular to the long trap axis. The lens NA 0.25 and 0.8 for coherent and incoherent light, respectively.
     }
    \label{fig: off-axis Tramissionl}
\end{figure}

The scattering in the off-axis case is markedly different to the on-axis case (Fig.~\ref{fig: off-axis Tramissionl}). While the coherent scattering is again strengthened by the DD interactions, the lineshape for the off-axis scattering becomes notably deformed, with double and triple peaks appearing in the coherent and incoherent scattering, respectively. The extra incoherent peak indicates the effect of the multilevel structure of the atoms, as the resonance is not captured by the lens for coherent scattering. 
\begin{figure}
    \centering
    \includegraphics[width=0.49\columnwidth]{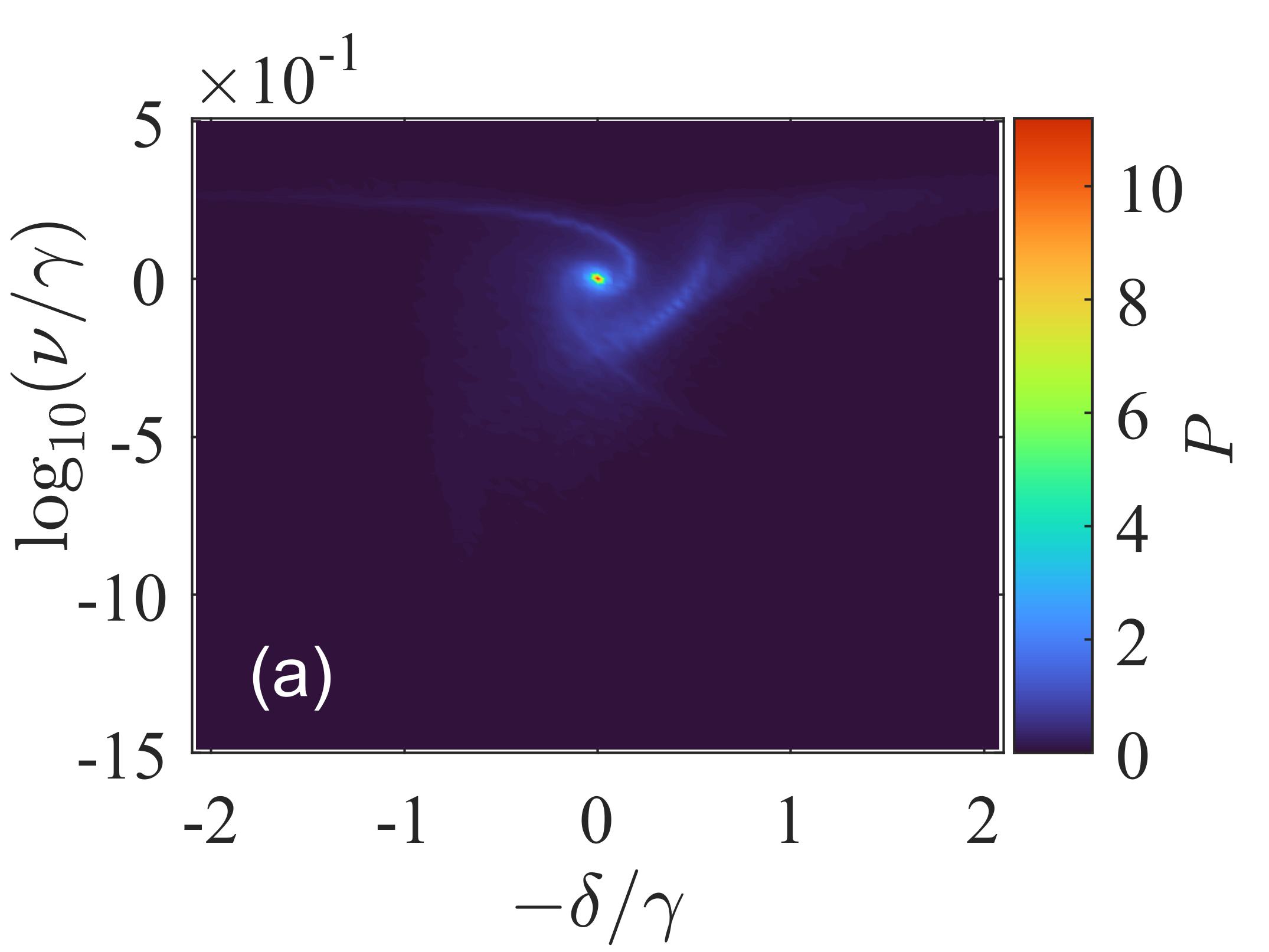}
\includegraphics[width=0.49\columnwidth]{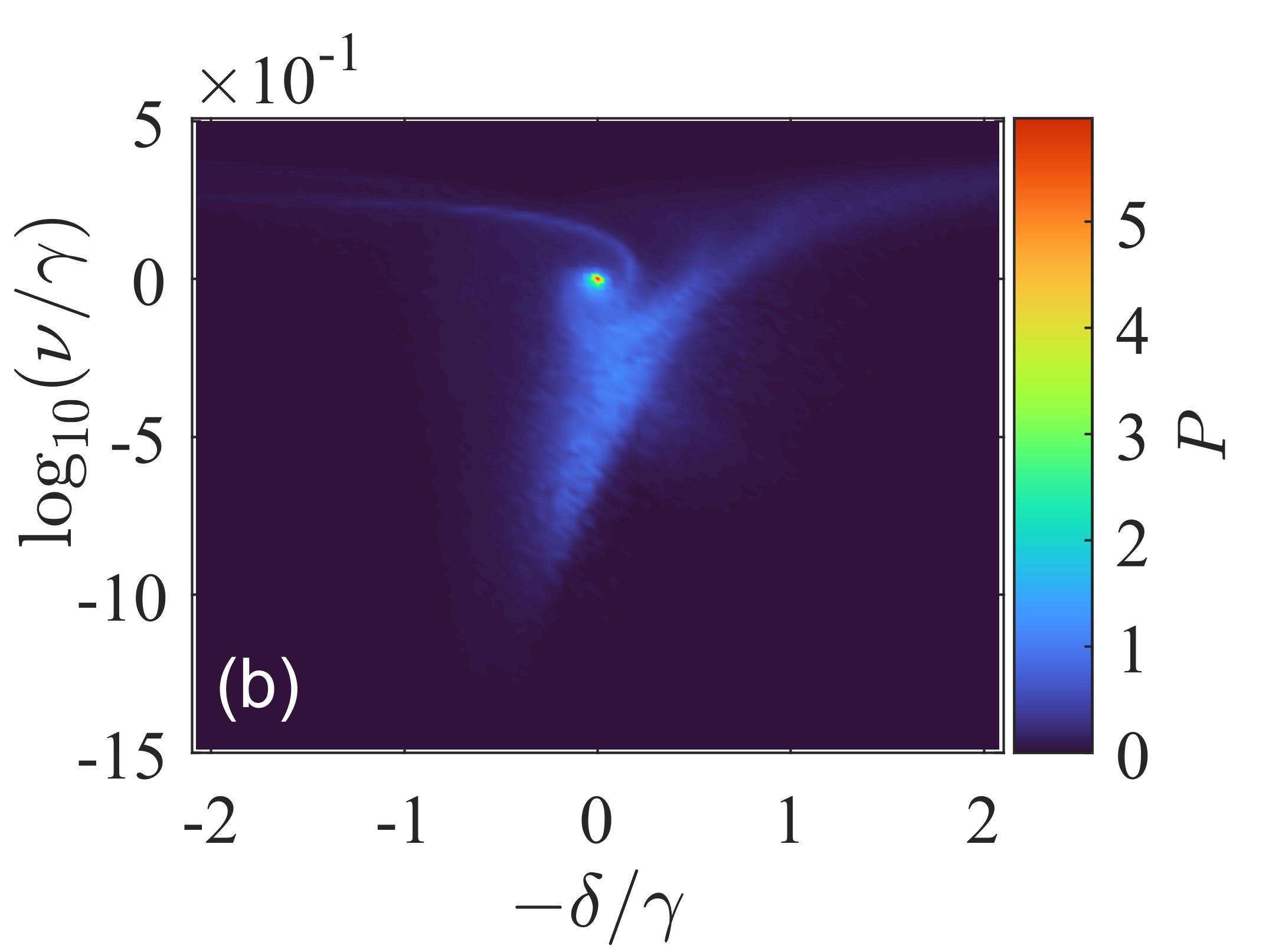}
\includegraphics[width=0.49\columnwidth]{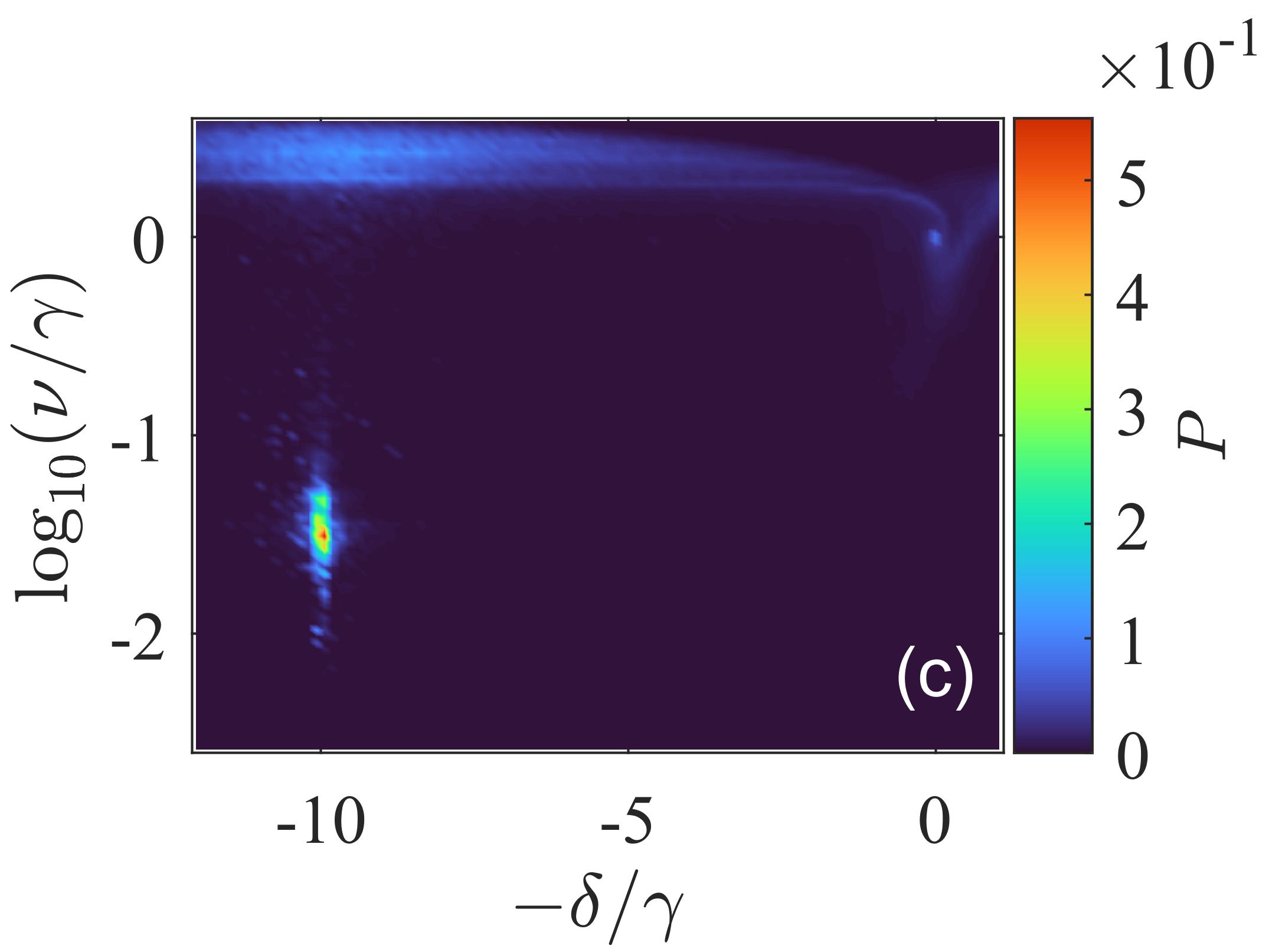}
\includegraphics[width=0.49\columnwidth]{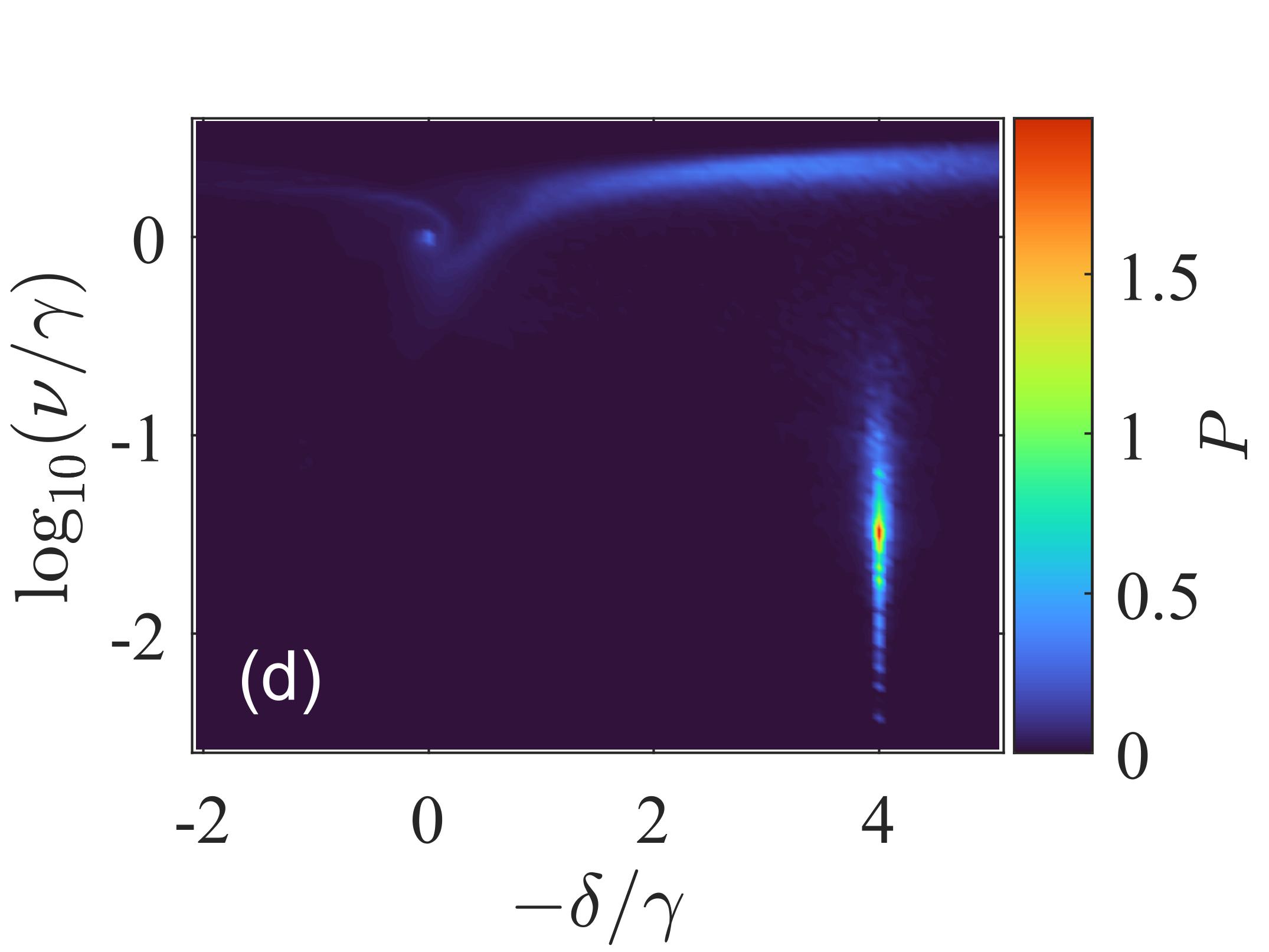}
    \vspace{-0.2cm}
\caption{Eigenmode occupations in Fig.~\ref{fig: off-axis Tramissionl} at the detuning (a), (b) $\Delta=0$; (c) \inline{-10\gamma}; and (d) \inline{4\gamma}.  (a) \inline{R_{\rm{dip}}k=0}  (b)-(d)  \inline{R_{\rm{dip}}k\simeq1}. 
The histogram bins \inline{[\Delta \log_{10} (\nu/\gamma), \Delta \delta/ \gamma] = [0.016,0.04]}, $[0.016,0.04]$, $[0.03,0.12]$, $[0.03,0.07]$, respectively.
}  
    \label{fig: off-axis eigenmode}     
\end{figure}

At the single-atom resonance, the static DD interactions induce subradiant excitations in Fig.~\ref{fig: off-axis eigenmode}, but the dominant excitations still exhibit the single-atom linewidth as in the independent-atom case. This unchanged dominant occupation across the interaction strengths explains the unshifted central peak in Fig.~\ref{fig: off-axis Tramissionl}.
The potential for targeted excitation of subradiant eigenmodes due to the static DD interactions is again shown in  Fig.~\ref{fig: off-axis eigenmode}(c), (d) at large detunings. However, the subradiant modes in the both cases co-exist with super-radiant eigenmodes. These super-radiant modes are visible in Fig.~\ref{fig: off-axis eigenmode}(c), (d) as distributions for $\nu>\gamma$ that extend over a wide range of resonance frequencies and can be excited even off resonance. It is these super-radiant eigenmodes that are also responsible for the additional resonances in the incoherent lineshape in Fig.~\ref{fig: off-axis Tramissionl}. Owing to their broad resonance linewidths, these modes show up in the lineshape profile.

\subsection{Localization of atomic polarization}
\label{sec:spots}

\begin{figure}[!hbtp]
    \centering
    \includegraphics[width=0.49\columnwidth]{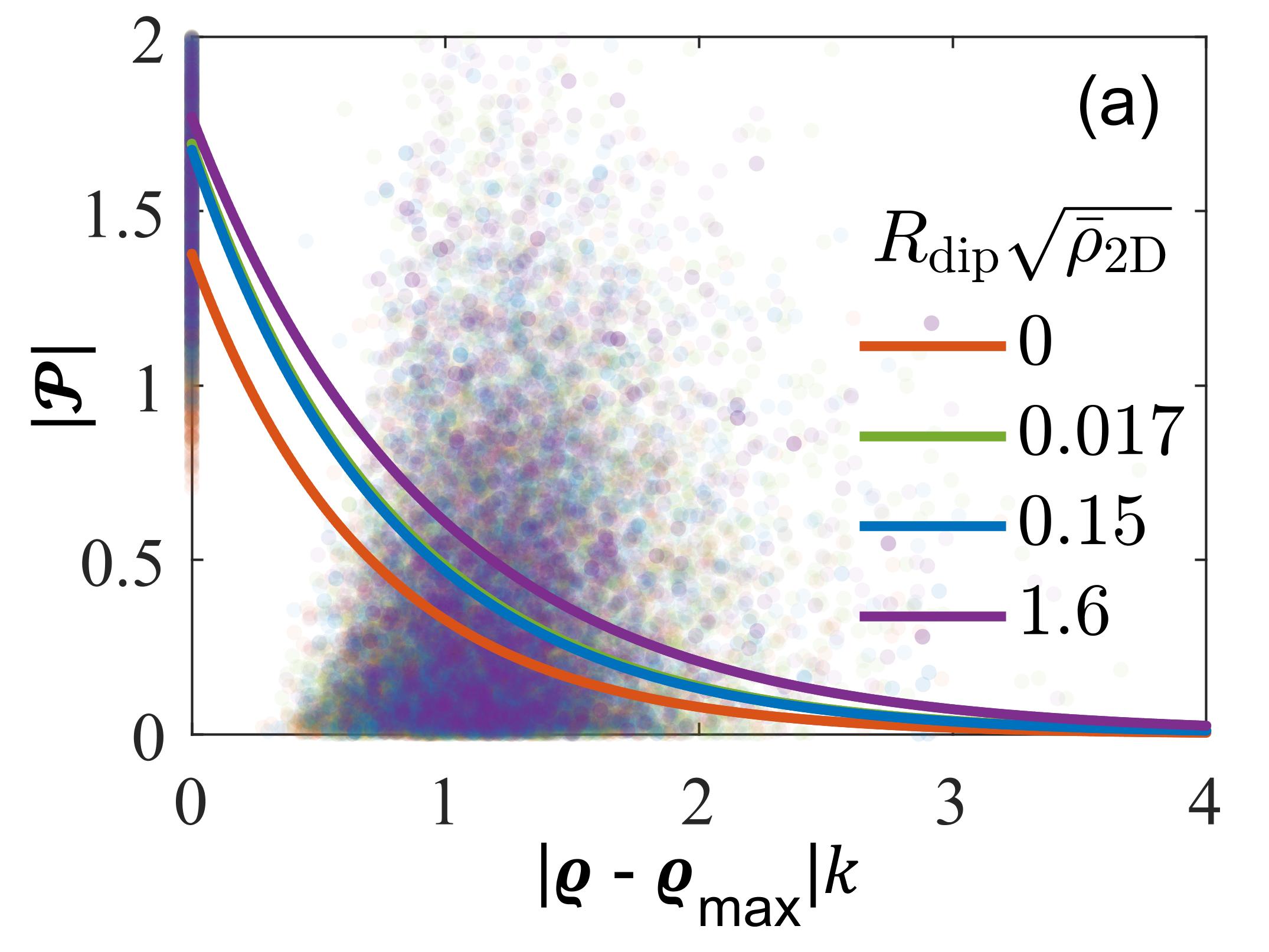}
    \includegraphics[width=0.49\columnwidth]{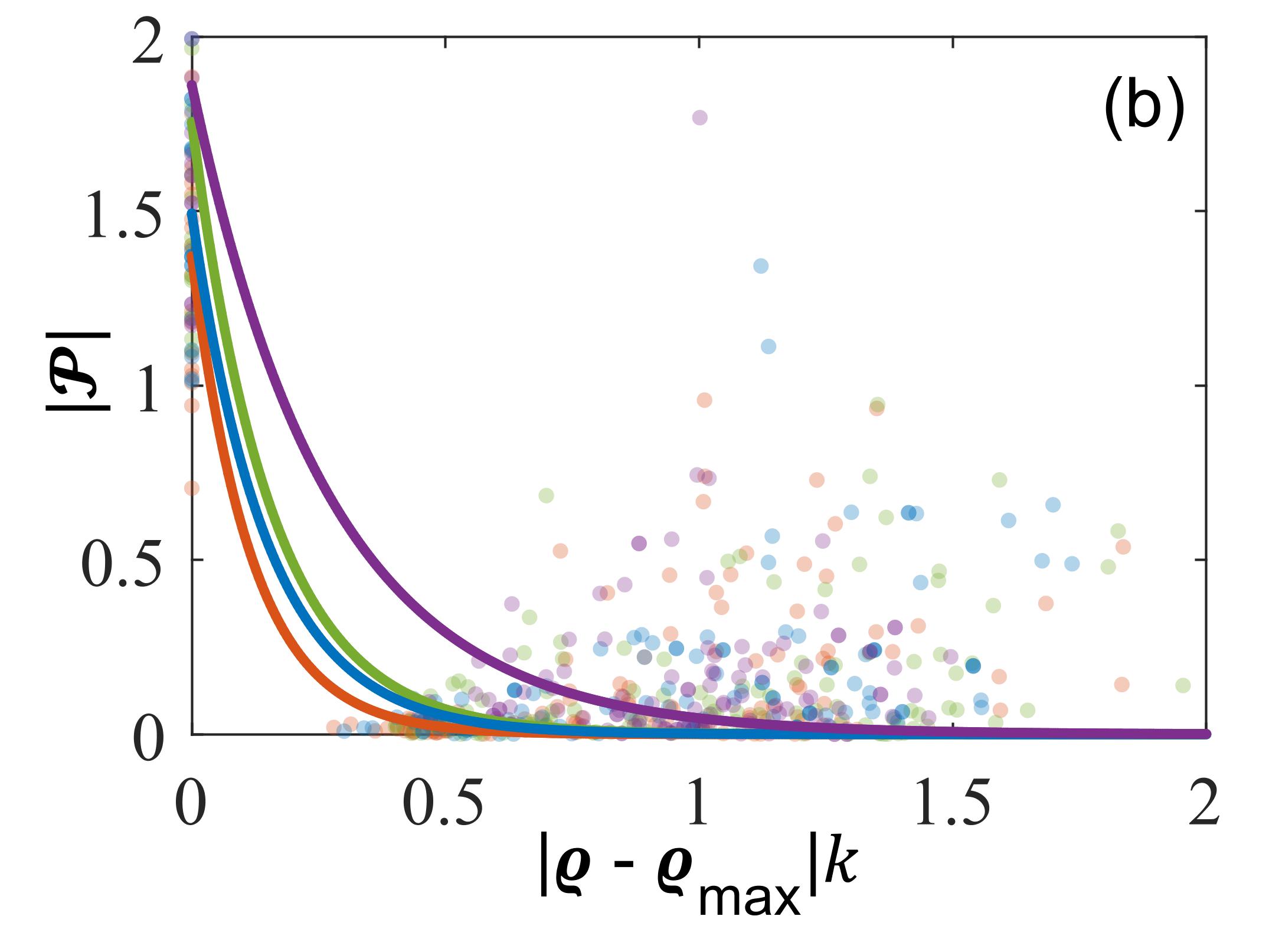}
    \includegraphics[width=0.49\columnwidth]{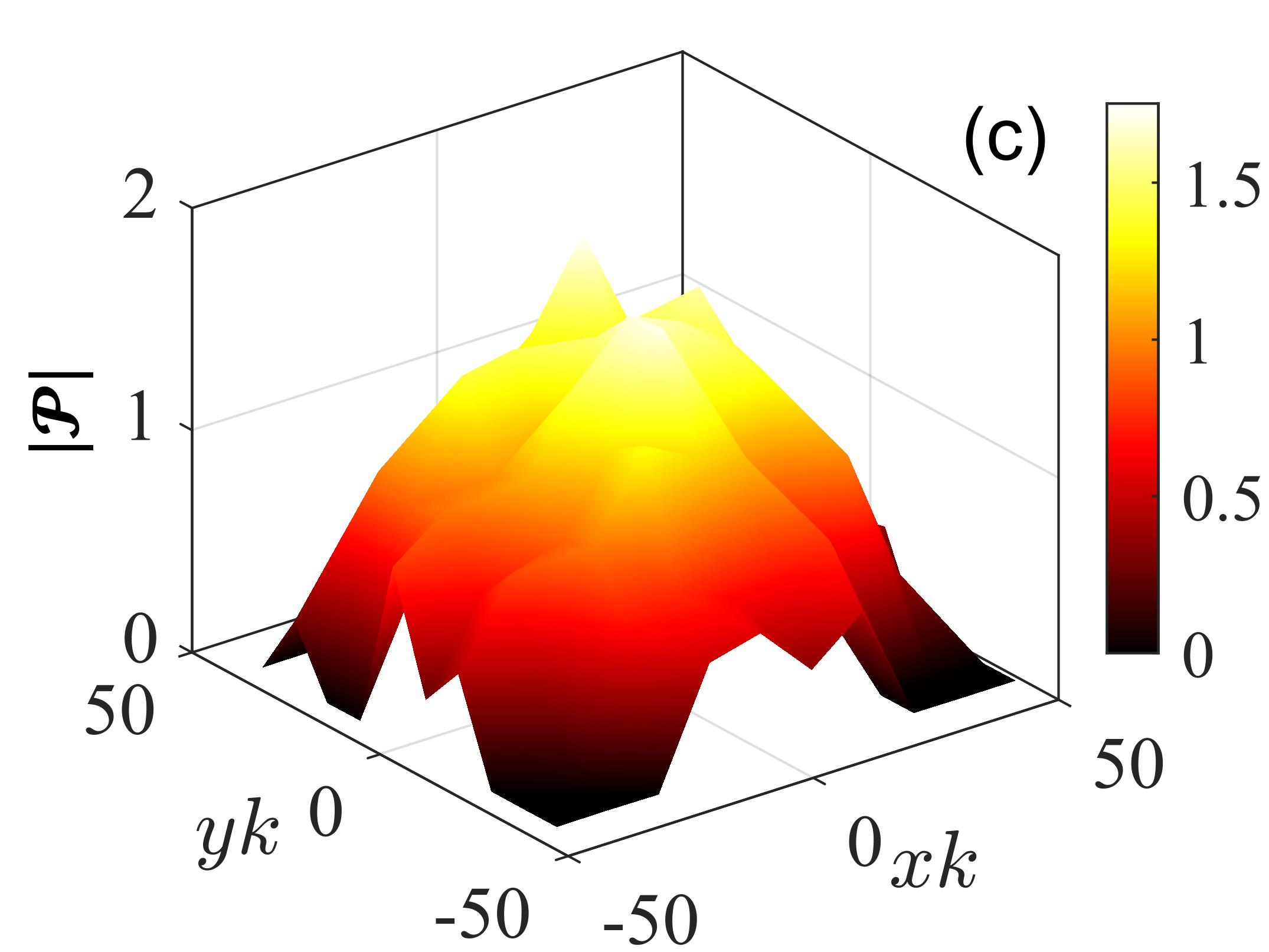}
    \includegraphics[width=0.49\columnwidth]{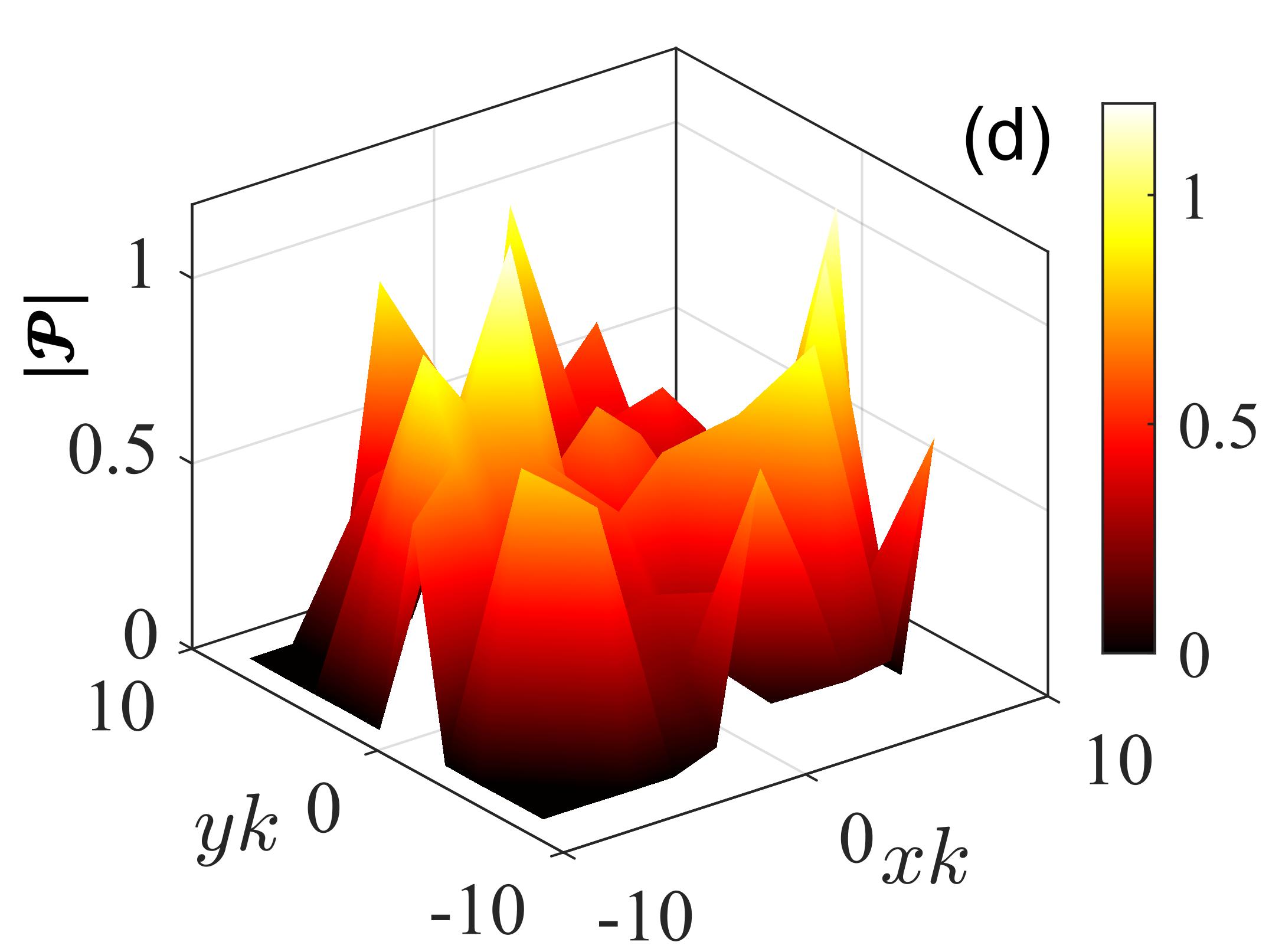}
        \vspace{-0.2cm}
    \caption{Localization of  excitations  \inline{|\boldvec{\Pc}|} [in units of \inline{\Dc \mathcal{E}/({\hbar \gamma})}] as a function of the distance from the most excited atom \inline{|\boldgreek{\varrho} - \boldgreek{\varrho}_{\rm{max}}|} for 200 atoms in an oblate trap at the peak density \inline{\bar{\rho}_{\rm{2D}}/k^2 \simeq 1}  for (a) all data, for (b) 2\% of the most localized cases (1000 realizations).    The corresponding profile in an individual stochastic realization with  \inline{R_{\rm{dip}}\sqrt{\bar{\rho}_{\rm{2D}}} \simeq 0.15} for (c) \inline{\bar{\rho}_{\rm{2D}}/k^2 \simeq 0.04}, (d) 1.} 
    \label{fig: localised pol 1}
\end{figure}
One of the immediate consequences of cooperative emitter responses is the resulting intricate interplay between the collective excitation eigenmodes and disorder in emitter positions. This can dramatically affect the near-field landscape of the optical response, resulting in the localization of eigenmodes and highly concentrated excitation energies~\cite{Gresillon99,Papasimakis19}.  In optics, this can be exploited, e.g., in achieving strong coupling between the excitation modes and a quantum emitter, thereby modifying its decay rate~\cite{Hein13}. A strongly localized excitation can effectively drive quantum emitters by acting as a cavity with its quality factor determined by the collective linewidth of the mode. We find that the near-field excitation landscape of the atoms exhibits strong localization of excitations on a subwavelength scale that depends on the atom density and the static interaction strength.

For each stochastic realization of the atomic configurations, we identify the most excited atom, \inline{{\rm{max}}(|\boldvec{\Pc}|)}, and its position \inline{\boldgreek{\varrho}_{\rm{max}}}. Then the 
distribution $ {\rm{max}}(|\boldvec{\Pc}|) \exp(-|\boldgreek{\varrho} -\boldgreek{\varrho}_{\rm{max}}|/\varrho_0)$ is determined from the amplitude \inline{|\boldvec{\Pc}|} of its five nearest neighbors, where  \inline{\varrho_0} is a fitting parameter for the localization. 
In Fig~\ref{fig: localised pol 1}, we show an example of such fitting for all the cases and for the 2\inline{\%} most localized realizations for the isotropic transition and positive circular light polarization.

\begin{figure}[!hbt]
    \centering
    \includegraphics[width=0.49\columnwidth]{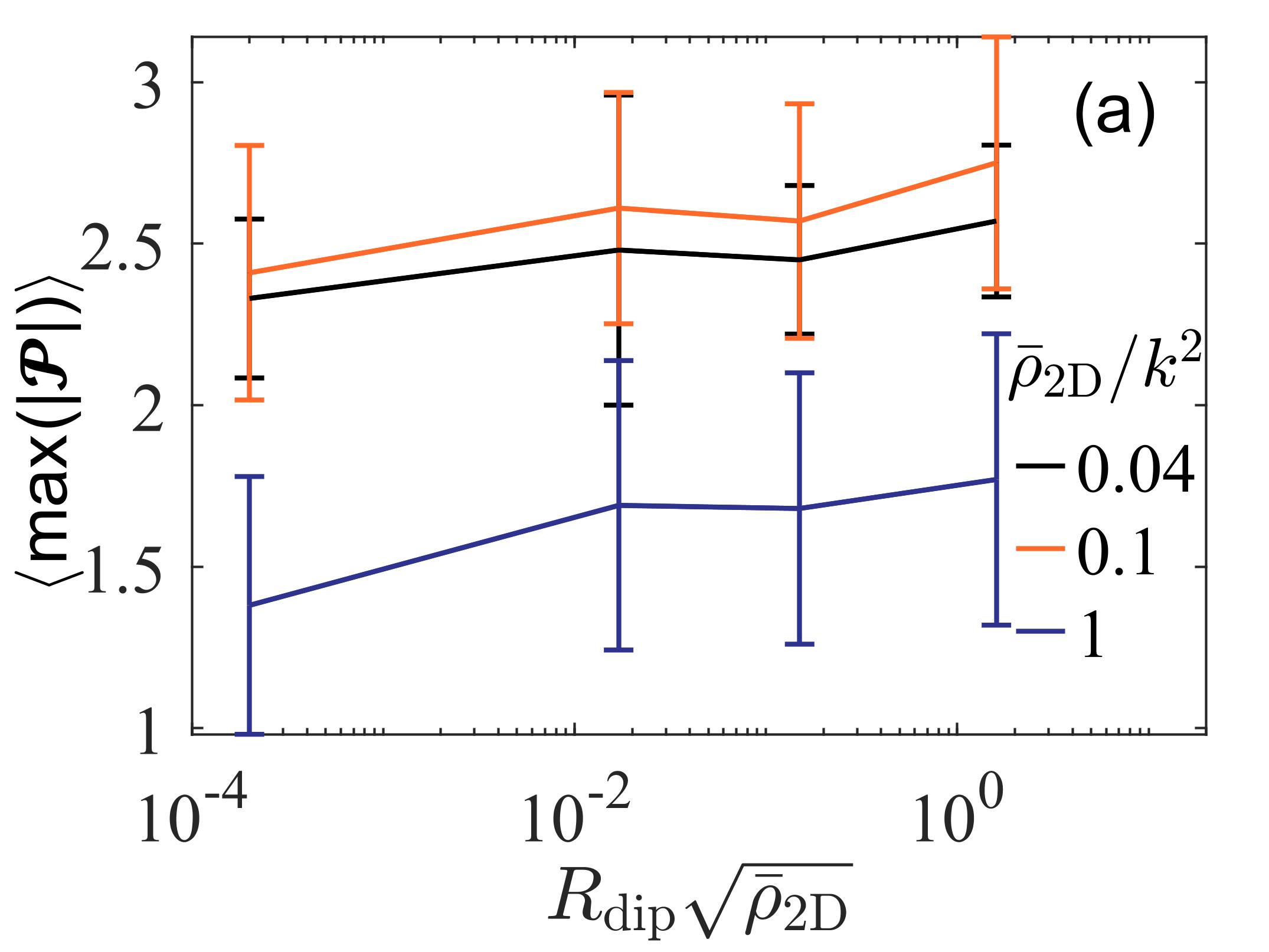}
        \includegraphics[width=0.49\columnwidth]{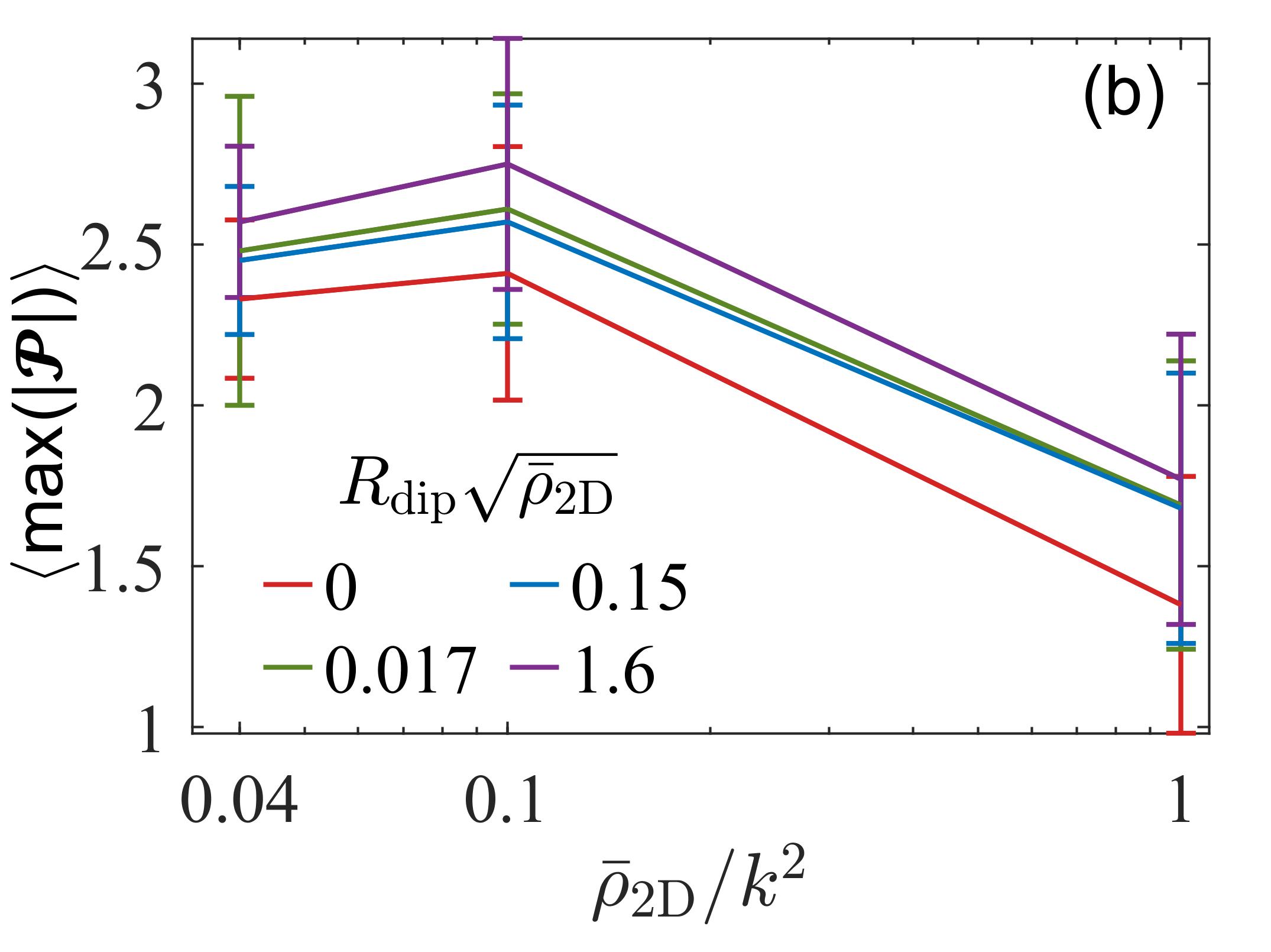}
        \includegraphics[width=0.49\columnwidth]{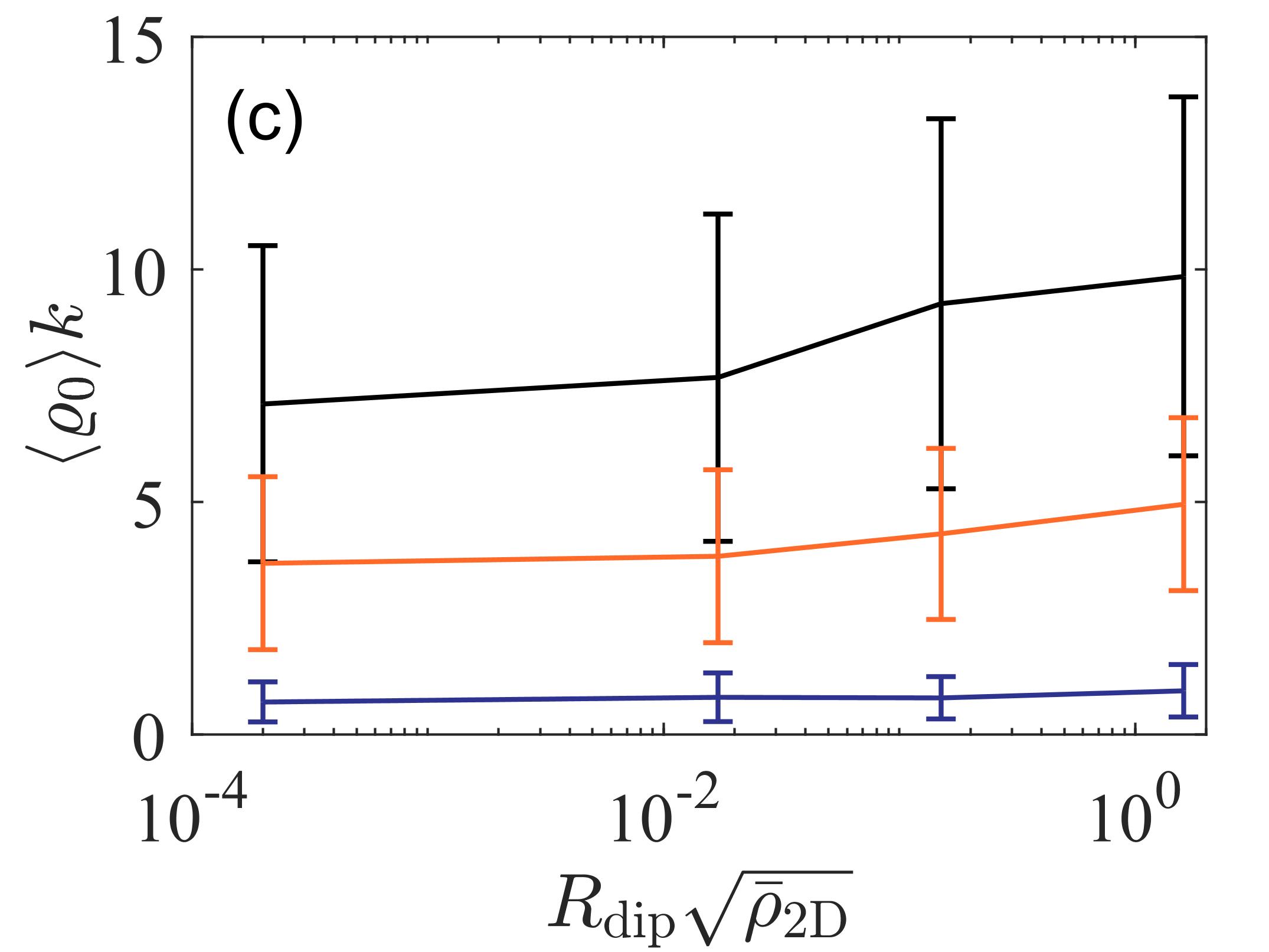}
         \includegraphics[width=0.49\columnwidth]{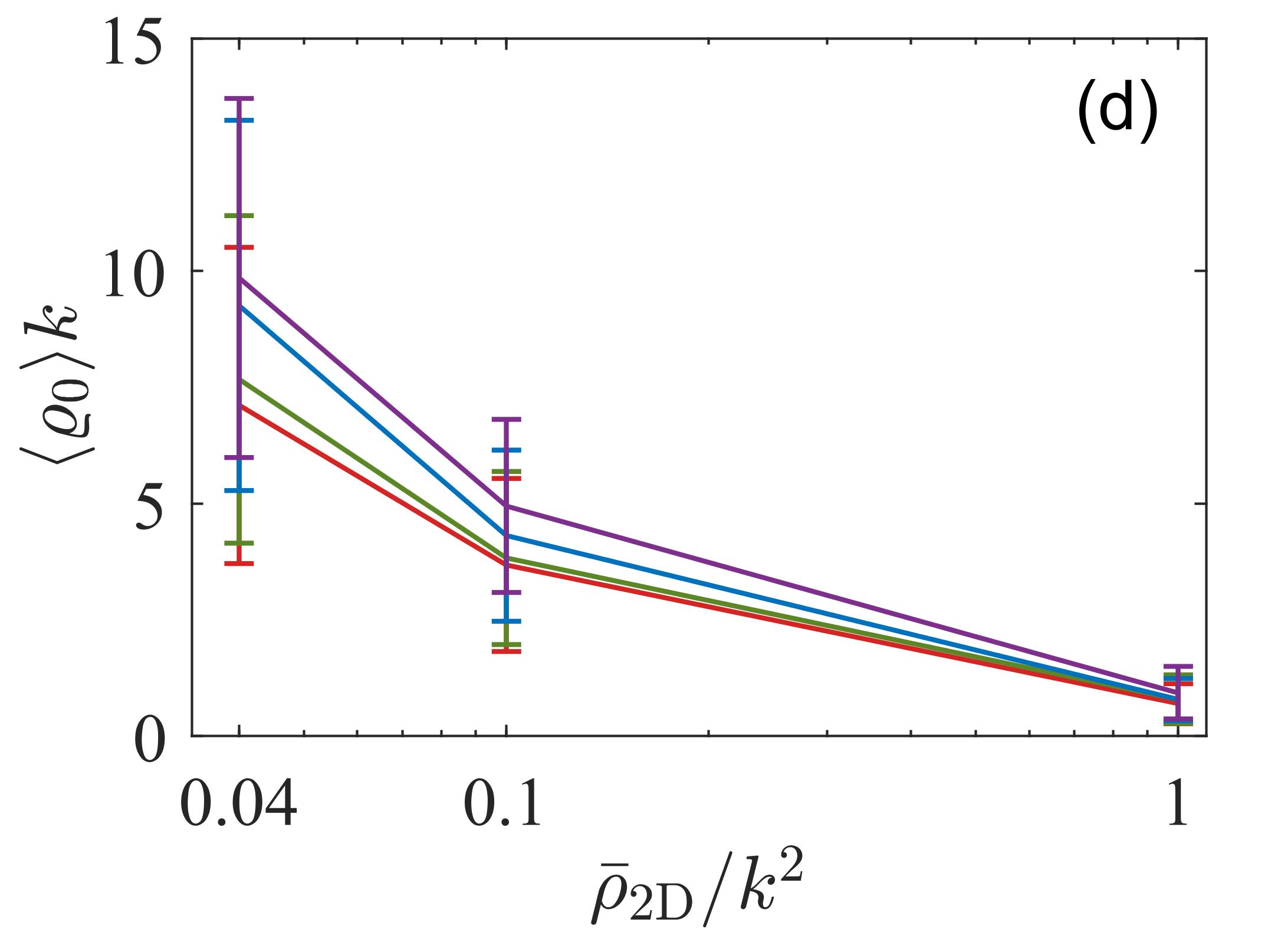}
    \includegraphics[width=0.49\columnwidth]{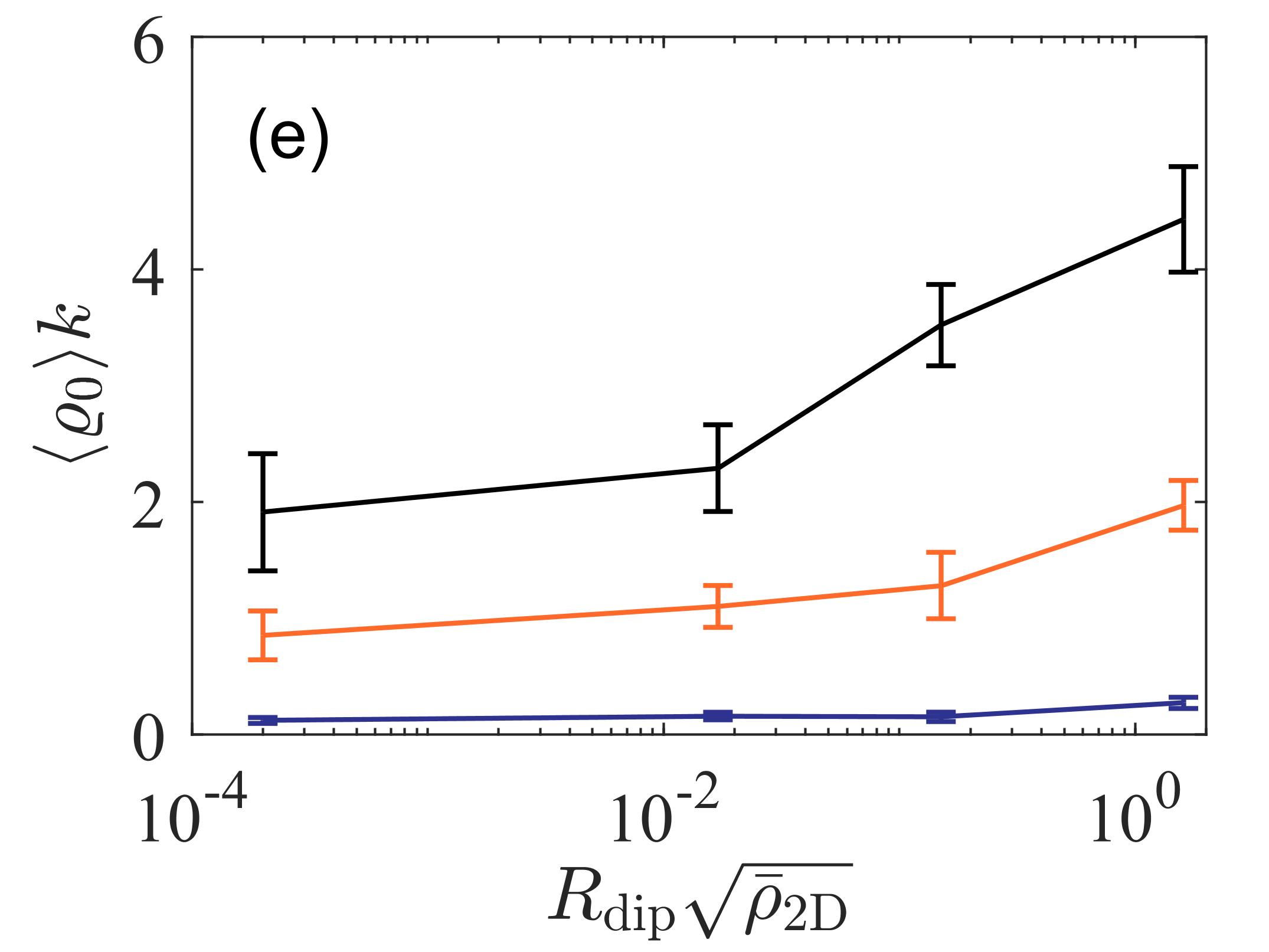}
    \includegraphics[width=0.49\columnwidth]{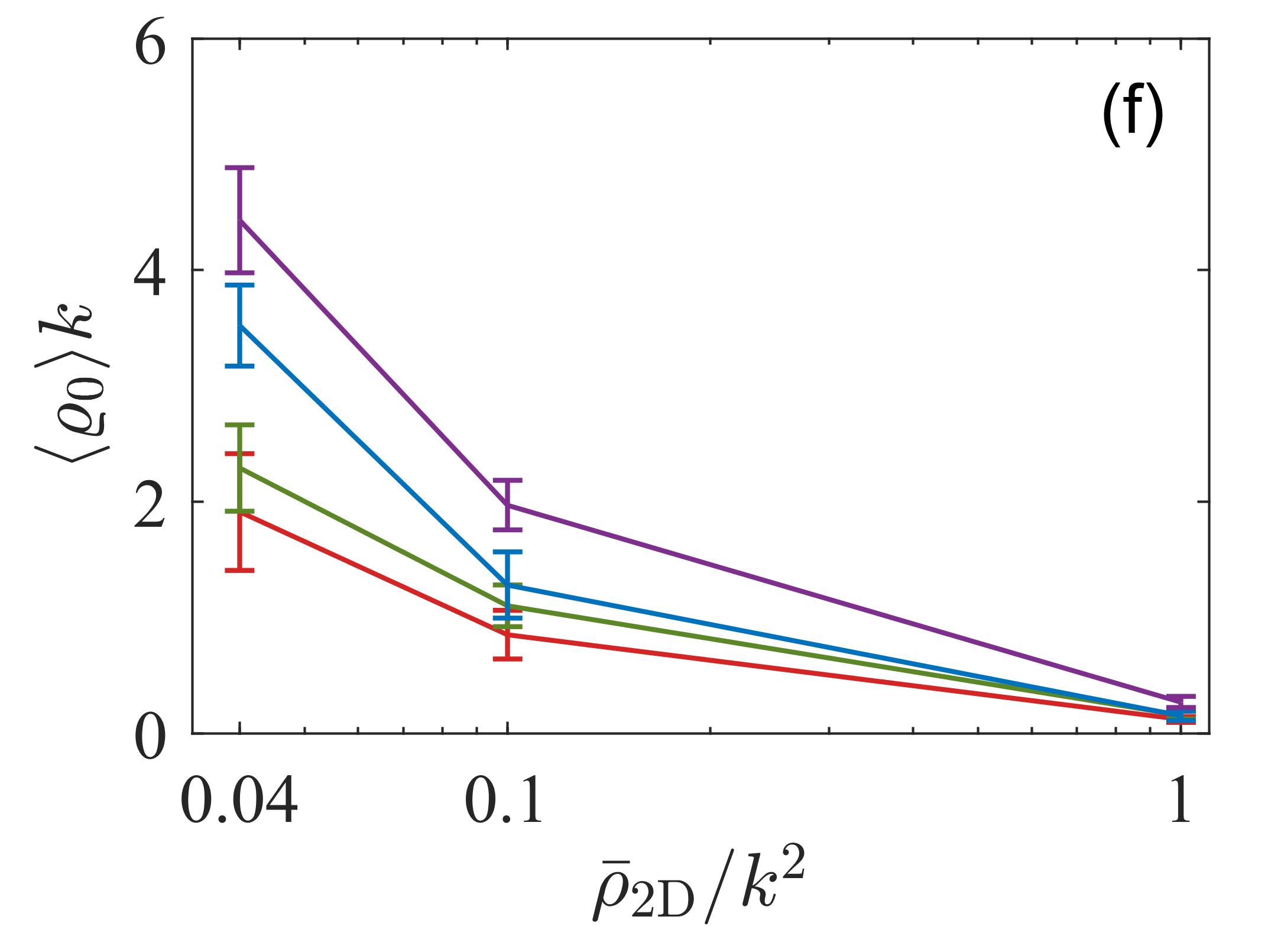}
    \vspace{-0.2cm}
    \caption{ (a), (b) Peak localized excitations \inline{\<{\rm{max}}(|\boldvec{\Pc}|)\>} [in units of \inline{\Dc \mathcal{E}/({\hbar \gamma})}] and  (c), (d) localization length \inline{\langle \varrho_0 \rangle} as a function of interaction strength \eqcaption[true]{\inline{R_{\rm{dip}}\sqrt{\bar{\rho}_{\rm{2D}}}}}{eq:Rdip} for 200 atoms in an oblate trap.  (e), (f) The localization length for the \inline{2 \%} most localized cases.  The values of (d) \inline{\<\rho_0\>k\simeq 1.28,1.69,1.68,1.77} and  (f) \inline{\<\rho_0\>k\simeq 0.27,0.15,0.16,0.199} for the density \inline{\bar{\rho}_{\rm{2D}}/k^2 \simeq1}}
    \label{fig: Localisation fit int}
\end{figure}
The localization rapidly increases with the atom density, independently of the static interaction.
Examples of single stochastic runs of atomic polarization density profiles are shown in Fig.~\ref{fig: localised pol 1}.  In the low density case, the profile is closer to the Gaussian, while the localized excitations are visible at high densities. 
As the static interaction is increased, the value of the peak excitation increases (Fig.~\ref{fig: Localisation fit int}). This is more pronounced for the most localized \inline{2 \%}.

\subsection{Varying the atom density in an oblate trap}
\label{sec:density}

\begin{figure}
    \centering
     \includegraphics[width=0.49\columnwidth]{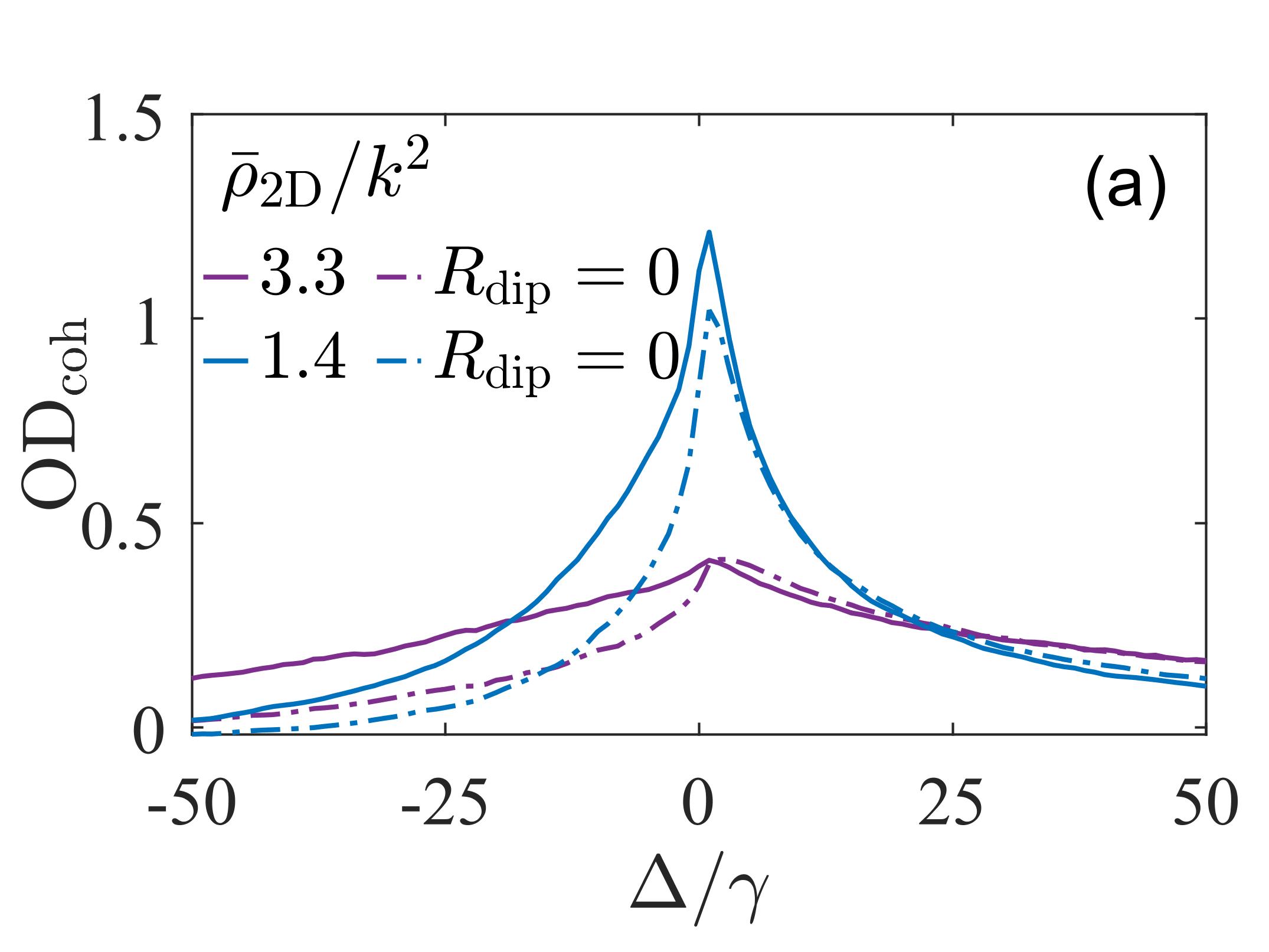}  
     \includegraphics[width=0.49\columnwidth]{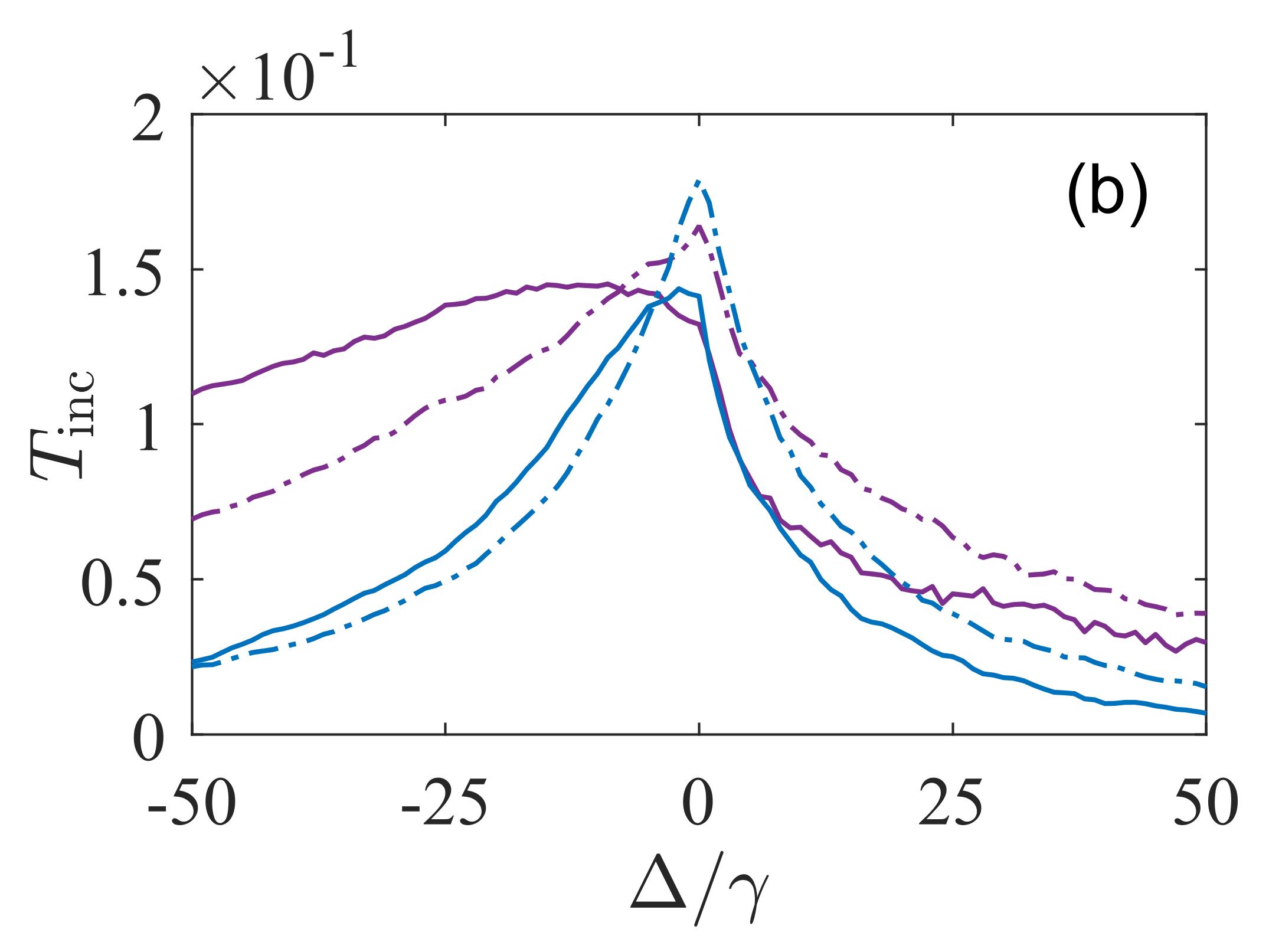}    
     \includegraphics[width=0.49\columnwidth]{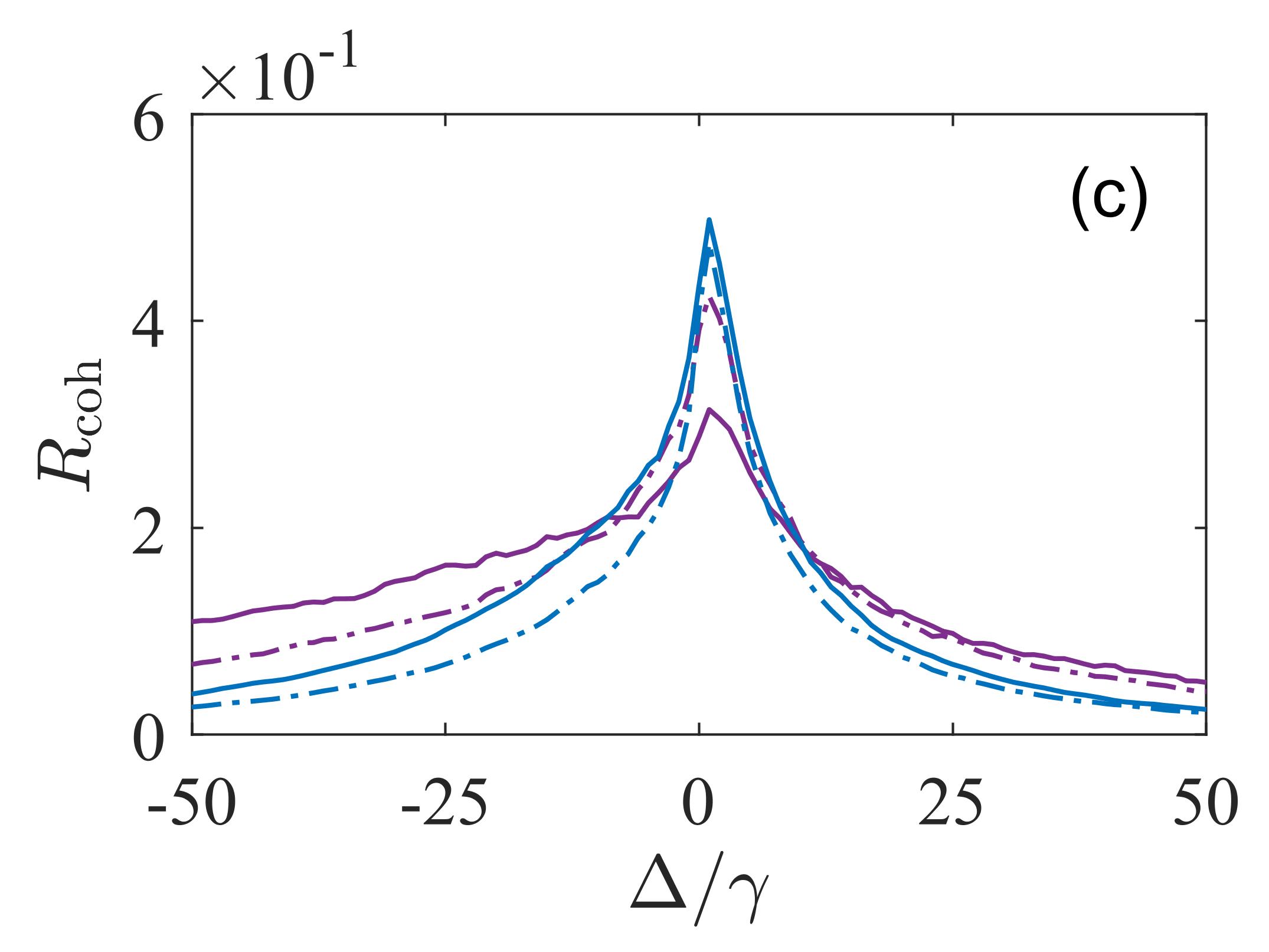}
     \includegraphics[width=0.49\columnwidth]{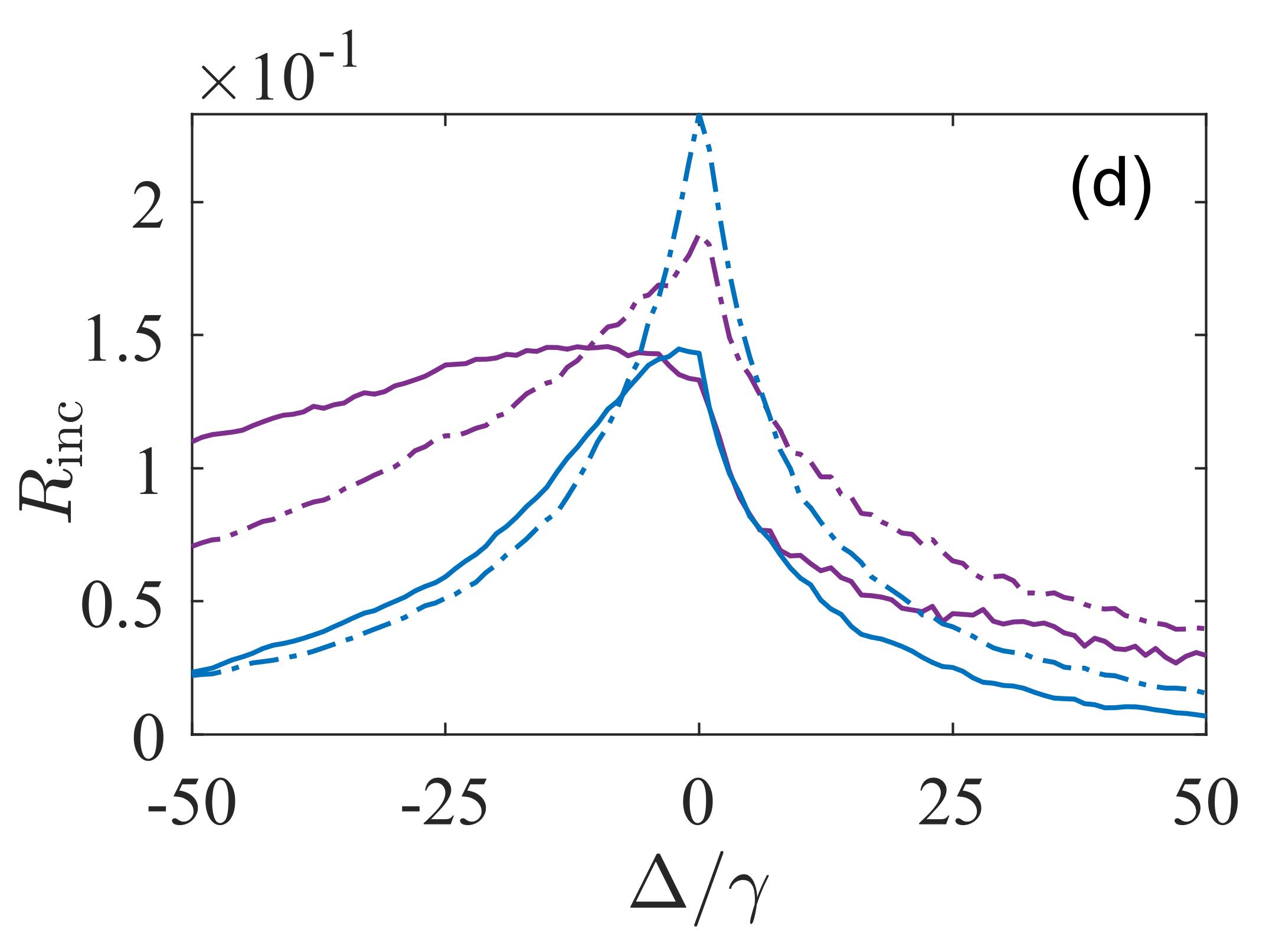}  
         \vspace{-0.2cm}
   \caption{Transmission and reflection from 200 two-level atoms in an oblate trap as a function of the laser detuning for different peak  densities \inline{\bar{\rho}_{\rm{2D}}/k^2}  with \inline{\ell_x/\ell_z=25}, for which  \inline{R_{\rm{dip}}\sqrt{\bar{\rho}_{\rm{2D}}} \simeq 0.12,0.18}, \inline{\ell_z k\simeq0.27,0.18, \ell_x k \simeq 6.7,4.4} (solid curves)  \inline{R_{\rm{dip}} = 0} (dashed curves).  (a) Coherent optical depth \inline{\rm{OD}_{\rm{coh}}},  (b) incoherent transmission \inline{T_{\rm{inc}}},  (c) coherent reflection \inline{R_{\rm{coh}}}, (d) incoherent reflection \inline{R_{\rm{inc}}}[Eqs.~\eqref{eq: Transmission}-\eqref{eq:Es expectation}]. The lens NA 0.45 and 0.8 for coherent and incoherent light, respectively.}
    \label{fig: ConstRdipTransmissionWide}
\end{figure}
\begin{figure*}
    \centering
   \includegraphics[width=0.4\columnwidth]{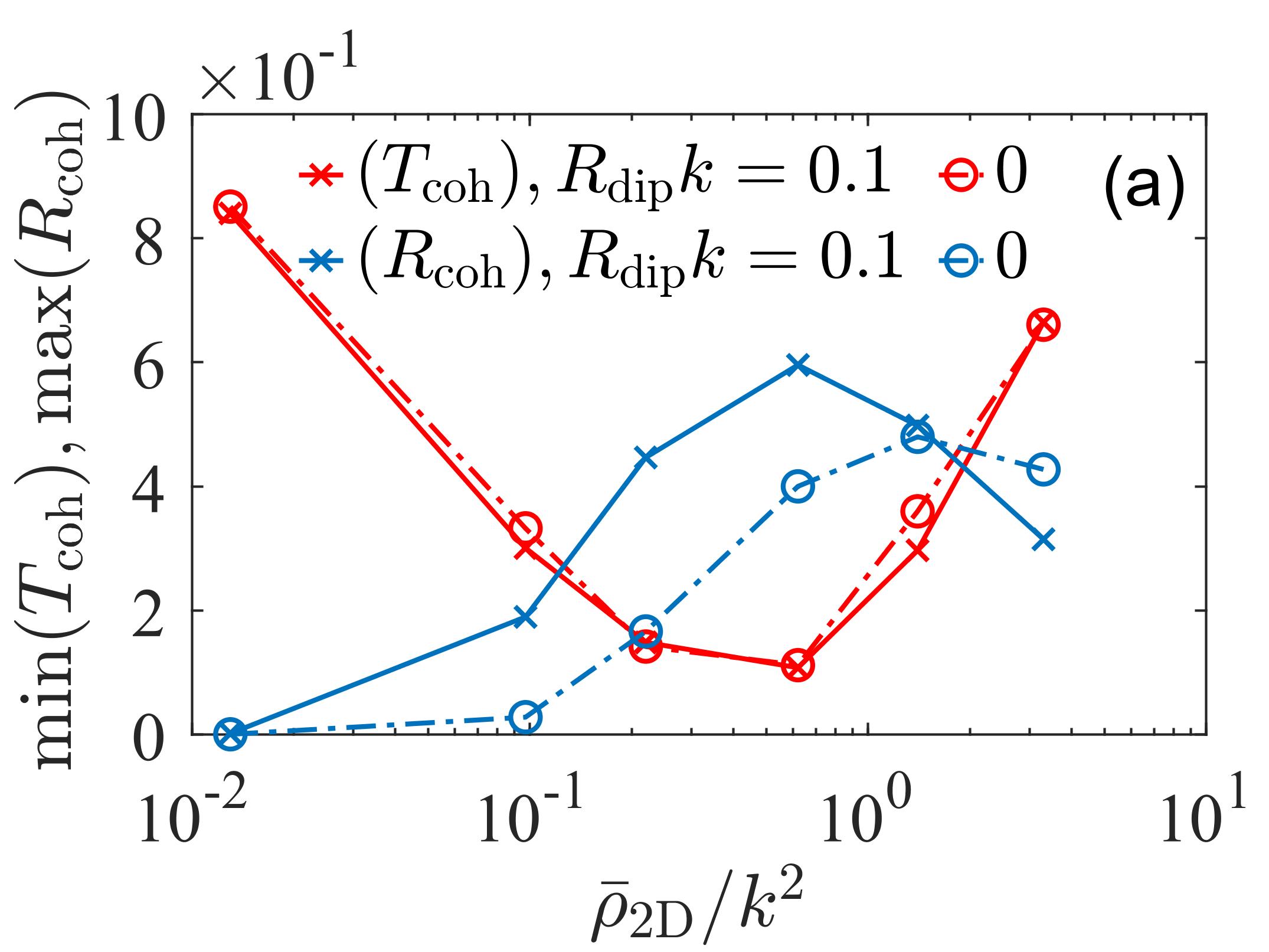}
   \includegraphics[width=0.4\columnwidth]{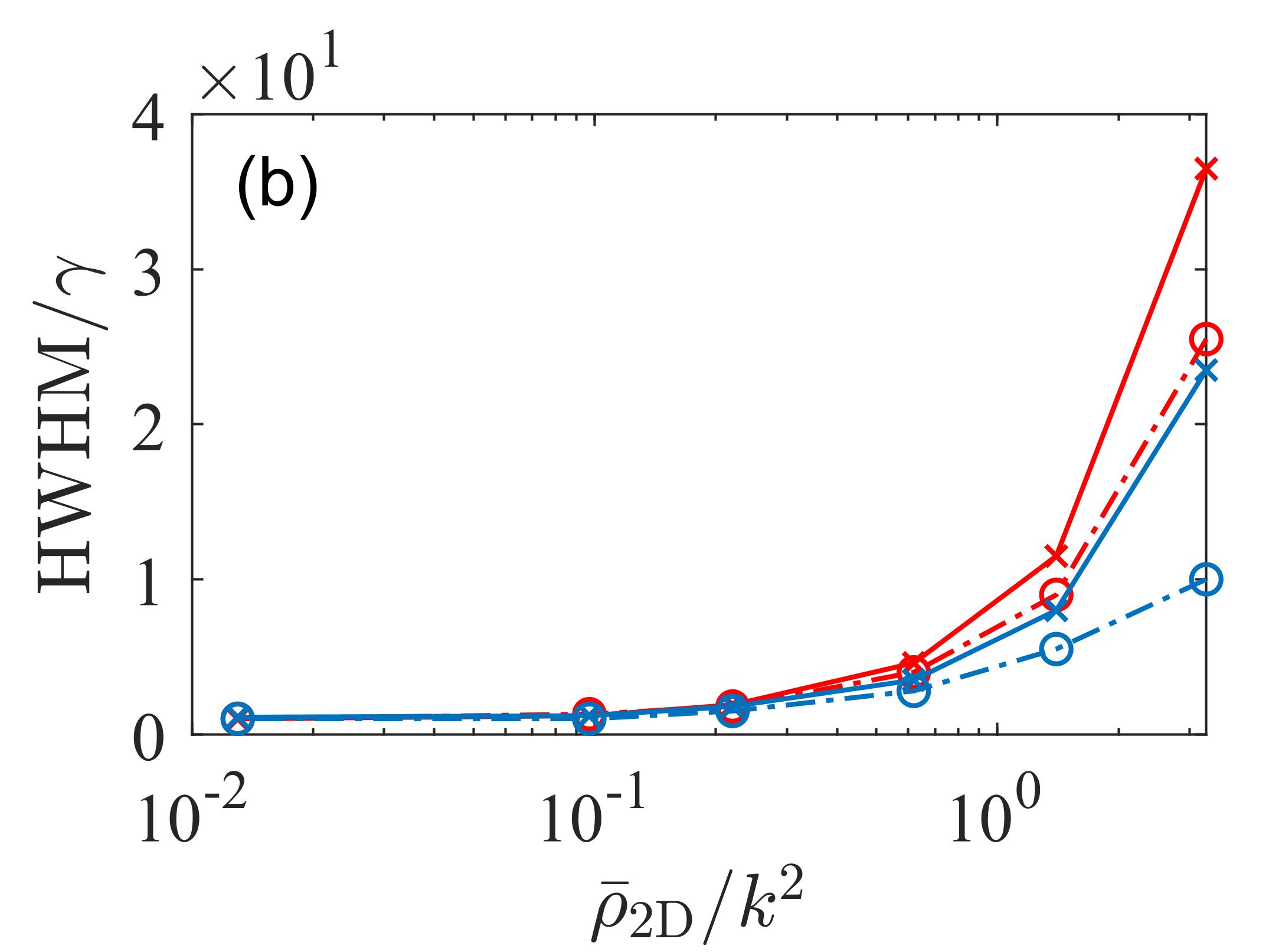}
   \includegraphics[width=0.4\columnwidth]{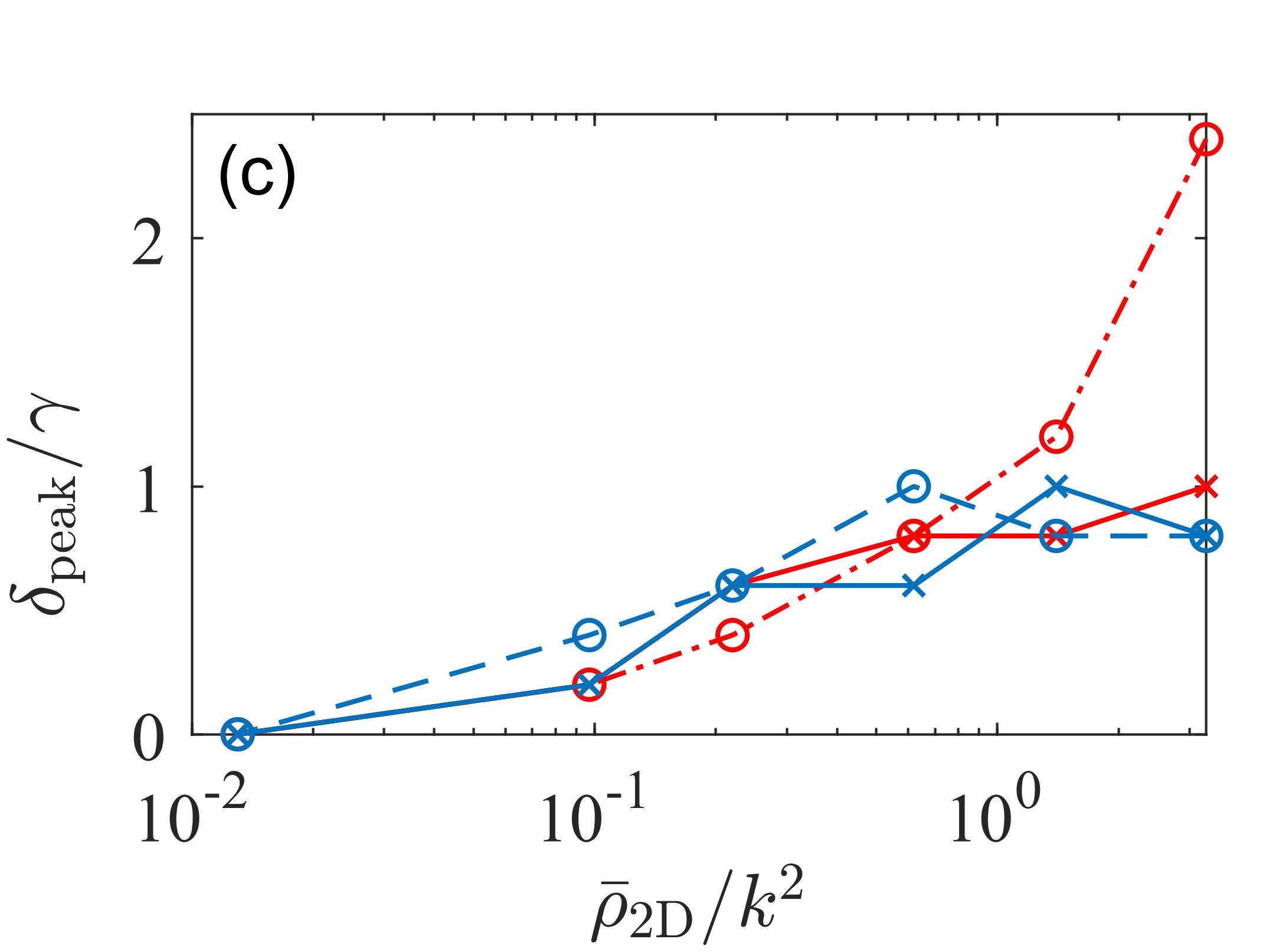}
   \includegraphics[width=0.4\columnwidth]{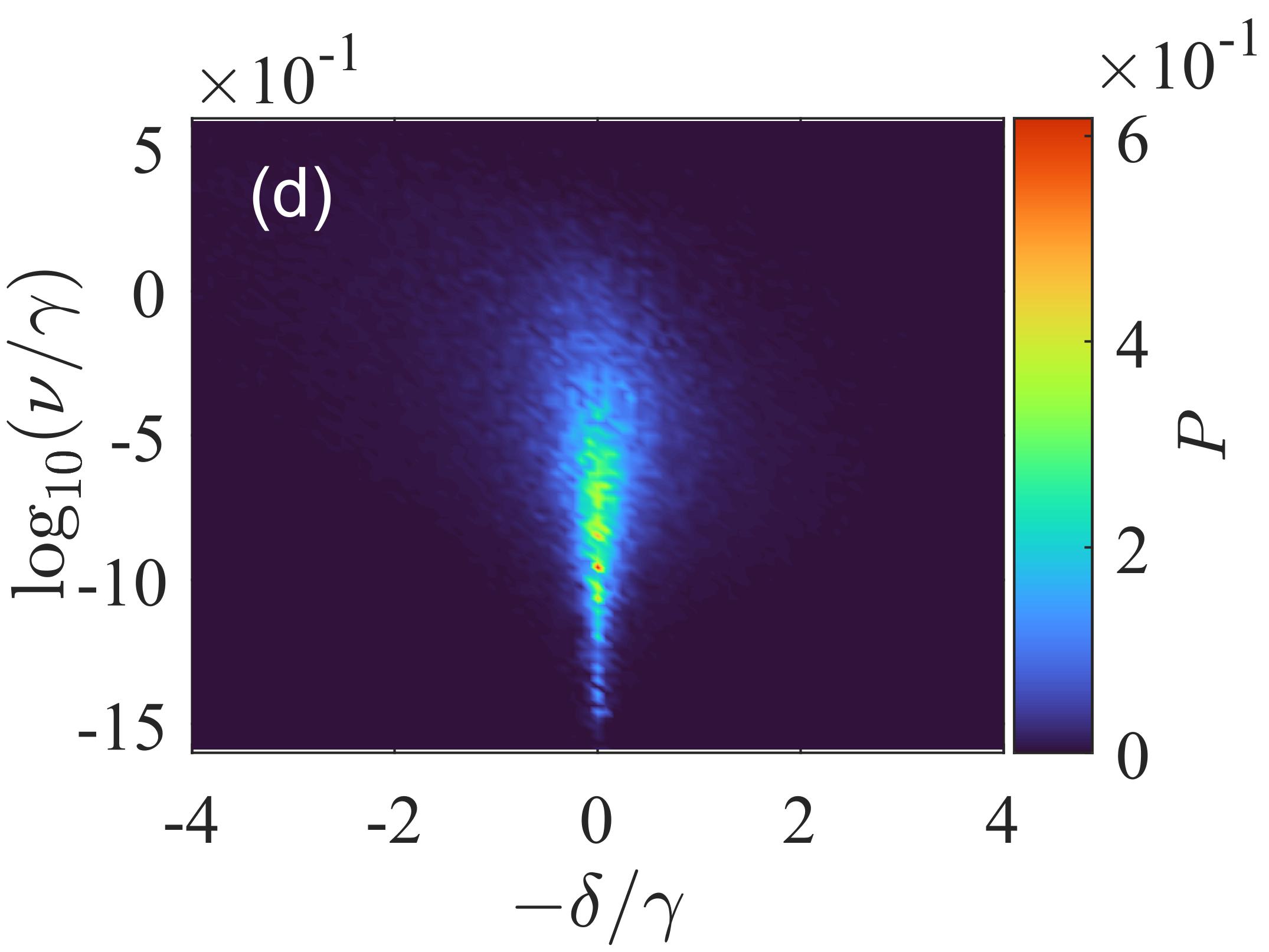}
   \includegraphics[width=0.4\columnwidth]{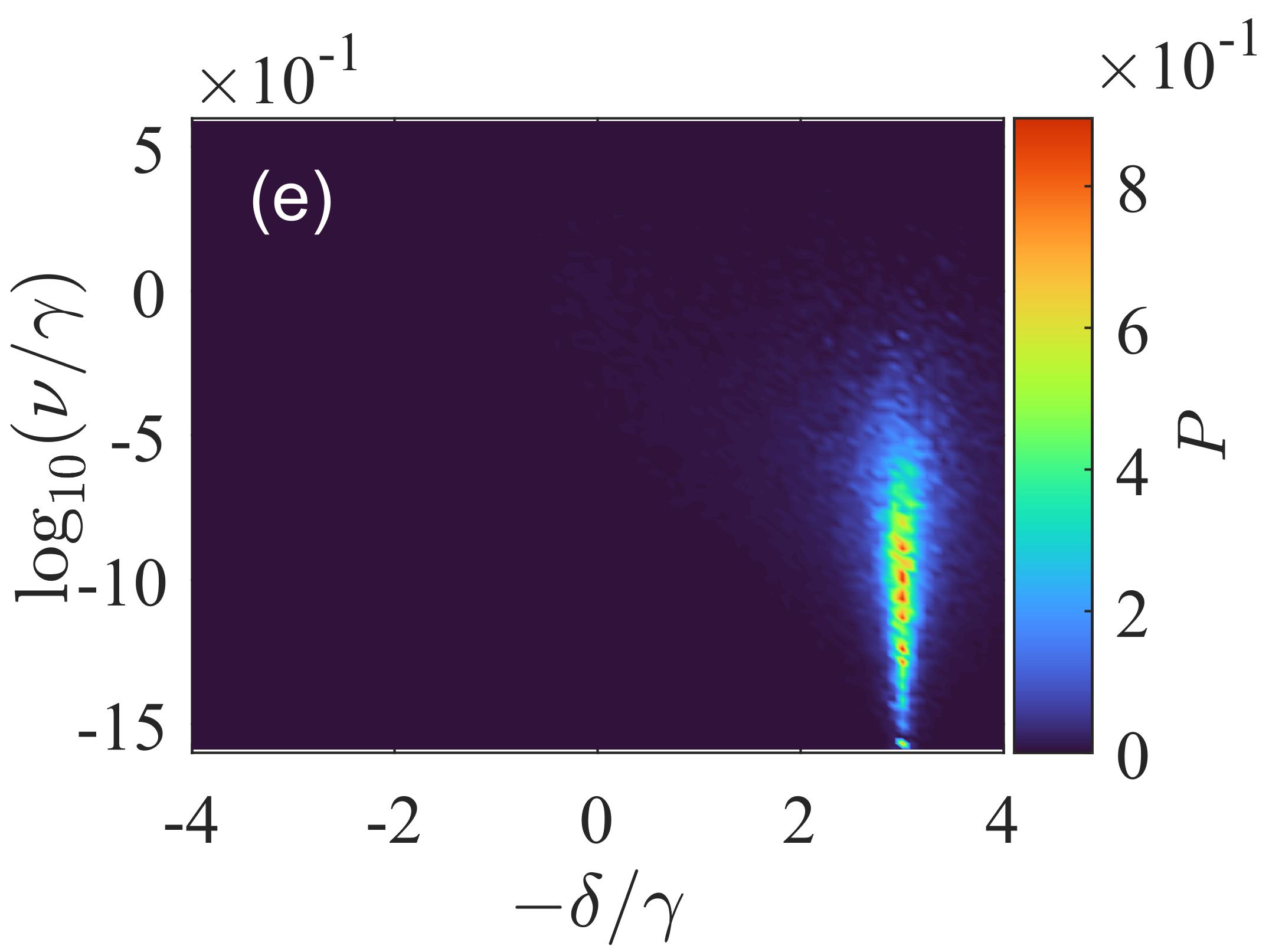}
   \caption{Variation of scattered light intensity extrema,  resonance width and shift  with peak density \inline{\bar\rho_{\rm{2D}} /k^2}, and eigenmode occupations  for the system shown in Fig.~\ref{fig: ConstRdipTransmissionWide}. (a) Peak coherent transmission \inline{\rm{min}(T_{\rm{coh}})} and reflection \inline{\max(R_{\rm{coh}})},  (c) HWHM, (d) shift. The independent-atom case (the static DD case) is represented by dashed (solid) lines. Eigenmode occupations for  \inline{\bar\rho_{\rm{2D}} /k^2\simeq3.3} (\inline{R_{\rm{dip}}\sqrt{\bar{\rho}_{\rm{2D}}} \simeq 0.18}) for (d)  $\Delta=0$, (e) $3\gamma$. The bin size \inline{[\Delta \log_{10} (\nu/\gamma), \Delta \delta/ \gamma] = [0.016,0.08]}.
   }
    \label{fig: ConstRdipWidthResonancePeakVariation}
\end{figure*}
Instead of fixing the peak atom density, we now study the optical response for fixed static DD interaction length in terms of the resonance wavenumber of light \inline{R_{\rm{dip}}k} [Eq.~\eqref{eq:Rdip}]. This allows to investigate how light transmission and reflection vary with the density by changing the trap frequency \inline{\ell_x k} for constant \inline{\ell_x/\ell_z=25} (Fig.~\ref{fig: ConstRdipTransmissionWide}). The setup is similar to the one studied in Sec.~\ref{sec:dipolarprolate} (see  Fig.~\ref{fig:trap}) and we consider the independent atoms \inline{R_{\rm{dip}}=0} and static dipoles with
\inline{R_{\rm{dip}}k\simeq0.1}. The independent-atom case behaves qualitatively similarly to the previous example where we only varied the atom number.
The lineshapes  at the peak densities \inline{\bar{\rho}_{\rm{2D}}/k^2 \simeq1.4} and 3.3 in Fig.~\ref{fig: ConstRdipTransmissionWide} are asymmetric in all the cases, displaying an increased optical depth for blue detuning and enhanced incoherent scattering for red detuning. The static interactions significantly increase the red-detuned incoherent scattering. 
The density-dependent resonance broadening  in Fig.~\ref{fig: ConstRdipWidthResonancePeakVariation} becomes more dramatic with the static DD interactions. 
The peak value of coherent reflection first significantly increases with density and then at high densities start decreasing again. 
This may be due to the eigenmode resonances becoming spectrally more distinguishable,
as illustrated for the eigenmode occupations in Fig.~\ref{fig: ConstRdipWidthResonancePeakVariation}. At high density \inline{\bar{\rho}_{\rm{2D}}/k^2\simeq3.3}, the occupation is prominent only around modes for which the laser frequency is resonant. There is little occupation of modes away from this resonance. This selectivity is responsible for the broad transmission resonances at high densities, 
as different eigenmodes can be excited at different frequencies and the mode resonances extend over a wide range of frequencies.

\section{Concluding remarks}
\label{sec:conc}

We successfully solved a challenging hierarchy of $N$ equations~\eqref{EXACT} for the correlation functions of $N$ atoms.  These equations encompass correlations among atoms in their electronic ground states as well as those involving both ground and excited states. This was made possible using stochastic electrodynamics simulations of coupled radiative dipole excitations where the positions of the dipoles are correlated by static repulsive DD interactions and sampled using quantum Monte Carlo methods. 
For the bosonic fluid-like states considered in this paper, ergodicity is not expected to pose a challenge when seeking the ground state using diffusion quantum Monte Carlo simulation, with the accuracy of the sampled distributions scaling linearly with the error in the trial wave function. As long as the sampling of positions is precise, the stochastic electrodynamics simulations of coherently scattered light converge to an exact solution for stationary, laser-driven atoms at arbitrary densities in the limit of LLI~\cite{Javanainen1999a,Lee16}.
The methodology can be extended to include also stronger static interactions and other strongly correlated ensembles. 

Our main findings showed how the repulsive static interactions lead to short-range ordering among the dipoles, which, in turn, curtails fluctuations in the light-induced resonant DD interactions (for an oblate trap in Sec.~\ref{sec:dipolaroblate} and in a prolate trap in Sec.~\ref{sec:dipolarprolate}). This phenomenon affects the resonance widths and shifts (Secs.~\ref{sec:dipolaroblate} and~\ref{sec:dipolarprolate}). Furthermore, it is identified in increased coherent reflection and optical depth that are accompanied by reduced incoherent scattering. The effects of the static interactions on the optical response can be analyzed in terms of the collective excitation eigenmodes (Secs.~\ref{sec:eigenmodes} and~\ref{sec:dipolarprolate}) and we found, e.g., that the presence of the static DD interactions enables much better targeted excitation of subradiant eigenmodes at high densities especially in a prolate trap, despite disordered atom distributions. For an isotropic level structure, the excited dipoles can exhibit non-negligible orientation even along the normal of an oblate trap (Sec.~\ref{sec:iso}). Intriguingly, the static interactions can also enhance the peak strengths of the disorder-induced excitation energy localization, providing control and manipulation of optical fields on a subwavelength scale (Sec.~\ref{sec:spots}). 

Optical responses in the presence of static DD interactions could be investigated, e.g., using atoms or polar molecules. The repulsive static DD interactions in a prolate and oblate trap can suppress inelastic losses in both systems even at high densities~\cite{Baranov12}.
Dy atoms have a large magnetic moment \inline{\mu\simeq 10\mu_B}, where $\mu_B$ is the Bohr magneton~\cite{Lu11}. For example, the 626nm transition has been used in magneto-optical trapping of $^{162}$Dy~\cite{Lucioni17}. For \inline{\bar\rho_{\rm{2D}} /k^2 \simeq 3.3}, this gives the average interatomic separation $0.09\lambda$ at the center of the trap and \inline{R_{\rm{dip}}\sqrt{\bar{\rho}_{\rm{2D}}} \simeq 0.4}, with  \inline{R_{\rm{dip}}\simeq 21}nm. 
If the static DD interactions strongly affect the optical transition frequencies, resonant emitters can experience them nonuniformly and become inhomogeneously broadened (Sec.~\ref{sec:Zeeman}). The strength of the level shifts in Eq.~\eqref{Eq: Zeeman Coupling} generally depends on the level structure, but similar estimates for Dy indicate a much weaker effect than any of the cases shown in Fig.~\ref{fig: Weak zeeman trans}, and therefore a negligible influence on the optical response.

For alkali-metal atoms the magnitude of the magnetic dipole $|\mu|\alt \mu_B$ and $R_{\rm dip}$ is more than two orders of magnitude smaller than for the case of Dy. The simulation results in Fig.~\ref{fig: TransmissionDen004} then indicate that the static magnetic DD interactions between the atoms could not contribute to observable effects, e.g., in the coherent light scattering experiments of Ref.~\cite{Jennewein_trans}. 

Heteronuclear polar molecules possess large electric dipole moments whose strength can be controlled by orienting the molecule by external electric field~\cite{Schmidt22}.
Also atomic Rydberg excitations under conditions of electromagnetically-induced transparency can be used to manipulate collective optical interactions~\cite{Pohl11,Mohl20} and the DD interaction between atoms in Rydberg states can be tuned to a weak-interaction regime~\cite{Kim21}. Rydberg transitions induce DD interactions resulting in level shifts that determine which atoms engage in resonant scattering of light.
Consequently, the positions of the atoms that are resonant with the incoming light can become correlated due to the Rydberg DD interactions.
Furthermore, in recent experiments,  it has been demonstrated that polar molecules can be utilized in effectively controlling atomic resonances~\cite{Guttridge23}.
Although the control in Ref.~\cite{Guttridge23} was obtained through Rydberg states, one could envisage a scenario in which an oblate trap of polar molecules exhibiting repulsive static DD interactions is placed on the top of a prolate atom trap. The interactions between these molecules and the atoms could potentially be harnessed to ascertain, through induced atomic level shifts, which atoms can engage in resonant interactions with light.
This selective process might ensure that only the atoms located atop the molecules are in resonance.  Consequently, the position correlations initially associated with the molecules would be transferred to the optically interacting atoms. 

\begin{acknowledgments}
We gratefully acknowledge Patrick Windpassinger for early engagement in the project and partial financial support of G.J.B\@. J.R.\ acknowledges support from EPSRC (Grant No.\ EP/S002952/1) and G.J.B.\  discussions with Marvin Proske and Ishan Varma. Computational resources were provided by Lancaster University's High-End Computing facility.
\end{acknowledgments}

\end{document}